\documentclass[a4paper, 11pt,notoc]{article}
\pdfoutput=1
\usepackage{jcappub}
\usepackage{graphicx}
\usepackage{booktabs}
\usepackage{verbatim}
\usepackage{caption}
\usepackage{xspace}
\usepackage{hyperref}
\usepackage{multirow}
\usepackage{placeins}
\usepackage{array}
\usepackage{subcaption}

\newcommand{\MET}{\ensuremath{E_T^\mathrm{miss}}\xspace}
\newcommand{\met}{\MET}

\newcommand{\mA}{\ensuremath{M_{A}}\xspace}
\newcommand{\ma}{\ensuremath{M_{a}}\xspace}
\newcommand{\mH}{\ensuremath{M_{H}}\xspace}
\newcommand{\mHc}{\ensuremath{M_{H^{\pm}}}\xspace}
\newcommand{\mh}{\ensuremath{M_{h}}\xspace}

\newcommand{\hdma}{\ensuremath{\textrm{2HDM+a}}\xspace}

\definecolor{cerulean}{RGB}{44,150,207}

\def\be   {\begin{equation}}   \def\ee   {\end{equation}}
\def\ba   {\begin{array}}      \def\ea   {\end{array}}
\def\bea  {\begin{eqnarray}}   \def\eea  {\end{eqnarray}}
\def\bean {\begin{eqnarray*}}  \def\eean {\end{eqnarray*}}

\allowdisplaybreaks

\RequirePackage[normalem]{ulem} %DIF PREAMBLE
\RequirePackage{color}\definecolor{RED}{rgb}{1,0,0}\definecolor{BLUE}{rgb}{0,0,1} %DIF PREAMBLE
\def\bm#1{\mbox{\boldmath$#1$\unboldmath}} 

\begin{document}
\title{\begin{boldmath} \huge LHC Dark Matter Working Group:  \\ Next-generation spin-0  dark matter models \vspace{7mm} \end{boldmath}}

%%%%%%

\author[1,2]{Tomohiro~Abe,}
\affiliation[1]{Institute for Advanced Research, Nagoya University, \\
Furo-cho Chikusa-ku, Nagoya, Aichi, 464-8602, Japan}
\affiliation[2]{Kobayashi-Maskawa Institute for the Origin of Particles and the Universe, \\
 Nagoya University, Furo-cho Chikusa-ku, Nagoya, Aichi, 464-8602, Japan}

\author[3]{Yoav Afik,} 
\affiliation[3]{Department of Physics, Technion: Israel Institute of Technology, Haifa, Israel}

\author[4]{Andreas~Albert,}
\affiliation[4]{III. Physikalisches Institut A, RWTH Aachen University, \\
Physikzentrum, Otto-Blumenthal-Stra{\ss}e, Aachen, Germany}

\author[5]{Christopher~R.~Anelli,}
\affiliation[5]{University of Victoria, Department of Physics and Astronomy, \\ 
Elliott Building, room 101, University of Victoria, Victoria, Canada}

\author[6]{Liron~Barak,}
\affiliation[6]{Tel Aviv University, Haim Levanon (Ramat Aviv), Tel Aviv 69978, Israel}

\author[7]{Martin~Bauer,}
\affiliation[7]{Institute for Particle Physics Phenomenology, Department of Physics, \\ 
Durham University, South Road, Durham DH1 3LE, UK}

\author[8]{J.~Katharina~Behr,}
\affiliation[8]{DESY, Notkestra{\ss}e 85, D-22607 Hamburg, Germany}

\author[9]{Nicole~F.~Bell,}
\affiliation[9]{ARC Centre of Excellence for Particle Physics at the Terascale \\
School of Physics, The University of Melbourne, Victoria 3010, Australia}

\author[10,*]{Antonio~Boveia,}
\affiliation[10]{Ohio State University and Center for Cosmology and Astroparticle Physics, \\
191 W. Woodruff Avenue Columbus, OH 43210, USA}

\author[11]{Oleg~Brandt,}
\affiliation[11]{Kirchhoff-Institut f{\"u}r Physik, Ruprecht-Karls-Universit{\"a}t Heidelberg, \\ 
Im Neuenheimer Feld 227, 69120 Heidelberg, Germany}

\author[9]{Giorgio~Busoni,}

\author[10]{Linda~M.~Carpenter,}

\author[8]{Yu-Heng~Chen,}

\author[12,*]{Caterina~Doglioni,}
\affiliation[12]{Fysiska institutionen, Lunds universitet, Professorsgatan 1, Lund, Sweden}

\author[13]{Alison~Elliot,}
\affiliation[13]{Department of Physics, Queen Mary University of London, \\ 
Mile End Rd, London E1 4NS, UK}

\author[14]{Motoko~Fujiwara,}
\affiliation[14]{Department of Physics, Nagoya University, Furo-cho Chikusa-ku, \\
Nagoya, Aichi, 464-8602, Japan}

\author[15]{Marie-Helene~Genest,}
\affiliation[15]{Univ. Grenoble Alpes, CNRS, Grenoble INP, LPSC-IN2P3, 38000 Grenoble, 
France}

\author[16]{Raffaele~Gerosa,}
\affiliation[16]{University California San Diego (UCSD), Department of Physics, \\
9500 Gilman Drive, La Jolla, CA 92093-0319, USA}

\author[17]{Stefania~Gori,}
\affiliation[17]{Santa Cruz Institute for Particle Physics, 1156 High Street, \\ 
Santa Cruz, CA 95064, USA}

\author[18]{Johanna~Gramling,}
\affiliation[18]{Department of Physics and Astronomy, University of California, \\ 
Irvine, California 92697, USA}

\author[8]{Alexander~Grohsjean,}

\author[19]{Giuliano~Gustavino,}
\affiliation[19]{University of Oklahoma, 440 W. Brooks St. Norman, OK 73019, USA}

\author[20,*]{Kristian~Hahn,}
\affiliation[20]{Department of Physics and Astronomy, Northwestern University, \\ 
Evanston, Illinois 60208, USA}

\author[21,22,23,*]{Ulrich~Haisch,}
\affiliation[21]{Max Planck Institute for Physics, F{\"o}hringer Ring 6,  80805 M{\"u}nchen, Germany}
\affiliation[22]{Rudolf Peierls Centre for Theoretical Physics, University of Oxford, \\ 
Oxford, OX1 3PN, UK}
\affiliation[23]{Theoretical Physics Department, CERN, CH-1211 Geneva 23, Switzerland}

\author[11]{Lars~Henkelmann,}

\author[2,14,24]{Junji~Hisano,}
\affiliation[24]{Kavli IPMU (WPI), UTIAS, University of Tokyo, Kashiwa, Chiba 277-8584, Japan}

\author[25]{Anders~Huitfeldt,}
\affiliation[25]{School of Physics, The University of Melbourne, Swanston St and Tin Alley, \\ 
Parkville, 3010, Victoria, Australia}

\author[26]{Valerio~Ippolito,} 
\affiliation[26]{Universit{\`a}  di Roma Sapienza and INFN, Piazza Aldo Moro, 2, 00185 Roma, Italy}

\author[27]{Felix~Kahlhoefer,}
\affiliation[27]{Institute for Theoretical Particle Physics and Cosmology (TTK), \\ 
RWTH Aachen University, D-52056 Aachen, Germany}

\author[28]{Greg~Landsberg,}
\affiliation[28]{Brown University, Dept. of Physics, 182 Hope St, Providence, RI 02912, USA}

\author[29]{Steven~Lowette,}
\affiliation[29]{Vrije Universiteit Brussel, Pleinlaan 2, 1050 Brussel, Belgium}

\author[30]{Benedikt~Maier,}
\affiliation[30]{Department of Physics, Massachusetts Institute of Technology, 77 Massachusetts Avenue, Cambridge, MA 02139-4307, USA}

\author[31]{Fabio~Maltoni,}
\affiliation[31]{Centre for Cosmology, Particle Physics and Phenomenology (CP3), \\ 
Universit{\'e} catholique de Louvain, B-1348 Louvain-la-Neuve, Belgium}

\author[32]{Margarete~Muehlleitner,}
\affiliation[32]{Institute for Theoretical Physics, Karlsruhe Institute of Technology, \\ Wolfgang-Gaede-Str. 1, 76131 Karlsruhe, Germany}

\author[33,34]{Jose~M.~No,}
\affiliation[33]{Departamento de Fisica Teorica and Instituto de Fisica Teorica, IFT-UAM/CSIC, \\ Universidad Autonoma de Madrid, Cantoblanco, 28049, Madrid, Spain}
\affiliation[34]{Department of Physics, King's College London, Strand, WC2R 2LS London, UK}

\author[8,35]{Priscilla~Pani,}
\affiliation[35]{DESY Zeuthen, Platanenallee 6, 15738 Zeuthen, Germany}

\author[36]{Giacomo~Polesello,}
\affiliation[36]{INFN, Sezione di Pavia, Via Bassi 6, 27100 Pavia, Italy}

\author[37]{Darren~D.~Price,}
\affiliation[37]{School of Physics and Astronomy, University of Manchester, Manchester M13 9PL, UK}

\author[38,39]{Tania~Robens,}
\affiliation[38]{MTA-DE Particle Physics Research Group, University of Debrecen, \\
4010 Debrecen, Hungary}
\affiliation[39]{Theoretical Physics Division, Rudjer Boskovic Institute, 10002 Zagreb, 
Croatia}

\author[40]{Giulia~Rovelli,}
\affiliation[40]{INFN Sezione di Pavia and Dipartimento di Fisica, Universit{\`a} di Pavia, Pavia, Italy.}

\author[3]{Yoram~Rozen,}

\author[9]{Isaac~W.~Sanderson,}

\author[41,42]{Rui~Santos,}
\affiliation[41]{ISEL - Instituto Superior de Engenharia de Lisboa, \\ 
Instituto Polit\'ecnico de Lisboa, 1959-007 Lisboa, Portugal}
\affiliation[42]{Centro de F\'{\i}sica Te\'{o}rica e Computacional, Faculdade de Ci\^{e}ncias, \\ 
Universidade de Lisboa, Campo Grande, Edif\'{\i}cio C8, 1749-016 Lisboa, Portugal}

\author[43]{Stanislava~Sevova,}
\affiliation[43]{Stanford University, 450 Serra Mall, Stanford, CA 94305, USA}

\author[44]{David~Sperka,}
\affiliation[44]{Boston University, 590 Commonwealth Avenue, 02215 Boston, MA, USA}

\author[20]{Kevin~Sung,}

\author[17,*]{Tim~M.~P.~Tait,}

\author[45]{Koji Terashi,}
\affiliation[45]{International Center for Elementary Particle Physics (ICEPP), School of Science, \\ 
The University of Tokyo 7-3-1 Hongo, Bunkyo-ku, Tokyo 113-0033, Japan}

\author[9]{Francesca~C.~Ungaro,}

\author[23]{Eleni~Vryonidou,}

\author[46]{Shin-Shan~Yu,}
\affiliation[46]{Department of Physics, National Central University No. 300, Zhongda Rd., \\ 
Zhongli District, Taoyuan City 32001, Taiwan} 

\author[47]{Sau~Lan~Wu,}
\affiliation[47]{Department of Physics, University of Wisconsin, Thomas C Chamberlin Hall, \\
1150 University Ave 2320, Madison, WI 53706, USA} 

\author[47]{and~Chen~Zhou.}

\affiliation[*]{DMWG organisers}

\hfill CERN-LPCC-2018-02

\abstract{Dark matter (DM) simplified models are by now commonly used by  the ATLAS and CMS Collaborations to interpret searches for missing transverse energy~($\MET$). The coherent use of these models sharpened the LHC DM search program, especially in the presentation of its results and their comparison to DM direct-detection (DD) and indirect-detection (ID) experiments.  However, the community has been aware of the limitations of the DM simplified models, in particular the lack of theoretical consistency of some of them and their restricted phenomenology leading to the relevance of only a small subset of~$\MET$ signatures. This document from the LHC Dark Matter Working Group identifies an example of a next-generation DM model, called \hdma, that provides the simplest theoretically consistent extension of the DM pseudoscalar simplified model. A~comprehensive study of the phenomenology of the \hdma model is presented, including a discussion of the rich and intricate pattern of mono-$X$ signatures and the relevance of other DM as well as non-DM experiments. Based on our discussions, a set of recommended scans are proposed to explore the parameter space of the $\hdma$ model through LHC searches. The exclusion limits obtained from the proposed  scans can  be consistently compared to the constraints on the  $\hdma$ model that derive from DD, ID and the DM relic density.}  

\maketitle

\newpage 

%%%%%%%%%%%%%%%%%%%%%%%%%%%%%%%%%%%%%%%%%%%%%%%%%%%%%%%%%%%%%%%%%%%%
%%%%%%%%%%%%%%%%%%%%%%%%%%%%%%%%%%%%%%%%%%%%%%%%%%%%%%%%%%%%%%%%%%%%
%%%%%%%%%%%%%%%%%%%%%%%%%%%%%%%%%%%%%%%%%%%%%%%%%%%%%%%%%%%%%%%%%%%%

\section{Introduction}
\label{sec:introduction}

Dark matter (DM) is one of the main search targets for LHC experiments (see for example~\cite{Kahlhoefer:2017dnp} for a recent review). Based on the assumption that DM is a weakly interacting massive particle~\cite{Bertone:2004pz}, the ATLAS and CMS Collaborations have searched for DM candidates manifesting as particles that escape the detectors, creating a sizeable transverse momentum imbalance ($\MET$). Therefore, the minimal experimental signature of DM production at a hadron collider consists  in an excess of events with a visible final-state object $X$ recoiling against the $\MET$, a so-called mono-$X$ signal.  The design of experimental searches for invisible particles can generally be kept independent from specific theoretical models, reflecting the lack of hints on the exact particle nature of DM. However, theoretical benchmarks are necessary to sharpen the regions of parameter space to which searches need to be optimised, to characterise a possible discovery and to define a theoretical framework for comparison with non-collider results. 

Originally, supersymmetry was the main theoretical framework used as a benchmark for many DM searches at the LHC.  Non-supersymmetric interpretations of the various~$\MET$ searches have  developed with time. At the start of data taking, DM effective field theories~(DM-EFTs)  were used due to their relative model independence~\cite{Cao:2009uw,Beltran:2010ww,Goodman:2010yf,Bai:2010hh,Goodman:2010ku,Fox:2011pm}.  DM simplified models, each representing a credible unit within a more complicated model and encapsulating the phenomenology of LHC DM interactions using a small set of parameters, provide more handles to study interactions when the momentum transfer of the collision is sufficient to probe the energy scale of a mediator particle. Further developments towards DM simplified models occurred before the start of the second LHC run~\cite{Abdallah:2015ter,Abercrombie:2015wmb}.  The coherent adoption of these DM simplified models by the LHC Collaborations focused the LHC DM search program, especially in the presentation of its results and their comparison to DM direct-detection (DD) and indirect-detection (ID) experiments~\cite{Boveia:2016mrp,Albert:2017onk}.  Throughout this time, the community has been aware of the shortcomings in DM simplified models, in particular the lack of theoretical consistency of some of them~\cite{Chala:2015ama,Bell:2015sza,Kahlhoefer:2015bea,Bell:2015rdw,Haisch:2016usn,Englert:2016joy,Ko:2016zxg} and their limited phenomenology leading to the relevance of only a small set of experimental signatures.  

With this white paper, we take a step beyond the proposed DM simplified models by identifying an example benchmark model and its parameters to be tested by LHC searches, with the following characteristics: 
\begin{itemize}
\item[(I)] the model should preferably be a theoretically consistent extension of one of the DM simplified models already used by the LHC Collaborations;
\item[(II)] the model should still be generic enough to be used in the context of broader, more complete theoretical frameworks;  
\item[(III)] the model should have a sufficiently varied phenomenology to encourage comparison of different experimental signals and to search for DM in new, unexplored channels;
\item[(IV)] the model should be of interest beyond the DM community, to the point that other direct and indirect constraints can be identified.
\end{itemize}

One of the models that meets these characteristics and is explored in this white paper,  referred to \hdma in what follows,  is a two-Higgs-doublet model (2HDM) containing an additional pseudoscalar boson which mediates the interactions between the visible and the dark sector.  The \hdma model is the simplest gauge-invariant and renormalisable extension of the  simplified pseudoscalar model recommended by the ATLAS/CMS DM Forum (DMF)~\cite{Abercrombie:2015wmb}. It includes a  DM candidate which is a singlet under the Standard Model (SM) gauge group~\cite{Ipek:2014gua,No:2015xqa,Goncalves:2016iyg,Bauer:2017ota,Tunney:2017yfp}.  Since the DD constraints are weaker for models with pseudoscalar mediators compared to models with scalar mediators, the observed DM relic abundance can be reproduced in large regions of parameter space, making LHC searches particularly relevant to test the \hdma or other pseudoscalar DM models.

In order to motivate the introduction of the \hdma model, we describe in Section~\ref{sec:evolution} the evolution of theories for LHC DM searches, focusing on the relevant case of pseudoscalar SM-DM interactions. A detailed description of the \hdma model and its parameters can be found in~Section~\ref{sec:modeldescription}. The constraints on the model parameters that arise from Higgs and flavour physics, LHC searches for additional spin-0 bosons, electroweak~(EW) precision measurements and vacuum stability considerations are summarised in Section~\ref{sec:constraints}. This section also provides guidance on the choice of benchmark parameters to be used by LHC searches. Section~\ref{sec:comparison} is dedicated to a short summary of other DM models that feature a 2HDM sector. 

The more phenomenological part of this work commences with Section~\ref{sec:experimentbasics}, where we describe the basic features of the most important mono-$X$ channels and identify the experimental observables that can be exploited to search for them. We discuss both resonant and non-resonant $\MET$ signatures, emphasising that only the latter type of signals is present in the DMF pseudoscalar model. The most important non-$\MET$ signatures that can be used to explore the \hdma parameter space  are examined in Section~\ref{sec:nonMET}. In Section~\ref{sec:sensitivitystudies} we then estimate the current experimental sensitivities  in the mono-Higgs and mono-$Z$ channel, which represent two of the most sensitive $\MET$ signatures for the \hdma model. The constraints  set on the parameter space of the \hdma model from DD and ID experiments, as well as its DM relic density, are summarised in Section~\ref{sec:DMdetection} and Section~\ref{sec:relic}, respectively. In Section~\ref{sec:scans} we conclude by proposing four parameter scans that highlight many of the features that are special in the \hdma model and showcase the complementarity of the various search strategies.  Additional material can be found in the Appendices~\ref{app:recast}, \ref{app:ttMETscalar}, \ref{app:mcgeneration} and~\ref{app:extramonoh}.

%%%%%%%%%%%%%%%%%%%%%%%%%%%%%%%%%%%%%%%%%%%%%%%%%%%%%%%%%%%%%%%%%%%%
%%%%%%%%%%%%%%%%%%%%%%%%%%%%%%%%%%%%%%%%%%%%%%%%%%%%%%%%%%%%%%%%%%%%
%%%%%%%%%%%%%%%%%%%%%%%%%%%%%%%%%%%%%%%%%%%%%%%%%%%%%%%%%%%%%%%%%%%%

\section{Evolution of theories for LHC DM searches} 
\label{sec:evolution}

The experimental results from DD and ID experiments are usually interpreted in the DM-EFT framework. The operators in these DM-EFTs are built from SM fermions and DM fields. Schematically, one has in the case of spin-0 interactions and Dirac fermion DM
\begin{align}\label{eq:EFT}
\mathcal{L}_\text{DM-EFT}= \sum_{f=u,d,s,c,b,t,e,\mu,\tau} \,\left(\frac{C_{1}^f}{\Lambda^2} \bar f f \bar \chi \chi  +\frac{C_{2}^f}{\Lambda^2} \bar  f \gamma_5 f \bar \chi\gamma_5 \chi \,+\ldots \right) \,, 
\end{align}
 where the ellipsis represents additional operators not relevant for the further discussion, the sum over $f=u,d,s,c,b,t,e,\mu,\tau$ includes all SM quarks and charged leptons, the DM candidate is called $\chi$  and $\gamma_5$ denotes the fifth Dirac matrix. The above DM-EFT is fully described by the parameters
\begin{align}\label{eq:EFTparams}
\big\{ m_\chi\,,\,\, C_n^f/\Lambda^2 \big\} \,.
\end{align}
Here $m_\chi$ is the mass of the DM candidate, $\Lambda$ is the suppression scale of the higher-dimensional operators and the $C_n^f$ are the so-called Wilson coefficients. It is important to note that $\Lambda$ and $C_n^f$ are not independent parameters but always appear in the specific combination given in~\eqref{eq:EFTparams}. 

\begin{figure}[t!]
\centering
\includegraphics[width=.85\textwidth]{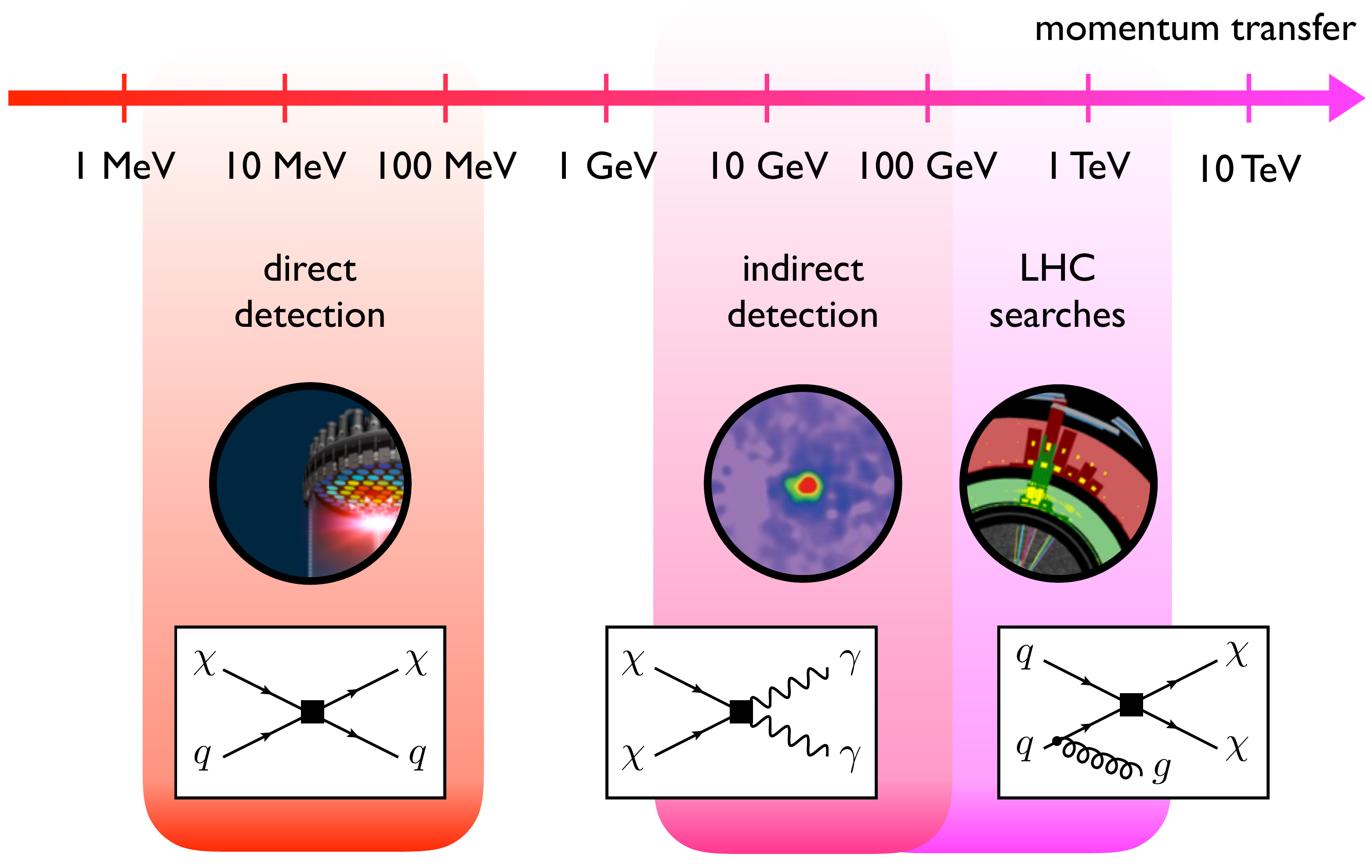}
\vspace{2mm}
\caption{\label{fig:momentumtransfer} Range of momenta probed in DD experiments, ID experiments and LHC searches. Prototypes of relevant Feynman diagrams are also shown. }
\end{figure}

The DM-EFT approach is justified for the small momentum transfer $q^2\ll \Lambda^2$ in DM-nucleon scattering (set by the non-relativistic velocities of DM in the halo) and in DM annihilation (set by the mass of the annihilating DM candidate). Figure~\ref{fig:momentumtransfer}  illustrates the relevant energy scales explored by DD, ID and collider experiments. Early studies~\cite{Cao:2009uw,Beltran:2010ww,Goodman:2010yf,Bai:2010hh,Goodman:2010ku,Fox:2011pm} of DM searches at colliders quantify the reach of the LHC in the parameter space in terms of~\eqref{eq:EFTparams} and similar operators. The momentum transfer at the LHC is however larger than the suppression scale,~i.e.~$q^2 \gg \Lambda^2$, for many theories of DM.  In this case, the mediator of the interaction between the dark sector and the SM can be resonantly produced and predictions  obtained using the DM-EFT framework often turn out to be inaccurate~(see for instance~\cite{Bai:2010hh,Fox:2011fx,Shoemaker:2011vi,Busoni:2013lha,Buchmueller:2013dya,Busoni:2014sya,Busoni:2014haa,Racco:2015dxa} and~\cite{Bruggisser:2016nzw,Bruggisser:2016ixa} for exceptions). 

The kinematics of on-shell propagators can be captured in DM simplified models, which aim to represent a large number of possible extensions of the SM, while keeping only the degrees of freedom relevant for LHC phenomenology~\cite{Abdallah:2015ter,Abercrombie:2015wmb}. In the case of a pseudoscalar mediator $a$, the relevant DM-mediator and SM-mediator interactions read
\begin{align}\label{eq:simp}
\mathcal{L}_\text{DM-simp}=-i g_\chi a\bar \chi \gamma_5 \chi -i a \sum_j \left(g_u y_j^u \bar u_j \gamma_5 u_j + g_d y_j^d \bar d_j \gamma_5 d_j + g_\ell y_j^\ell \bar \ell_j\gamma_5 \ell_j  \right) \,,
\end{align}
with  $j$ representing a flavour index.  Since the mediator $a$ is a singlet, it can also couple to itself and to $H^\dagger H$, where $H$ denotes  the SM Higgs doublet. The most general renormalisable scalar potential for a massive $a$ is therefore
\begin{align}\label{eq:VaH}
V_\text{DM-simp} =\frac{1}{2}m_a^2 a^2 +  b_a a^3 + \lambda_a a^4 + b_{H} a H^\dagger H +\lambda_{H} a^2H^\dagger H \,.
\end{align}
The parameters $ b_{H}$  and $\lambda_{H}$ determine the couplings between the $a$ and the $H$ fields, thereby altering  the interactions  of the SM-like scalar $h$ at $125 \, {\rm GeV}$ as well as giving rise to possible new decay channels such as $h \to aa$ (see~\cite{Curtin:2013fra,Haisch:2018kqx} for details on the LHC phenomenology). Avoiding the resulting strong constraints for any choice of~$m_a$, requires that  $b_H \ll m_a $ and $\lambda_H \ll 1$. While the former requirement can be satisfied by imposing a $Z_2$ symmetry $a \to -a$, in the latter case one has to assume that $\lambda_H$ is accidentally small if $m_a \lesssim 100 \, {\rm GeV}$ (cf.~the related discussion on invisible decays of the Higgs boson in Section~\ref{sec:invisiblehiggs}). Under such an assumption and noting that the self-couplings $b_a$ and $\lambda_a$ are largely irrelevant for collider phenomenology, the DM simplified model is  fully described by the parameters 
\begin{align}
\big\{ m_\chi, \,\, m_a\,,\,\, g_\chi\,, \,\, g_u\,,\,\,g_d\,,\,\, g_\ell \big\}\,. 
\end{align}
In fact, in the  limit of infinite mediator mass $m_a \to \infty$,  the DM-simp Lagrangian~\eqref{eq:simp} matches onto the DM-EFT Lagrangian~\eqref{eq:EFT}. The corresponding tree-level matching conditions are $C^f_2/\Lambda^2 = g_\chi g_f y_f  /m_a^2$ and $C_n^f=0$ for all other Wilson coefficients. Here $y_f$~denotes the Yukawa couplings of the fermions $f$ entering~\eqref{eq:simp}.

\begin{figure}[t!]
\centering
\includegraphics[width=.75\textwidth]{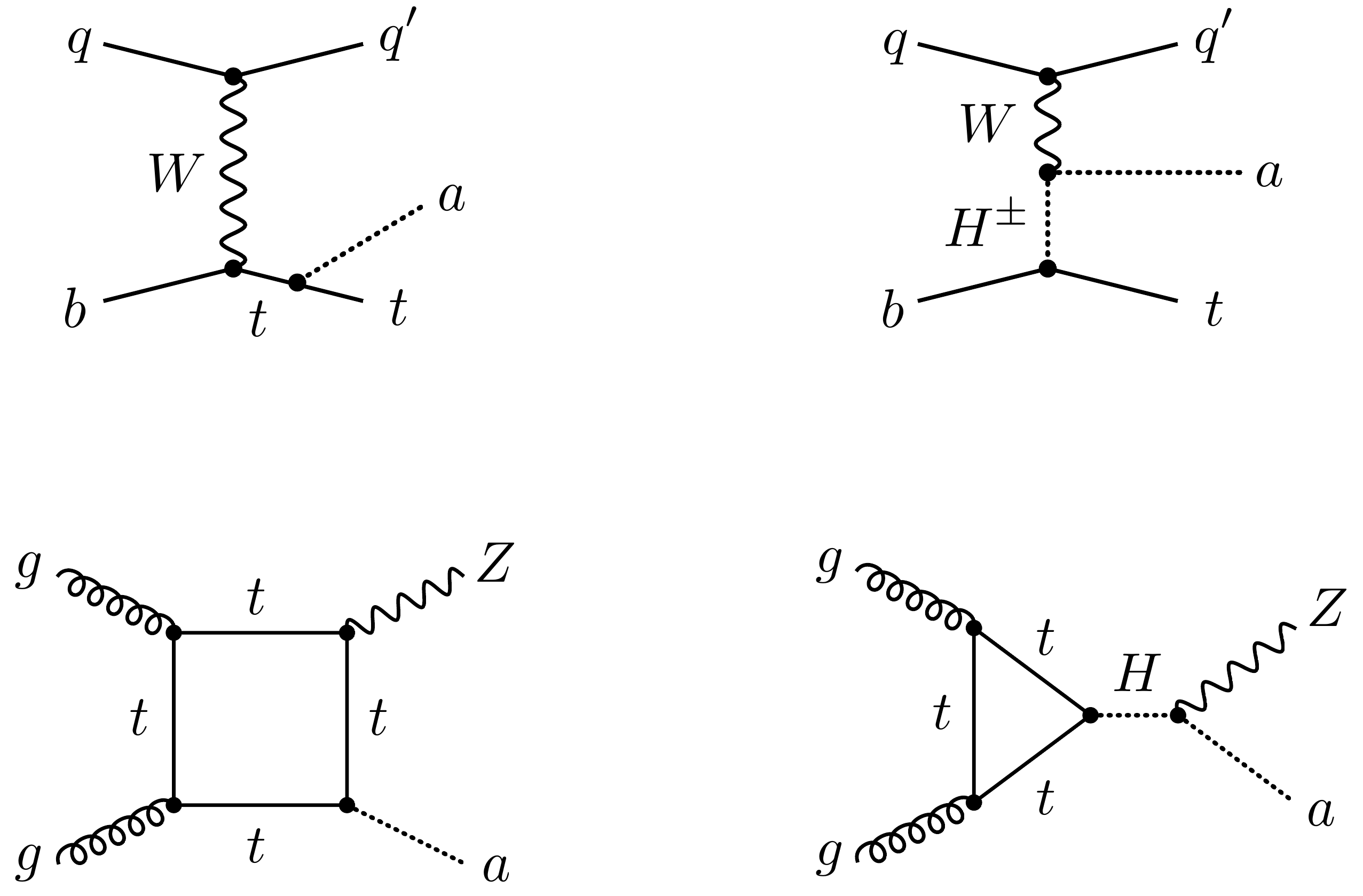}
\vspace{6mm}
\caption{\label{fig:diagrams}  Diagrams contributing to the $q b \to q^\prime t a$~(upper row) and $gg \to Za$~(lower row) scattering processes.  Only the graphs on the left-hand side appear in the DM simplified model with a pseudoscalar, while in the \hdma model in addition the diagrams on the right-hand side are present. See text for further details.}
\end{figure}

Unfortunately, the operators in both $\mathcal{L}_\text{DM-EFT}$ and $\mathcal{L}_\text{DM-simp}$ violate gauge invariance, because the left- and right-handed SM fermions belong to different representations of the SM gauge group. In the case of the DM-EFT this suggests the Wilson coefficients~$C_n^f$ introduced in~\eqref{eq:EFT} actually scale as $C_n^f = c_n^f m_{f_i}/\Lambda$~\cite{Bell:2015sza}, whereas for the DM simplified model restoring gauge invariance requires the embedding of the mediator $a$ into an EW multiplet. The absence of gauge invariance leads to unitarity-violating amplitudes in DM simplified models~(cf.~\cite{Bell:2015sza,Bell:2015rdw,Haisch:2016usn,Englert:2016joy,Maltoni:2001hu,Farina:2012xp}). In the case of the DM simplified model described by~\eqref{eq:simp}, one can show~e.g.~that the amplitudes ${\cal A} ( q b \to q^\prime t a) \propto \sqrt{s}$ and ${\cal A} ( g g \to Z a ) \propto \ln^2 s$ diverge in the limit of large center-of-mass energy $\sqrt{s}$. The Feynman diagrams that lead to this behaviour are depicted on the left-hand side in Figure~\ref{fig:diagrams}. Similar singularities appear in other single-top processes and in the mono-Higgs case.  Since the divergences are not power-like, weakly-coupled realisations of~\eqref{eq:simp} do not break down for the energies accessible at the LHC. The appearance of the $\sqrt{s}$ and $\ln^2 s$ terms, however, indicates the omission of diagrams that would be present in any gauge-invariant extension that can be approximated by $\mathcal{L}_\text{DM-EFT}$ in the limit where all additional particles $X$ are heavy (i.e.~$M_X \gg \sqrt{s}$).   For example, the $pp \to tj a$ cross section is made finite by the exchange of a charged Higgs~$H^\pm$, while in the case of $pp \to Za$  an additional scalar~$H$ unitarises the amplitude. The corresponding diagrams are displayed on the right in Figure~\ref{fig:diagrams}. The cancellation of unitarity-violating terms among the diagrams of the latter figure is not at all accidental, but a direct consequence of the local gauge invariance of the underlying model.

The additional degrees of freedom necessary to unitarise the amplitudes may change substantially the phenomenology of the DM simplified model. In fact, as shown by Figure~\ref{fig:diagrams}, the presence of the $H^\pm$ ($H$) allows to produce a mono-top (mono-$Z$)  signal resonantly. Since resonant production is strongly enhanced compared to initial-state radiation (ISR), the  importance of the various mono-$X$ signals in the extended DM model may then differ from the simplified model predictions~\cite{Goncalves:2016iyg,Bauer:2017ota,Pani:2017qyd}. In fact, we will see that in a specific extension of~\eqref{eq:simp}  called \hdma model, the mono-Higgs, mono-$Z$ and $t X + \MET$ signals can be as or even more important than the $t \bar t + \MET$ and mono-jet channel, which are  the leading $\MET$ signatures in the DM simplified pseudoscalar model~\cite{Haisch:2012kf,Fox:2012ru,Buckley:2014fba,Harris:2014hga,Haisch:2015ioa,Mattelaer:2015haa,Backovic:2015soa,Neubert:2015fka,Arina:2016cqj}. We emphasise that the embedding of~\eqref{eq:simp} is not unique, since  both the mediator and the DM particle can belong to different EW multiplets. In this white paper, we consider the simplest embedding with a single SM-singlet DM candidate,  but we will briefly comment on other possible embeddings and related DM models in Section~\ref{sec:comparison}.  

%%%%%%%%%%%%%%%%%%%%%%%%%%%%%%%%%%%%%%%%%%%%%%%%%%%%%%%%%%%%%%%%%%%%
%%%%%%%%%%%%%%%%%%%%%%%%%%%%%%%%%%%%%%%%%%%%%%%%%%%%%%%%%%%%%%%%%%%%
%%%%%%%%%%%%%%%%%%%%%%%%%%%%%%%%%%%%%%%%%%%%%%%%%%%%%%%%%%%%%%%%%%%%

\section{Description of the \hdma model}
\label{sec:modeldescription}

The \hdma model is a 2HDM that contains, besides the Higgs doublets $H_1$ and $H_2$, an additional pseudoscalar singlet $P$. It is the simplest renormalisable extension of~\eqref{eq:simp} with an SM-singlet DM candidate~\cite{Ipek:2014gua,No:2015xqa,Goncalves:2016iyg,Bauer:2017ota,Tunney:2017yfp}. The gauge symmetry is made manifest by coupling the $P$ to the dark Dirac fermion  $\chi$ via
\begin{equation} \label{eq:Lx}
{\cal L}_\chi = - i y_\chi P \hspace{0.25mm} \bar \chi \hspace{0.25mm} \gamma_5 \hspace{0.1mm} \chi \,,
\end{equation}
while the Higgs doublets couple to the SM fermions through
\begin{equation} \label{eq:LY}
{\cal L}_{Y} = - \sum_{i=1,2} \left ( \bar Q Y_u^i \tilde H_i u_R  + \bar Q Y_d^i H_i d_R   + \bar L Y_\ell^i H_i \ell_R  + {\rm h.c.}  \right ) \,.
\end{equation}
Here $y_\chi$ is a dark-sector Yukawa coupling, $Y_f^i$ are Yukawa matrices acting on the three fermion generations (where indices concerning the flavour of the fermion are suppressed), $Q$ and $L$ are left-handed quark and lepton doublets, while $u_R$, $d_R$ and $\ell_R$ are right-handed up-type quark, down-type quark and charged lepton singlets, respectively. Finally, $\tilde H_i = \epsilon H_i^\ast$ with $\epsilon$ denoting the  two-dimensional antisymmetric tensor.

The particle that mediates the interactions between the dark sector and the SM is a superposition of the CP-odd components of $H_1$, $H_2$ and $P$. We impose a $Z_2$ symmetry under which $H_1\to H_1$ and $H_2\to -H_2$, such that only one Higgs doublet couples to a certain fermion in ${\cal L}_{Y}$. The different ways to construct these terms result in different Yukawa~structures and in this white paper we will, for concreteness, consider only the so-called type-II~2HDM. This specific choice corresponds to setting $Y_u^1  = Y_d^2 = Y_\ell^2 =0$ in~\eqref{eq:LY} --- see~for example~Section~2.2~of~\cite{Bauer:2017ota} for further explanations.  The $Z_2$ symmetry is the minimal condition necessary to guarantee the absence of flavour-changing neutral currents at tree level~\cite{Glashow:1976nt,Paschos:1976ay} and such a symmetry is realised at in many well-motivated complete ultraviolet~(UV) theories in the form of supersymmetry, a $U(1)$ symmetry or  a discrete symmetry acting on the Higgs doublets. The fields $P$ and $\chi$ are $Z_2$-even and $Z_2$-odd, respectively,~i.e.~they transform as $P \to P$ and $\chi \to -\chi$. For these choices, the coupling introduced in~\eqref{eq:Lx} is the only DM Yukawa coupling that is allowed by symmetry, since  a term of the form $\bar L  \tilde H_1 \chi_R + {\rm h.c.}$ is forbidden. 

In addition, all parameters in the scalar potential are chosen to be real, such that CP eigenstates are identified with the mass eigenstates,~i.e.~two scalars $h$ and $H$, two pseudoscalars $A$ and $a$ and a charged scalar~$H^\pm$. Under these conditions, the most general renormalisable scalar potential can be written as 
\begin{align} \label{eq:V2HDMa}
V=V_H+V_{HP}+V_P\,,
\end{align}
with the potential for the two Higgs doublets
\begin{equation}\label{eq:VH}
\begin{split}
V_{H} & = \mu_1 H_1^\dagger H_1 + \mu_2 H_2^\dagger H_2 + \left ( \mu_3  H_1^\dagger H_2 + {\rm h.c.} \right ) + \lambda_1  \hspace{0.25mm} \big ( H_1^\dagger H_1  \big )^2  + \lambda_2  \hspace{0.25mm} \big ( H_2^\dagger H_2 \big  )^2  \\
& \phantom{xx} +  \lambda_3 \hspace{0.25mm} \big ( H_1^\dagger H_1  \big ) \big ( H_2^\dagger H_2  \big ) + \lambda_4  \hspace{0.25mm} \big ( H_1^\dagger H_2  \big ) \big ( H_2^\dagger H_1  \big ) + \left [ \lambda_5   \hspace{0.25mm} \big ( H_1^\dagger H_2 \big )^2 + {\rm h.c.} \right ]  \,,
\end{split}
\end{equation}
where the terms $\mu_3  H_1^\dagger H_2 + {\rm h.c.}$  softly break the  $Z_2$ symmetry. The potential terms which connect doublets and singlets are 
\begin{equation} \label{eq:VHP}
\begin{split}
V_{HP}  = P \left ( i  b_P   H_1^\dagger H_2 + {\rm h.c.} \right ) + P^2 \left (  \lambda_{P1}  H_1^\dagger H_1 +   \lambda_{P2}    H_2^\dagger H_2 \right )  \,,
\end{split} 
\end{equation}
where the first term breaks the  $Z_2$ symmetry softly.  The singlet potential is given by 
\begin{equation} \label{eq:VP}
V_{P}  =  \frac{1}{2}  m_P^2  P^2  \,.
\end{equation}
Notice  that compared to~\cite{Ipek:2014gua,No:2015xqa,Goncalves:2016iyg,Tunney:2017yfp}, which include only the trilinear portal coupling $b_P$, we  follow~\cite{Bauer:2017ota} and also allow for quartic portal interactions proportional to $\lambda_{P1}$ and  $\lambda_{P2}$. A~quartic self-coupling $P^4$ has not been included in~\eqref{eq:VP}, because such a term would not lead to any relevant effect in the $\MET$ observables studied in this white paper.

Upon rotation to the mass eigenbasis, we trade the five dimensionful and the eight dimensionless parameters in the potential for physical masses, mixing angles and four quartic couplings:
\begin{align} \label{eq:inputparameters}
\left\{ \,\,\begin{matrix}
\mu_1,\,\mu_2,\,\mu_3,\,b_P,\,m_P,\,m_\chi\\[3pt]
y_\chi,\,\lambda_1,\,\lambda_2,\,\lambda_3,\,\lambda_4,\,\lambda_5,\\
\lambda_{P1},\,\lambda_{P2}
\end{matrix}\,\,\right\}\quad  \longleftrightarrow  \quad \left\{ \,\,\begin{matrix}
v,\,M_h,\,M_A,\,M_H,\,M_{H^\pm},\,M_a,\,m_\chi \\[3pt]
\cos(\beta-\alpha),\,\tan \beta,\,\sin  \theta,\\[3pt]
y_\chi,\,\lambda_3,\,\lambda_{P1},\,\lambda_{P_2}
\end{matrix}\,\,\right\}\,.
\end{align}
Here $\alpha$ denotes the mixing angle between the two CP-even weak spin-0 eigenstates, $\tan \beta$ is the ratio of the vacuum expectation values~(VEVs) of the two Higgs doublets and $\theta$~represents the mixing angle of the two CP-odd weak spin-0 eigenstates. The parameters shown on the right-hand side of~\eqref{eq:inputparameters} will be used as input in the following sections. Out of these  parameters, the EW VEV $v \simeq 246 \, {\rm GeV}$ and the mass of the SM-like CP-even mass eigenstate $M_h \simeq 125 \, {\rm GeV}$ are already fixed by observations. The experimental and theoretical constraints on the remaining parameter space will be examined in the next section. 

%%%%%%%%%%%%%%%%%%%%%%%%%%%%%%%%%%%%%%%%%%%%%%%%%%%%%%%%%%%%%%%%%%%%
%%%%%%%%%%%%%%%%%%%%%%%%%%%%%%%%%%%%%%%%%%%%%%%%%%%%%%%%%%%%%%%%%%%%
%%%%%%%%%%%%%%%%%%%%%%%%%%%%%%%%%%%%%%%%%%%%%%%%%%%%%%%%%%%%%%%%%%%%

\section{Constraints on the \hdma parameter space}
\label{sec:constraints}

In the following we examine the constraints on the input parameters~\eqref{eq:inputparameters} that arise from Higgs and flavour physics, LHC searches for additional spin-0 bosons, EW precision measurements and vacuum stability considerations. The discussed constraints will motivate certain parameter benchmarks, which will be summarised at the end of the section. 

\subsection[Constraints on $\cos (\beta - \alpha)$]{Constraints on $\bm{\cos (\beta - \alpha)}$}
%\subsection{Constraints on $\bm{\cos (\beta - \alpha)}$}

The mixing angle $\alpha$ between the CP-even scalars $h$ and $H$ is constrained by Higgs coupling strength measurements and we display the regions in the $\cos(\beta-\alpha)\hspace{0.5mm}$--$\hspace{0.5mm}\tan\beta$ plane that are allowed by the LHC Run-I combination~\cite{Khachatryan:2016vau} in  the left panel of Figure~\ref{fig:higgsflavourfit}. See \cite{ATLAS-CONF-2018-031,CMS-PAS-HIG-17-031} for the latest 13~TeV LHC results.  The  95\%~confidence level (CL) contour shown has been obtained in the type-II~2HDM. For arbitrary values of $\tan \beta$, only parameter choices with $\cos(\beta-\alpha) \simeq 0$ are experimentally allowed.  Additional exclusion limits in the $\cos(\beta-\alpha)\hspace{0.5mm}$--$\hspace{0.5mm}\tan\beta$ plane arise from searches for $A \to hZ$ \cite{Aaboud:2017cxo,CMS-PAS-HIG-18-005}. To avoid the constraints from Higgs physics and to simplify the further analysis, we will concentrate in this white paper on the so-called alignment limit of the 2HDM where $\cos (\beta - \alpha) = 0$~\cite{Gunion:2002zf}, treating $\tan \beta$ as a free parameter. In this limit the constraints from $A \to hZ$ are satisfied as well because the $AhZ$ coupling scales as $g_{AhZ} \propto \cos (\beta - \alpha)$. 

\subsection[Constraints on $\tan \beta$]{Constraints on $\bm{\tan \beta}$}
%\subsection{Constraints on $\bm{\tan \beta}$}
\label{sec:flavour}

\begin{figure}[t!]
\centering
\includegraphics[width=0.435\textwidth]{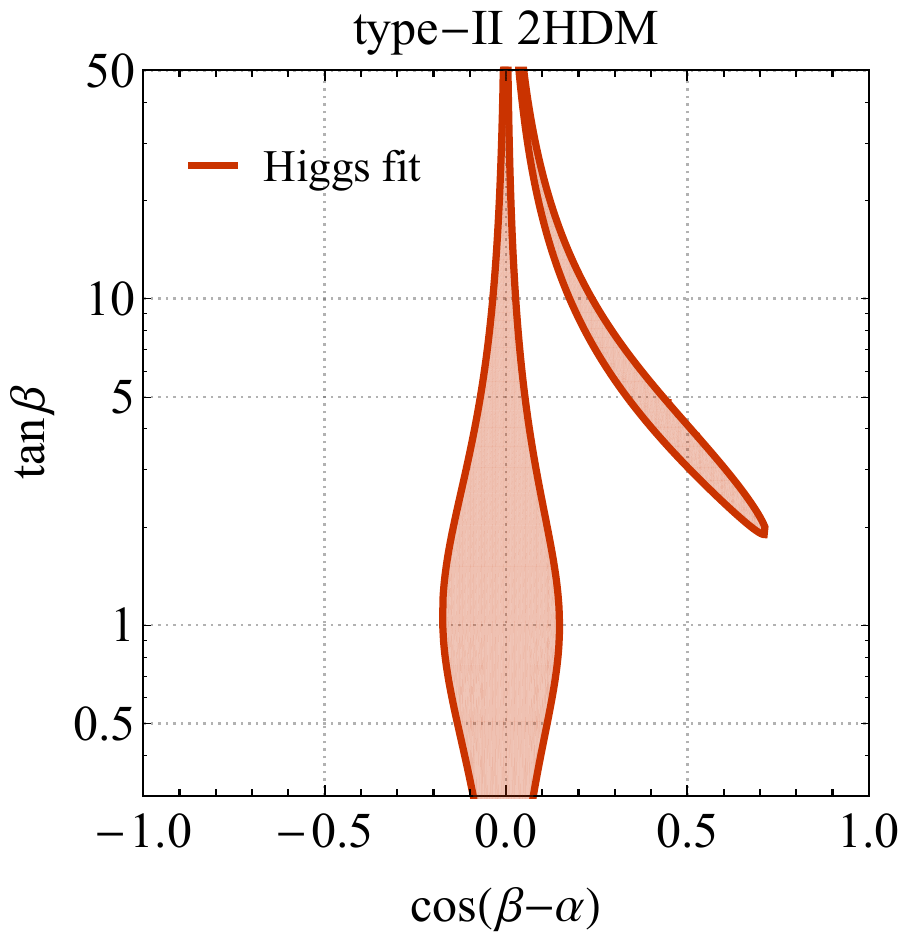} \qquad 
\includegraphics[width=0.45\textwidth]{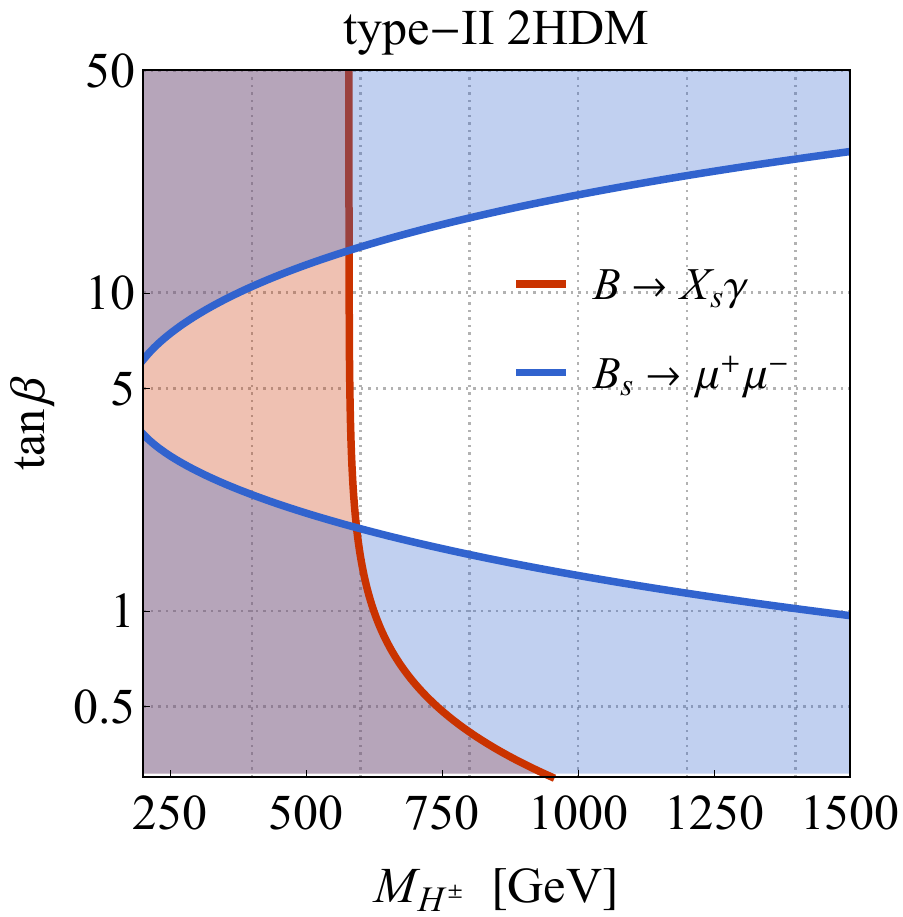}
\vspace{4mm}
\caption{\label{fig:higgsflavourfit} Left: Parameter space allowed, at 95\% CL, by a global fit to the LHC Run-I Higgs coupling strength measurements in the context of a 2HDM type-II scenario. Right: Parameter space in the $M_{H^\pm}\hspace{0.5mm}$--$\hspace{0.5mm}\tan \beta$ plane that is disfavoured by the flavour observables $B \to X_s \gamma$ (red) and $B_s \to \mu^+ \mu^-$ (blue). The open region in the center of the plot is allowed at 95\%~CL. }
\end{figure}

Indirect constraints on $\tan \beta$ as a function of $M_{H^\pm}$ arise from $B \to X_s \gamma$~\cite{Hermann:2012fc,Misiak:2015xwa,Misiak:2017bgg}, $B$-meson mixing~\cite{Abbott:1979dt,Geng:1988bq,Buras:1989ui,Kirk:2017juj} as well as  $B_s \to \mu^+ \mu^-$~\cite{Skiba:1992mg,Logan:2000iv,Chankowski:2000ng,Bobeth:2001sq,Bobeth:2013uxa,CMS:2014xfa,Aaij:2017vad}, but also follow from $Z \to b \bar b$~\cite{Denner:1991ie,Haisch:2007ia,Freitas:2012sy}. For the case of the type-II 2HDM, the most stringent constraints on the $M_{H^{\pm}}\hspace{0.5mm}$--$\hspace{0.5mm}\tan \beta$ plane are depicted in the right panel of Figure~\ref{fig:higgsflavourfit}. From the shown results it is evident that $B \to X_s \gamma$ provides a lower limit on the charged Higgs mass of $M_{H^\pm} > 580 \, {\rm GeV}$ that is practically independent of $\tan \beta$ for $\tan \beta \gtrsim 2$, while $B_s \to \mu^+ \mu^-$ is the leading constraint for very heavy charged Higgses, excluding for instance values of $\tan \beta <  1.3$ and $\tan \beta > 20$   for $M_{H^\pm} = 1 \, {\rm TeV}$.  Since the  indirect constraints   arise from loop corrections they can in principle be weakened by the presence of additional particles that are too heavy to be produced at the LHC. We thus consider the bounds from flavour only as indicative, and will not directly impose them on the parameter space of the \hdma in what follows. The constraints on $\tan \beta$ that follow from the existing LHC searches for heavy spin-0 bosons (see for instance~\cite{Aaboud:2017sjh,Sirunyan:2018zut,Aaboud:2017hnm,Sirunyan:2017roi,Aaboud:2018xuw}) will be discussed in Section~\ref{sec:nonMET}.

\subsection[Constraints on $\sin \theta$]{Constraints on $\bm{\sin \theta}$}
%\subsection{Constraints on $\bm{\sin \theta}$}
\label{sec:EWPO}

EW precision measurements constrain the differences between the masses of the additional scalar and pseudoscalar particles $M_H, M_A, M_{H^\pm}$ and~$M_a$, because the exchange of spin-0 states modifies the propagators of the $W$- and $Z$-bosons at the one-loop level and beyond. For $M_H=M_{H^\pm}$ and $\cos(\beta-\alpha)=0$, these corrections vanish due to a custodial symmetry in the tree-level potential $V_H$~\cite{Haber:1992py,Pomarol:1993mu,Gerard:2007kn,Grzadkowski:2010dj,Haber:2010bw} introduced in~\eqref{eq:VH} and the masses of the CP-odd mass eigenstates can be treated as free parameters. This custodial symmetry is also present if $M_A=M_{H^\pm}$ and $\cos(\beta-\alpha)=0$, but the presence of the pseudoscalar mixing term in~\eqref{eq:VHP}  breaks this symmetry softly~\cite{Bauer:2017ota}. As a result, the pseudoscalar mixing angle $\theta$ and the mass splitting between $M_H$, $M_A$ and $M_a$ are constrained in such a case. An illustrative example of the resulting constraints is given in the left panel of Figure~\ref{fig:EWVAC}. To keep $\sin \theta$ and $M_a$ as free parameters, we consider below only \hdma model configurations in which the masses of the $H$, $A$ and $H^\pm$ are equal. The choice $M_H = M_A = M_{H^\pm}$ is also adopted  in some 2HDM interpretations of the searches for heavy spin-0 resonances performed at the LHC~(cf.~\cite{Aaboud:2017gsl,Aaboud:2017rel,Sirunyan:2018taj}~for example). 

\subsection[Constraints on $M_a$]{Constraints on $\bm{M_a}$}
%\subsection{Constraints on $\bm{M_a}$}
\label{sec:invisiblehiggs}

Invisible decays of the Higgs boson allow to set a lower limit on the mass of  the~pseudoscalar~$a$ in \hdma scenarios with light DM~\cite{Bauer:2017ota}. In the case of $m_\chi = 1 \, {\rm GeV}$, it turns out for instance that mediator masses $M_a \lesssim 100 \, {\rm GeV}$ are excluded by imposing the 95\%~CL limit on the branching ratio ${\rm BR} (h \to  \text{invisible}) \lesssim 25\%$~\cite{Aad:2015pla,Khachatryan:2016whc}.  This limit is largely independent of the choices of the other parameters since ${\rm BR} (h \to  \text{invisible}) \simeq {\rm BR} (h \to a a^\ast \to 2 \chi 2 \bar \chi) \simeq 100\%$ for sufficiently light DM, unless  the $haa$ coupling, which for $\cos (\beta-\alpha) = 0$ and $M_H = M_{H^\pm}$ takes the following form~\cite{Bauer:2017ota}
\begin{equation}
\begin{split}
g_{haa} & = \frac{1}{M_h v}  \, \Big [ \left ( M_h^2  + 2 M_H^2  -  2 M_a^2 - 2 \lambda_3 v^2 \right) \sin^2 \theta \\[1mm] &  \hspace{1.5cm} - 2 \left (  \lambda_{P1} \cos^2 \beta + \lambda_{P2} \sin^2 \beta  \right ) v^2 \cos^2 \theta \, \Big ]  \,, 
\end{split}
\end{equation}
is sufficiently suppressed by tuning,~i.e.~$|g_{haa}| \ll 1$. To~evade the limits from invisible Higgs decays, we consider in this white paper only $M_a$ values larger than $100 \, {\rm GeV}$ when studying $\MET$ signatures at the LHC. 

\subsection[Constraints on $\lambda_3$]{Constraints on $\bm{\lambda_3}$}
%\subsection{Constraints on $\bm{\lambda_3}$}

\begin{figure}[t!]
\centering
\includegraphics[height=.45\textwidth]{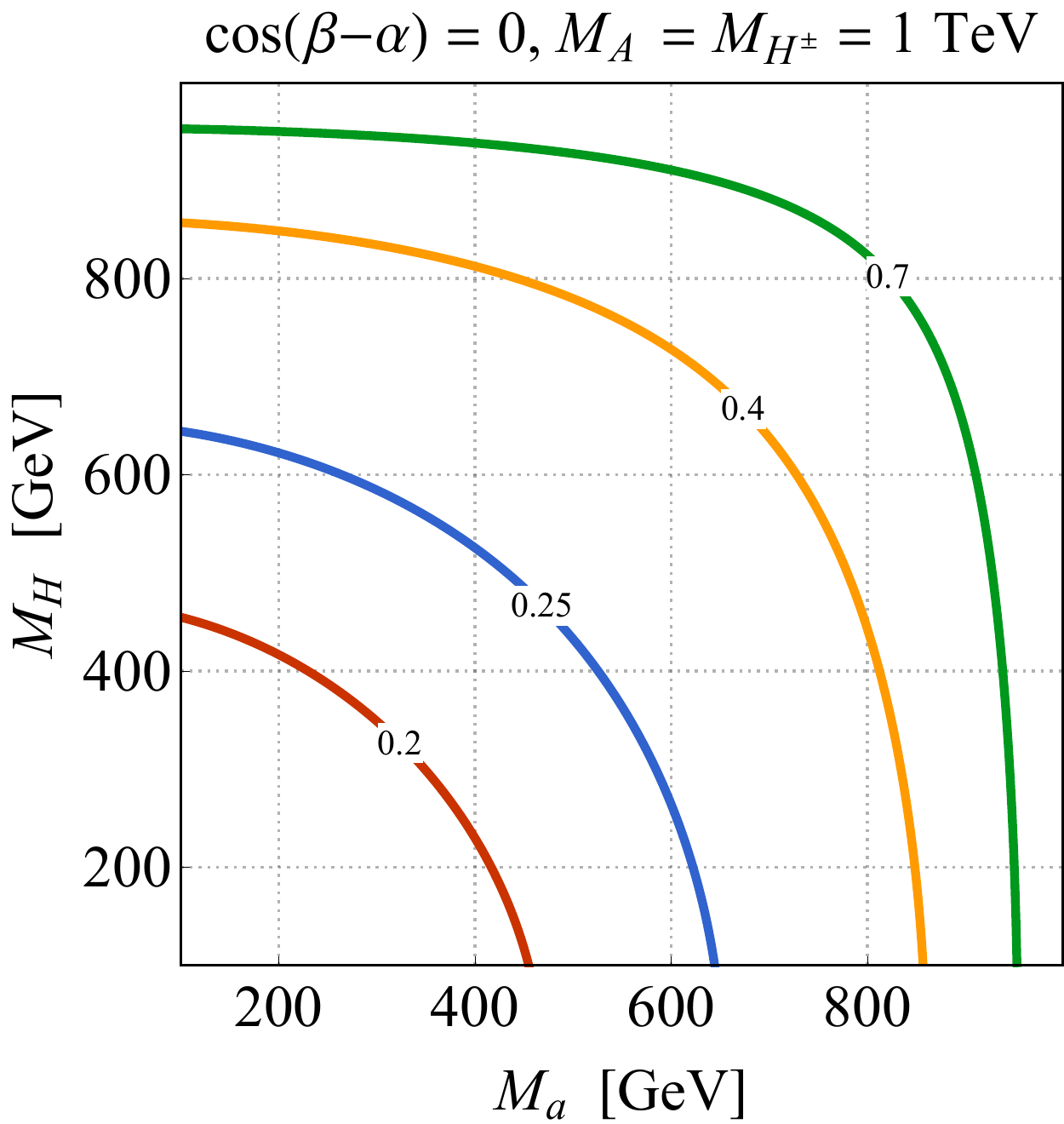} \qquad 
\includegraphics[height=.45\textwidth]{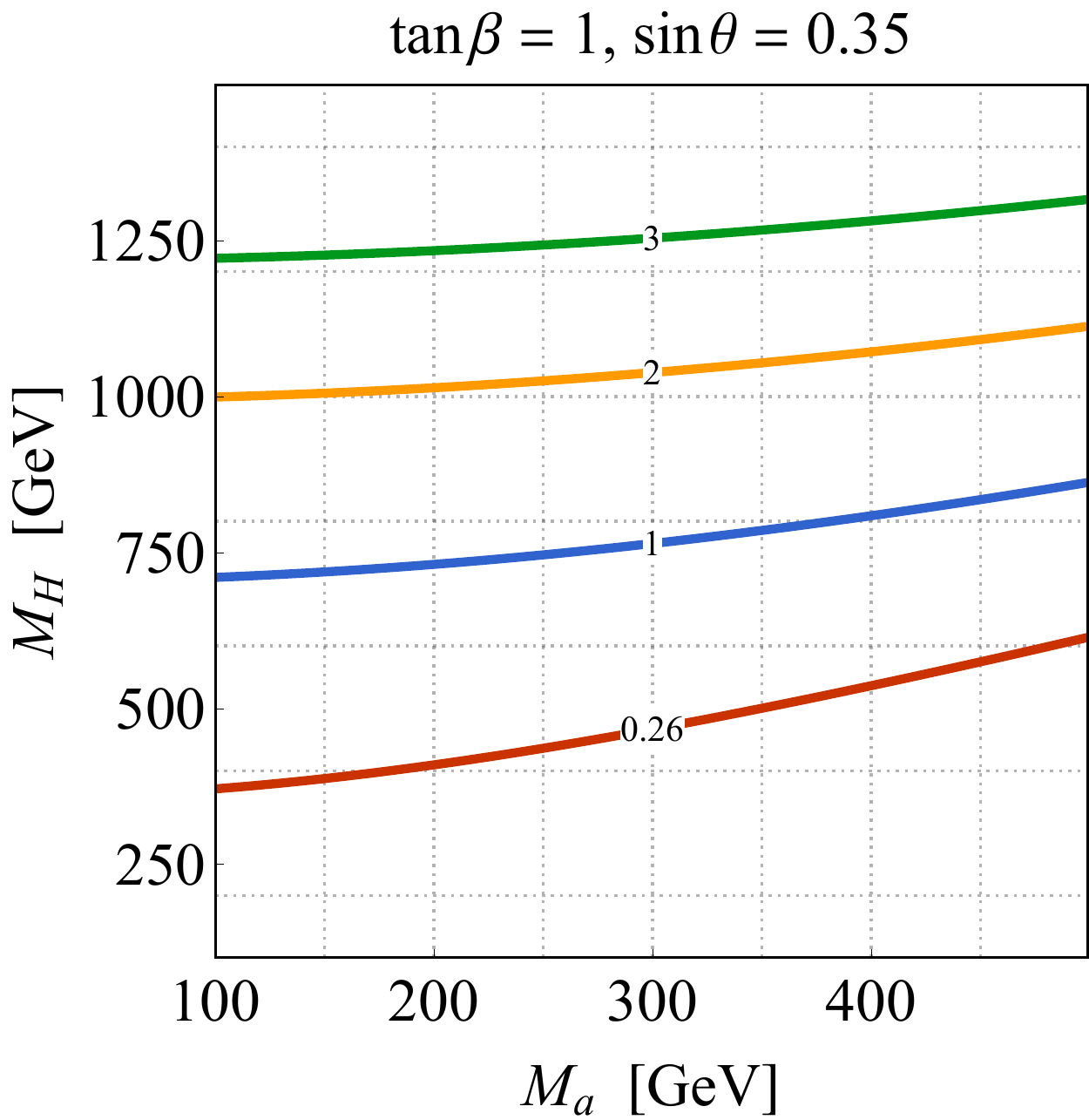}
\vspace{4mm}
\caption{\label{fig:EWVAC} Left: Values of $M_a$ and $M_H$ allowed by EW precision constraints assuming $\cos(\beta-\alpha)=0$, $M_A= M_{H^\pm}=1 \, {\rm TeV}$ and four different values of $\sin \theta$, as indicated by the contour labels.  The parameter space below and to the left of the contours is excluded. Right: Constraints in the $M_{a}\hspace{0.5mm}$--$\hspace{0.5mm} M_H$ plane following from the  BFB requirement. The  results shown correspond to $\tan \beta = 1$, $\sin \theta = 0.35$ and degenerate heavy spin-0 boson masses $M_H = M_A = M_{H^\pm}$. The region above  each contour  is excluded for the indicated value of the quartic coupling~$\lambda_3$.}  
\end{figure}

The requirement that the scalar potential~\eqref{eq:V2HDMa} of the \hdma  is bounded from below~(BFB) restricts the possible choices of the spin-0 boson masses, mixing angles and quartic couplings. Assuming that $\lambda_{P1}, \lambda_{P2} > 0$, the  BFB conditions in the~\hdma model turn out to be identical to those found in the pure 2HDM~\cite{Gunion:2002zf}. For our choice $M_H = M_A = M_{H^\pm}$ of heavy spin-0 boson masses, one finds that the tree-level BFB conditions can be cast into  two inequalities. The first inequality connects $\lambda_3$ with the cubic SM Higgs self-coupling $\lambda = M_h^2/(2 v^2) \simeq 0.13$ and simply reads   
\begin{align} \label{eq:BFB1}
\lambda_3 > 2 \lambda  \,.
\end{align}
The second BFB condition relates $\lambda_3$ with $\tan \beta$, $\sin \theta$, the common heavy  spin-0 boson  mass $M_H$ and $M_a$. In the limit $M_H \gg M_h, M_a$ it takes a rather simple form that we quote here for illustration: 
\begin{align} \label{eq:BFB2}
\lambda_3 > \frac{M_H^2 -M_a^2}{v^2} \sin^2 \theta  - 2 \lambda \cot^2 (2 \beta )  \,.
\end{align}
This formula implies that large values of $M_H^2/v^2 \sin^2 \theta$ are only compatible with the requirements from BFB if the quartic coupling $\lambda_3$ is sufficiently large.  The interplay between BFB and perturbativity of $\lambda_3$,~i.e.~$\lambda_3 < 4 \pi$, leads to a non-decoupling of $H, A$ and $H^\pm$ for $|M_H - M_a| \neq 0$ and $\sin \theta  \neq 0$~\cite{Goncalves:2016iyg} such that the spin-0 states are potentially within LHC reach. The right plot in~Figure~\ref{fig:EWVAC} which shows the constraints in the $M_{a}\hspace{0.5mm}$--$\hspace{0.5mm} M_H$ plane that derive from the exact version of~\eqref{eq:BFB2} confirms the latter statement. For $\tan \beta = 1$, $\sin \theta = 0.35$ and $M_H = M_A = M_{H^\pm}$, values of $\lambda_3 \gtrsim 2$ are needed in order for $M_H \simeq 1 \, {\rm TeV}$ to be allowed by BFB.  Due to the~$\sin^2 \theta$ dependence in~\eqref{eq:BFB2},   a common 2HDM  spin-0 boson  mass of $M_H = M_A = M_{H^\pm} \simeq 1 \, {\rm TeV}$ would only be viable for $\sin \theta = 0.7$ if the quartic coupling $\lambda_3$ takes close to non-perturbative values $\lambda_3 \gtrsim 8$. In order to allow for heavy Higgs above $1 \, {\rm TeV}$ to be acceptable while keeping $\lambda_3$ perturbative, we will choose $\sin \theta = 0.35$ and $\lambda_3 = 3$ as our benchmark in this white paper.  

\subsection[Constraints on $\lambda_{P1}$ and $\lambda_{P2}$]{Constraints on  $\bm{\lambda_{P1}}$ and $\bm{\lambda_{P2}}$}
%\subsection{Constraints on $\bm{\lambda_{P1}}$ and $\bm{\lambda_{P2}}$}

The quartic couplings $\lambda_3$, $\lambda_{P1}$ and $\lambda_{P2}$ affect all  cubic Higgs interactions. In the case of the $Haa$ and $Aha$ couplings, one obtains under the assumption that  $\cos (\beta-\alpha) = 0$ and $M_H = M_A = M_{H^\pm}$, the following expressions~\cite{Bauer:2017ota}
\begin{eqnarray} \label{eq:cubic}
\begin{split}
g_{Haa}   & = \frac{1}{M_H v}  \Big [  \cot \left ( 2 \beta \right) \left (  2 M_h^2 - 2 \lambda_3 v^2 \right )  \sin^2 \theta + \sin \left ( 2 \beta \right ) \left (\lambda_{P1}-\lambda_{P2} \right ) v^2 \cos^2 \theta \Big  ] \,, \\[2mm]
g_{Aha}   & = \frac{1}{M_H v} \, \Big [ M_h^2  + M_H^2   -  M_a^2 - 2 \lambda_3 v^2 + 2 \left (  \lambda_{P1} \cos^2 \beta + \lambda_{P2} \sin^2 \beta  \right ) v^2 \Big ] \sin \theta \cos \theta \,. \hspace{6mm}
\end{split}
\end{eqnarray}
Because $\Gamma (H \to aa) \propto g_{Haa}^2$ and $\Gamma (A \to ha) \propto g_{Aha}^2$, the relations~\eqref{eq:cubic} imply  that  in order to keep the total widths $\Gamma_H$ and $\Gamma_A$ small, parameter choices of the form $\lambda_3 = \lambda_{P1} = \lambda_{P2}$ are well suited. 

\subsection{Benchmark parameter choices}

The above discussion motivates the following choice of parameters
\begin{gather} 
 M_H  = M_A = M_{H^\pm} \,, \quad m_\chi = 10 \, {\rm GeV} \,,\notag  \\[1mm]
\cos (\beta - \alpha) = 0\,, \quad   \tan \beta = 1\,, \quad  \sin\theta = 0.35\,, \label{eq:benchmark} \\[1mm]
y_\chi  = 1\,, \quad \lambda_3 =  \lambda_{P1} = \lambda_{P2} =3 \,. \notag 
\end{gather}
For the choices $m_\chi = 10 \, {\rm GeV}$ and $y_\chi  = 1$ the  branching ratio ${\rm BR} (a \to \chi \bar \chi)$ is sizeable for all values of $M_a$ considered in this white paper,~i.e.~$M_a > 100 \, {\rm GeV}$. For masses below the top threshold of around $350 \, {\rm GeV}$, $a \to t \bar t$ is  kinematically forbidden and therefore ${\rm BR} (a \to \chi \bar \chi)$ can be as large as 100\%.  The choice of~$y_\chi  = 1$ is thereby largely arbitrary for the mono-$X$ phenomenology, which is not the case for the DD and ID cross sections where the magnitude of $y_\chi$ plays an important role. This feature  has to be  kept in mind when performing a comparison between LHC and DD/ID constraints. Concerning the mono-Higgs and mono-$Z$ signals in the \hdma model it is furthermore important to realise that the relevant couplings scale as $g_{Aha} \propto \sin \theta \cos \theta$  (cf.~\eqref{eq:cubic}) and $g_{HZa} \propto \sin \theta$. Since in addition $g_{t \bar t a} \propto \sin \theta$,  it follows that in the limit~$\sin \theta \to 0$ all mono-$X$ signatures vanish. In order to obtain detectable LHC signals involving $\MET$, we have chosen $\sin\theta = 0.35$ in the above benchmark parameter scenario. We furthermore add that since $\tan \beta$ has been set equal to 1 in~\eqref{eq:benchmark}, most of the results presented in this white paper are independent of the type chosen for the \hdma Yukawa sector. 

In the  type-II \hdma  benchmark scenario \eqref{eq:benchmark} the only free parameters are~$M_H$ and~$M_a$. We will study the sensitivity of the existing mono-$X$ searches in the corresponding two-dimensional parameter plane in Section~\ref{sec:sensitivitystudies}. Parameter scans in the $M_a\hspace{0.5mm}$--$\hspace{0.5mm}\tan \beta$ plane can also be found in this section.  In these latter scans, the choices \eqref{eq:benchmark} are adopted except for $\tan \beta$, which is not fixed to 1 anymore but allowed to vary freely, as well as    
\begin{equation} \label{eq:matbscan}
M_H  = M_A = M_{H^\pm} =  600 \, {\rm GeV} \,.
\end{equation} 
Since the $g_{b \bar b A}$ and $g_{b \bar b a}$ couplings are $\tan \beta$-enhanced in the type-II~\hdma model,  effects from $b \bar b$-initiated production can be relevant  for $\tan \beta \gg 1$. Such $\tan \beta$-enhanced contributions will be included in our sensitivity studies of the mono-Higgs and mono-$Z$ channels to be presented in Section~\ref{sec:sensitivitystudies}. 

At this point it is worthwhile to add that the mono-$X$ signatures that are most sensitive to the mass splitting between the $H$ and the $A$, the parameter $\sin \theta$ and the quartic couplings $\lambda_{3}$, $\lambda_{P1}$, $\lambda_{P2}$ turn out to be  the mono-Higgs and mono-$Z$ channels (see~Section~\ref{sec:experimentbasics} for details). Four benchmark scenarios that illustrate these model dependencies have been proposed and studied in~\cite{Bauer:2017ota}.  We believe that the specific benchmarks  chosen in~\eqref{eq:benchmark} and~\eqref{eq:matbscan}   exemplify  the rich structure of $\MET$ signatures in the \hdma model, and they should therefore serve well as a starting point for further more detailed experimental and theoretical investigations. 

\begin{figure}[t!]
\centering
\includegraphics[height=.45\textwidth]{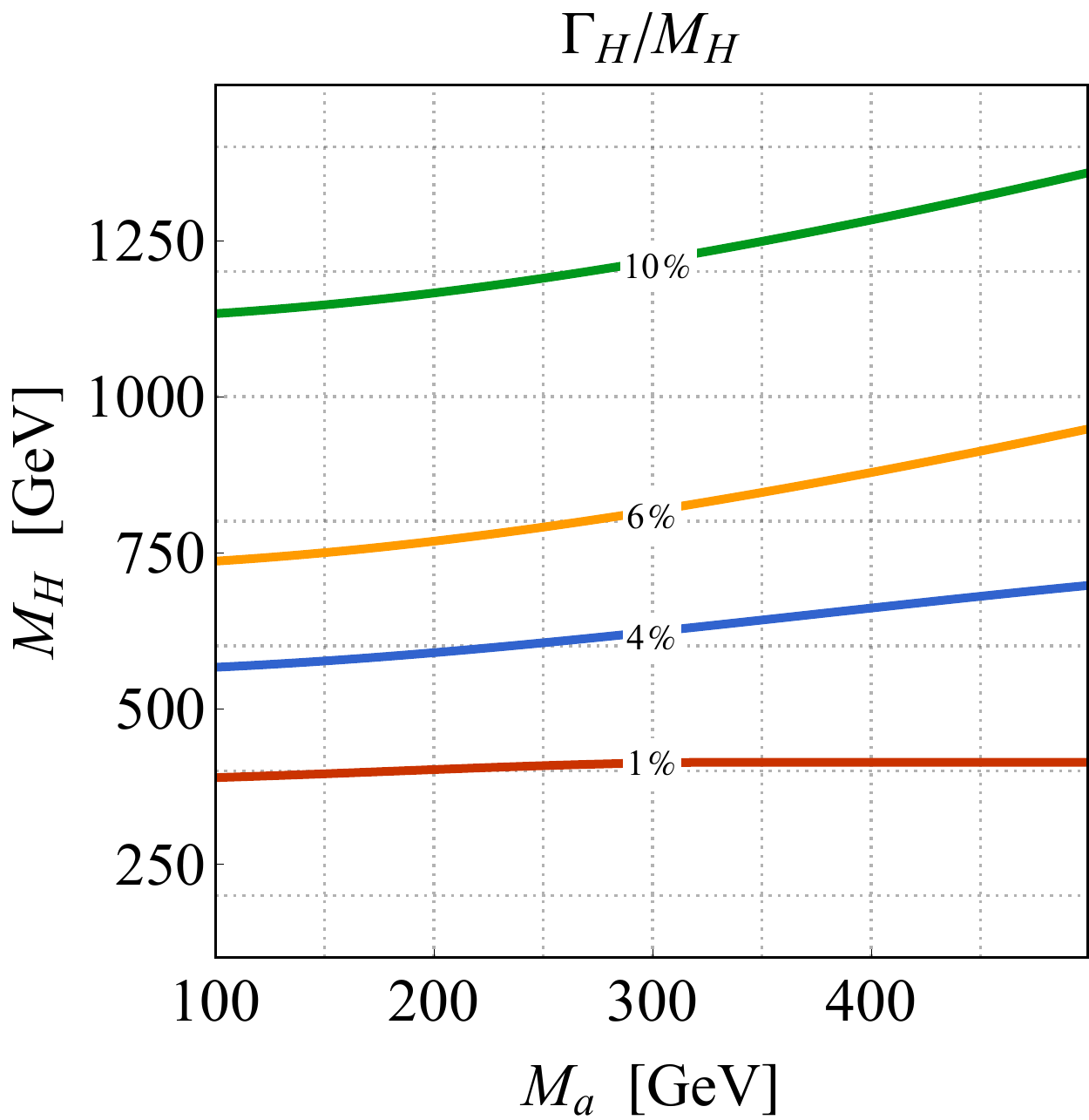} \qquad 
\includegraphics[height=.45\textwidth]{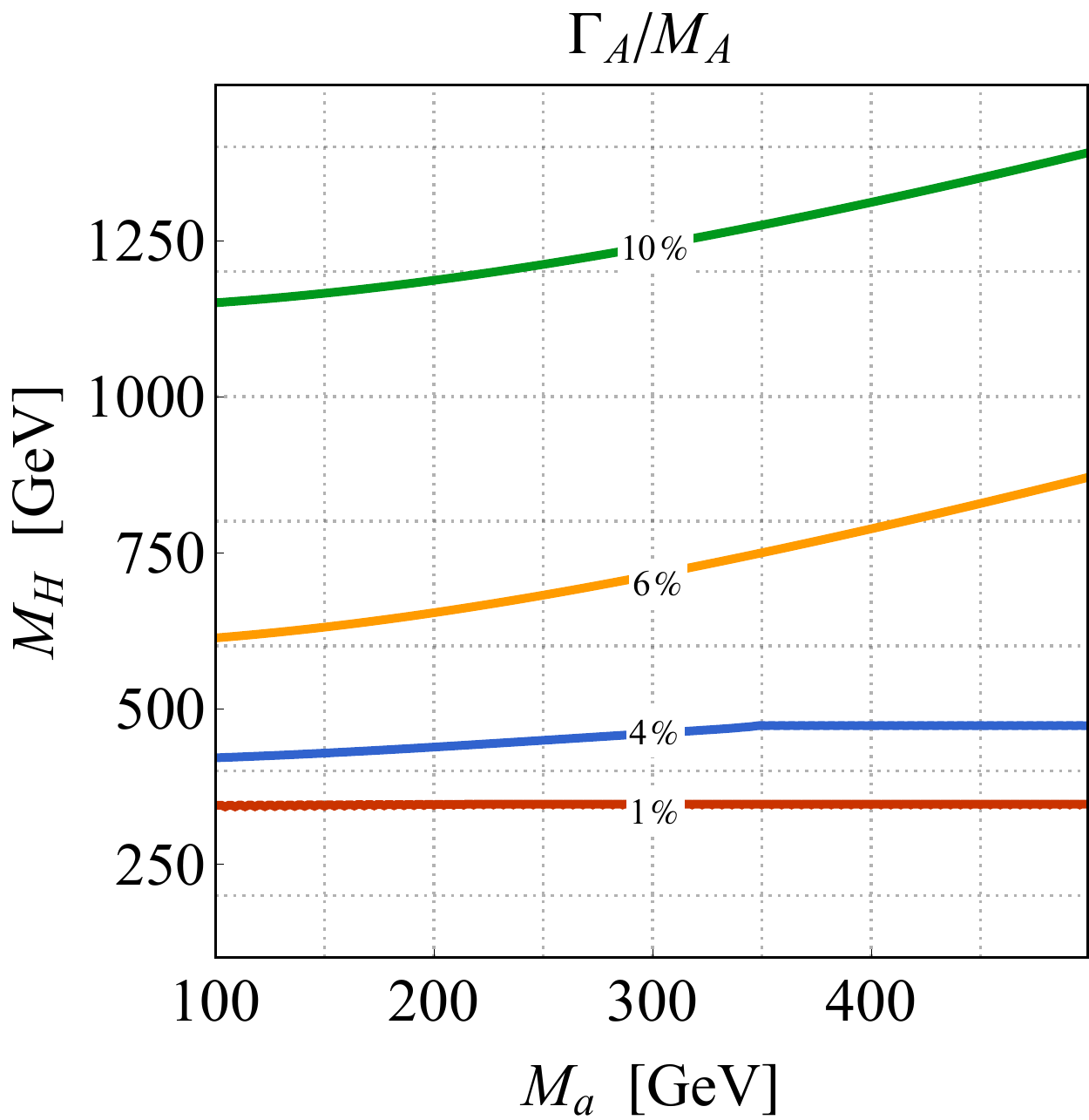}
\vspace{4mm}
\caption{\label{fig:Gammas}  Predictions for  $\Gamma_H/M_H$~(left panel) and $\Gamma_A/M_A$~(right panel). The results shown correspond to the type-II \hdma benchmark parameter choices given in~\eqref{eq:benchmark}.}
\end{figure}

As a final validation (or first application) of the proposed benchmark scenario, we present in  Figure~\ref{fig:Gammas} the predictions for the ratios $\Gamma_H/M_H$~(left) and $\Gamma_A/M_A$~(right). We see that the heavy neutral Higgs states $H$ and $A$ are relatively narrow even for values $M_H > 1 \, {\rm TeV}$ and $M_a = 100 \, {\rm GeV}$.  The narrow width assumption is thus justified in the entire parameter space considered in our $M_a\hspace{0.5mm}$--$\hspace{0.5mm}M_H$ scans. 

%%%%%%%%%%%%%%%%%%%%%%%%%%%%%%%%%%%%%%%%%%%%%%%%%%%%%%%%%%%%%%%%%%%%
%%%%%%%%%%%%%%%%%%%%%%%%%%%%%%%%%%%%%%%%%%%%%%%%%%%%%%%%%%%%%%%%%%%%
%%%%%%%%%%%%%%%%%%%%%%%%%%%%%%%%%%%%%%%%%%%%%%%%%%%%%%%%%%%%%%%%%%%%

\section{Comparison to other DM models}
\label{sec:comparison}

In this section we briefly discuss DM models that also feature a 2HDM sector.  Our discussion will focus on the similarities and differences between these scenarios and the \hdma model  concerning the mono-$X$ phenomenology. 

\subsection{2HDM with an extra scalar singlet}
\label{sec:2HDMs}

Instead of mixing an additional CP-odd singlet $P$ with the pseudoscalar $A$, as done in~\eqref{eq:VHP}, it is also possible to consider the mixing of a  scalar singlet~$S$ with  the CP-even spin-0 states~$h,H$. Detailed studies of the DD and relic-density phenomenology of this so-called 2HDM+s model have been presented in~\cite{Bell:2016ekl,Bell:2017rgi}.  Like in the case of the \hdma~model, the presence of  non-SM Higgs bosons in the 2HDM+s model can lead to novel $\MET$ signatures that are not captured by a DM simplified model with just a single scalar mediator. In the pure alignment limit,~i.e.~$\cos ( \beta - \alpha) = 0$, the most interesting collider signals are mono-Higgs, mono-$Z$ and the $t X + \MET$  channels, because  these signatures can all arise resonantly.  In fact, the relevant one-loop diagrams are precisely those that  lead to the leading mono-$X$ signals in the \hdma model~(see Figure~\ref{fig:resonant}), and in consequence resonant $\MET$ searches that can constrain the \hdma model could also be interpreted in the 2HDM+s context. Away from alignment, the scalar mediator couples to the EW gauge bosons and as a result it may also be possible to have a sizeable amount of $\MET$ in association with a $Z$ or $W$ boson or in  vector boson fusion (VBF). Due to the CP properties of the $a$, the latter tree-level $\MET$ signatures are not present in the \hdma model. 

\subsection{2HDM with singlet-doublet DM}

In both the \hdma and the 2HDM+s model the DM particle is  an EW singlet. The DM particle may, however, also be a mixture of an EW singlet and doublet(s)~\cite{Mahbubani:2005pt,Enberg:2007rp,Cohen:2011ec,Cheung:2013dua}, as in the minimal supersymmetric SM with both bino and higgsino components. Generically, this model is referred to as singlet-doublet DM. The phenomenology of 2HDM models with singlet-doublet DM has been discussed in~\cite{Berlin:2015wwa,Arcadi:2018pfo}, where only the $b+\MET$ and $t \bar t+\MET$ signatures have been considered and found to provide only weak constraints. Additionally, a recent study~\cite{Bauer:2017fsw} suggests that $b+\MET$ and $tX+\MET$ may give stronger constraints in the 2HDM with singlet-doublet DM for scenarios in which the additional scalars have a mass not too far above the pseudoscalar mass. 

\subsection{2HDM with  higher-dimensional couplings to DM}

A gauge-invariant DM model where a pseudoscalar is embedded into a 2HDM that has renormalisable couplings to SM fields but an effective coupling to DM via the dimension-five operator $H_1^\dagger H_2 \bar \chi \gamma_5 \chi$ has been  discussed in~\cite{Bauer:2017fsw}. It has been shown that  such an  effective DM coupling can be obtained in different UV completions such as the \hdma model or  a 2HDM with singlet-doublet DM by integrating out heavy particles. Apart from the $t X+\MET$ signatures, the whole suite of mono-$X$ signals has been considered in~\cite{Bauer:2017fsw}. It was found that a resonant mono-$Z$ signal via $pp \to H \to AZ \to  Z + \chi \bar \chi$ is a universal prediction in all DM pseudoscalar mediator models, while other signatures such as mono-Higgs are model dependent. Given that a sizeable $H^\pm \to A W$ rate is also a generic feature of DM pseudoscalar models if $M_{H^\pm} > M_A + M_W$, channels like $tW+\MET$~\cite{Pani:2017qyd} should also  provide relevant constraints on the DM model introduced in~\cite{Bauer:2017fsw}. 

\subsection{Inert doublet model}

In the scenarios discussed so far the DM particle has always been a fermion. The so-called inert doublet model~(IDM)~\cite{Deshpande:1977rw,Barbieri:2006dq,Cao:2007rm} is a DM model based on a 2HDM sector that can provide DM in the form of the spin-0 resonances $H, A$.  The presence of a $Z_2$ symmetry renders the DM candidate stable and also implies  that the bosonic states originating from the second~(dark) Higgs doublet can only be pair-produced. Since the dark scalars do not couple to the SM fermions, $H,A,H^\pm$ production arises in the IDM dominantly from Drell-Yan processes. The IDM offers a rich spectrum of LHC $\MET$ signatures that ranges from mono-jet, mono-$Z$, mono-$W$, mono-Higgs to  ${\rm VBF}+\MET$~\cite{Dolle:2009ft,Miao:2010rg,Gustafsson:2012aj,Belanger:2015kga,Ilnicka:2015jba,Poulose:2016lvz,Datta:2016nfz,Hashemi:2016wup,deFlorian:2016spz,Belyaev:2016lok,Dutta:2017lny,Wan:2018eaz}.  While the prospects to probe the IDM parameter space via the mono-jet channel seem to be limited~\cite{Belyaev:2016lok}, LHC searches for multiple leptons~\cite{Dolle:2009ft,Miao:2010rg,Gustafsson:2012aj,Belanger:2015kga,Datta:2016nfz,Hashemi:2016wup}, multiple jets~\cite{Poulose:2016lvz,Dutta:2017lny}  or a combination thereof~\cite{Hashemi:2016wup,Wan:2018eaz} are expected to probe the IDM parameter space in regions that are not accessible by DD experiments of~DM or measurements of the invisible decay width of the SM Higgs. Furthermore, in scenarios in which the mass of DM is almost degenerate with~$M_{H^\pm}$, searches for disappearing charged tracks provide a rather unique handle on the IDM high-mass regime~\cite{Belyaev:2016lok}. While the IDM can lead to the same $\MET$ signals  as the \hdma model, the resulting kinematic distributions will in general not be the same, due to the different production mechanisms and decay topologies in the two models. Selection cuts that are optimised for a \hdma interpretation of a given mono-$X$ search will thus often not be ideal in the IDM context. Dedicated ATLAS and CMS  analyses of the mono-$X$ signatures in the IDM do unfortunately not exist at the moment. Such studies would, however, be highly desirable.

\subsection{2HDM with an extra scalar mediator and scalar DM}

 Like the 2HDM+s model, the DM scenario proposed in~\cite{vonBuddenbrock:2016rmr} contains an extra scalar singlet, which, however, does not couple to a fermionic DM current $\bar \chi \chi$ but to the scalar operator~$\chi^2$. The latter work focuses on the parameter space of the model where the mediator $s$ is dominantly produced via either  $pp \to H + j \to 2s + j \to j + 4 \chi$ or $pp \to H \to sh \to h + 2\chi$. The resulting mono-jet and mono-Higgs cross sections, however, turn out to be safely below the existing experimental limits. In case the mass hierarchy  $M_A > M_H + M_Z$ is realised, the channel $pp \to A \to HZ$ is also interesting, since it either leads to a mono-$Z$ or a $hZ+\MET$ signature, depending on whether $H \to 2 s \to 4 \chi$ or $H \to h s \to h \chi^2$ is the leading decay. We add that an effective version of the model brought forward in~\cite{vonBuddenbrock:2016rmr}  has already been constrained by ATLAS~\cite{Aaboud:2017uak} using the mono-Higgs channel.  

%%%%%%%%%%%%%%%%%%%%%%%%%%%%%%%%%%%%%%%%%%%%%%%%%%%%%%%%%%%%%%%%%%%%
%%%%%%%%%%%%%%%%%%%%%%%%%%%%%%%%%%%%%%%%%%%%%%%%%%%%%%%%%%%%%%%%%%%%
%%%%%%%%%%%%%%%%%%%%%%%%%%%%%%%%%%%%%%%%%%%%%%%%%%%%%%%%%%%%%%%%%%%%

\section{$\bm{E_T^{\rm miss}}$ signatures and parameter variations in the \hdma model}
\label{sec:experimentbasics}

The mono-$X$ phenomenology in the  \hdma model is determined by the values of the parameters introduced in~\eqref{eq:inputparameters}. These model parameters can affect the total signal cross sections of the $\MET$ signatures, their kinematic distributions, or both. In this section we will discuss the basic features of the most important mono-$X$ channels and identify the experimental observables that can be exploited to search for them. Our discussion will  mainly focus on the benchmark~\eqref{eq:benchmark} but we will also present results for other parameter choices to illustrate how a given parameter affects a certain $\MET$ signature.  All results in this section are obtained at the parton level (i.e.~they are fixed-order predictions that do not include the effects of a parton shower) and employ no or only minimal selection requirements. The signal samples have been generated using an {\tt UFO}~\cite{Degrande:2011ua} implementation of the~type-II~\hdma model~\cite{hdmaUFO} together with {\tt MadGraph5\_aMC@NLO}~\cite{Alwall:2014hca}. Further details on the Monte Carlo (MC) simulations can be found in Appendix~\ref{app:mcgeneration}.

\subsection[Resonant $E_T^{\rm miss}$ signatures]{Resonant $\bm{E_T^{\rm miss}}$ signatures}
\label{sec:resonant}

\begin{figure}[t!]
\centering
\includegraphics[width=.8\textwidth]{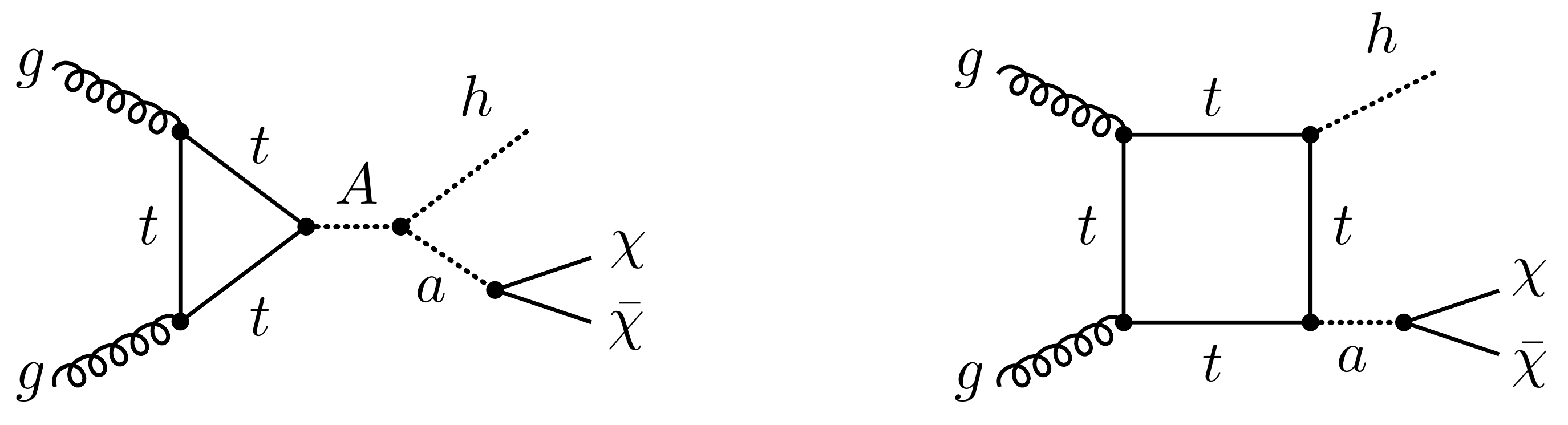}

\vspace{4mm}

\includegraphics[width=.8\textwidth]{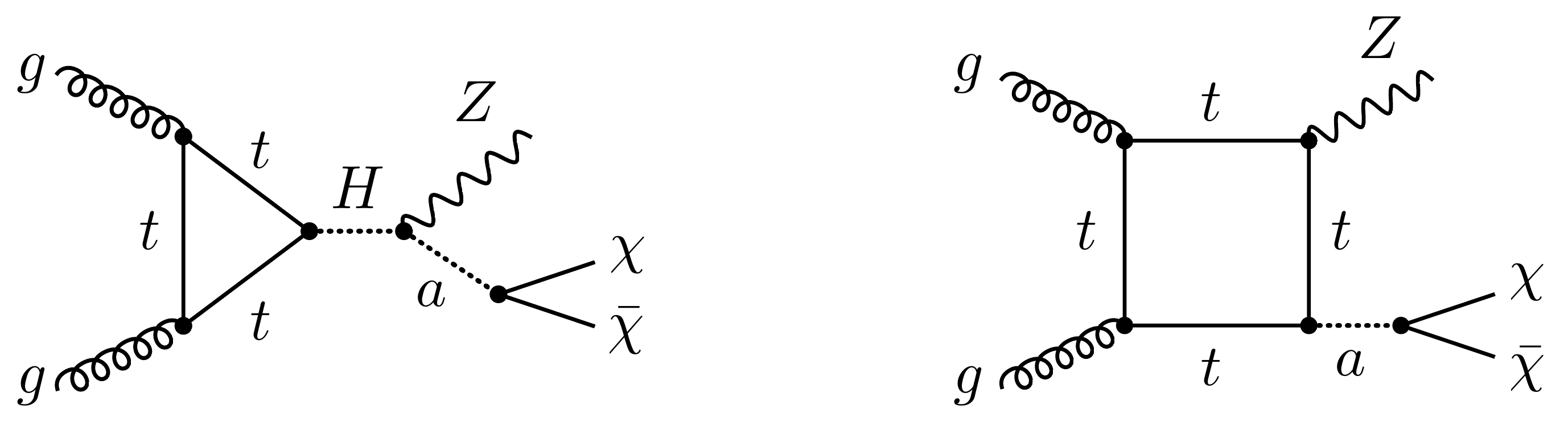}

\vspace{5mm}

\includegraphics[width=.8\textwidth]{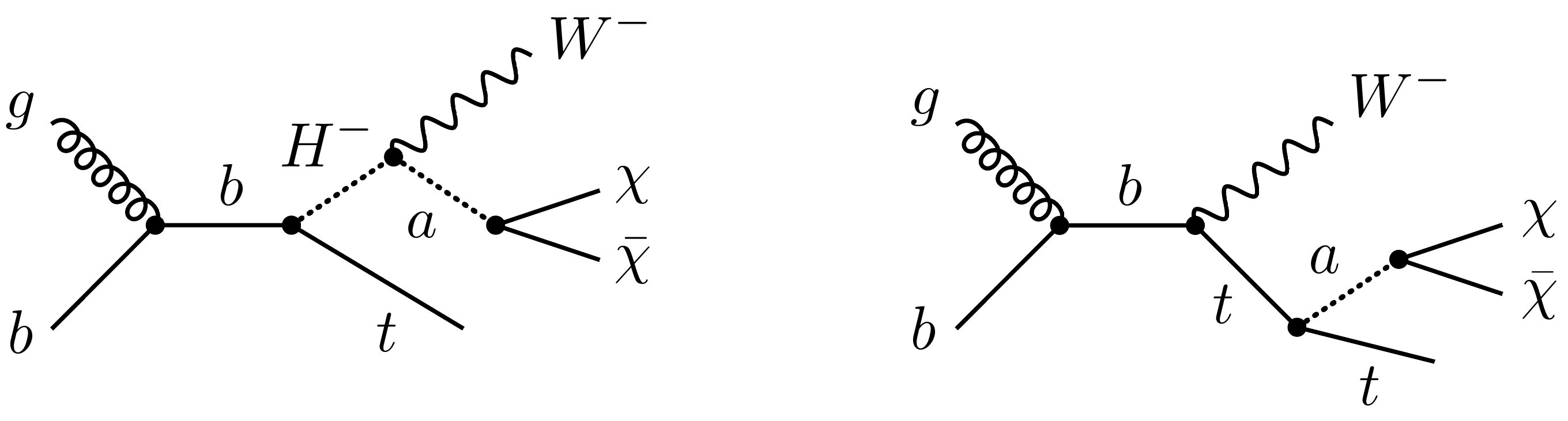}

\vspace{4mm}
\caption{\label{fig:resonant} Example diagrams that give rise to an $h+\MET$~(upper row), $Z+\MET$~(middle row) and $tW + \MET$ (lower row) signal in the \hdma model. For further details consult the main text. }
\end{figure}

In the \hdma model there are broadly speaking two different kinds of $\MET$ signatures. In the first case, the spin-0 mediator can be resonantly produced as in Figure~\ref{fig:resonant} depicting relevant Feynman diagrams. Channels such as $h+\MET$, $Z+ \MET$ and $tW+\MET$ belong to this class. In the case of the mono-Higgs signature, it is evident from the figure that for $M_A > M_h + M_a$ the triangle  graph shown on the left in the upper row  allows for resonant mono-Higgs production.  Similar resonance enhancements arise from the diagram on the left-hand side for the mono-$Z$ (middle row) and $tW+\MET$ (lower row) channel if $M_H > M_Z + M_a$ and $M_{H^\pm} > M_W + M_a$, respectively. The interference between the box diagram and the resonant production is further described in Section~\ref{sec:parameter_variations}. Resonant $h+\MET$, $Z+\MET$ and $tW+\MET$ production is not allowed in the spin-0 DM models proposed by the DMF because the mediators couple only to fermions at tree level. As a result only diagrams of the type shown on the right-hand side of Figure~\ref{fig:resonant} are present in these models. 

\subsubsection{Mono-Higgs signature}

\begin{figure}[t!]
\centering
\includegraphics[height=0.45\textwidth]{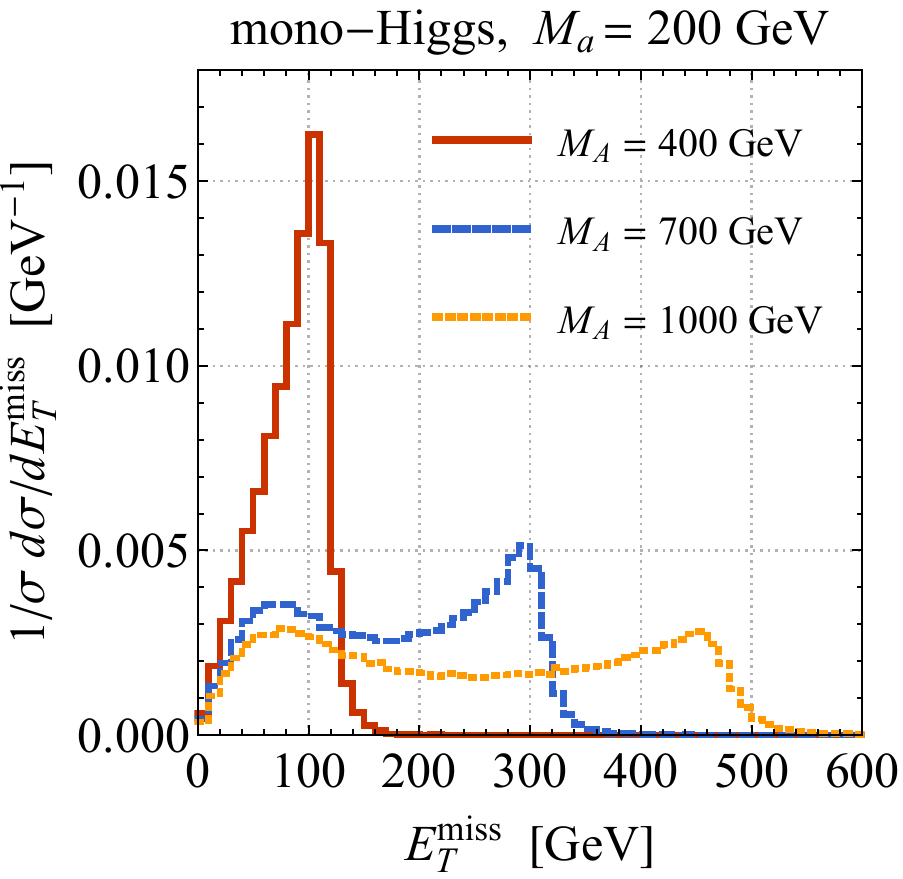}	\qquad 
\includegraphics[height=0.45\textwidth]{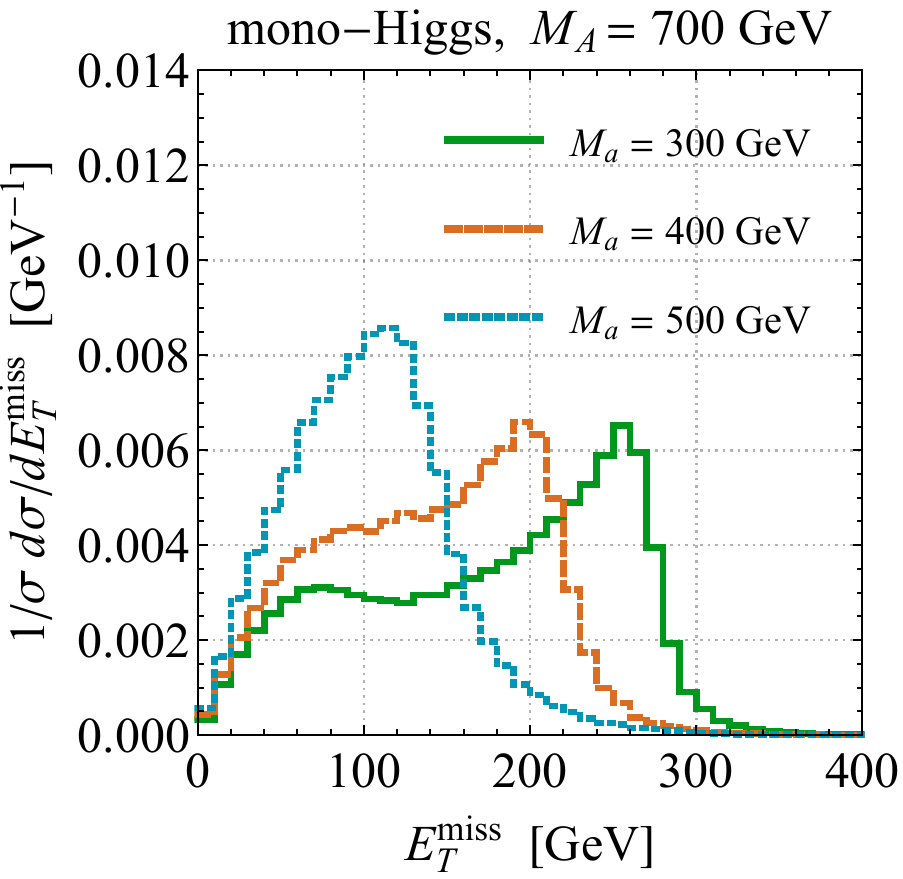}
\vspace{2mm}
\caption{\label{fig:hMET} Normalised $\MET$ distributions of mono-Higgs production in the \hdma model for different values of $\mA$ and $\ma$ as indicated in the legends. The  results shown correspond to the benchmark parameter choices introduced in~\eqref{eq:benchmark}. }  
\end{figure}

Processes that are resonantly enhanced in the \hdma model have in common that they involve the on-shell decay of a heavy Higgs $H,A,H^\pm$ to a SM particle and the mediator~$a$, which    subsequently decays to a pair of DM particles. The kinematics of the process $A \to B C$ is governed by the two-body phase space for three massive particles 
\begin{equation} \label{eq:twophasespace}
\lambda(m_A, m_B, m_C) = (m_A^2 -m_B^2-m_C^2)^2 -  4 \hspace{0.25mm} m_B^2 \hspace{0.25mm}  m_C^2 \,,
\end{equation}
and this quantity determines the characteristic shape of resonant $\MET$ signals in the context of the \hdma model. For instance, in the case of the mono-Higgs signal the $\MET$ spectrum  will have a Jacobian peak with an endpoint at~\cite{No:2015xqa,Bauer:2017ota}
\begin{equation} \label{eq:monoHMETpeak}
E^{\rm{miss}}_{T, {\rm max} } \simeq \frac{\lambda^{1/2}(M_A, M_h, M_a)}{2 M_A}\,, 
\end{equation}
for all mass configurations that satisfy $M_A > M_h + M_a$. 

In Figure~\ref{fig:hMET} we show the predictions for the normalised $\MET$ distribution of $h+\MET$ production in the \hdma model for different spin-0 boson masses $\mA$ and $\ma$. Besides the indicated values of $\mA$ and $\ma$ the parameters  used  are those given in~\eqref{eq:benchmark}. Increasing~$\mA$ ($\ma$) shifts the endpoint of the Jacobian peak to higher~(lower) $\MET$  values as expected from~\eqref{eq:monoHMETpeak}. A~second feature that is also visible is that for large mass splittings~$\mA - \ma$, the $\MET$ spectra develop a pronounced low-$\MET$ tail. The events in these tails arise dominantly from the box diagram shown on the right in the upper row of~Figure~\ref{fig:resonant}. It can also be noted that these non-resonant contributions interfere with the resonant contributions that stem from triangle graphs. Due to the interplay of resonant and non-resonant contributions,  the exact shape of the $\MET$ distribution is away from the endpoint~\eqref{eq:monoHMETpeak} a non-trivial function of the \hdma parameters~\eqref{eq:inputparameters}.  

 At the LHC a mono-Higgs signal has so far been searched for in the $h \to \gamma \gamma$, $h \to b \bar b$ and $h \to \tau^+ \tau^-$ channel (see~\cite{Aaboud:2017uak,ATLAS-CONF-2018-039,CMS-PAS-EXO-16-050,CMS:2018yme,Sirunyan:2018qob}  for the latest ATLAS and CMS results).  While all searches use $\MET$ as the main selection variable to discriminate signal from background,  the  $h \, (\gamma \gamma) + \MET$  channel  is sensitive to lower $\MET$ values than the $h \, (b \bar b) + \MET$  channel, because events can be selected (triggered) based on the presence of photons, and data recording occurs at a sustainable rate at a lower $\MET$ threshold. The $h \, (b \bar b) + \MET$  channel has instead the advantage that it is more sensitive to smaller $h + \MET$ production cross sections. These features make the two modes complementary, as models with  small  splittings $\mA - \ma$ are best probed in the former channel, while realisations with a larger mass hierarchy can be better probed via the $h \, (b \bar b) + \MET$ final state. We add that the CMS Collaboration has very recently provided first constraints on the \hdma model using the $h \, (b \bar b) + \MET$ signal~\cite{CMS-PAS-EXO-16-050}. The results obtained are compatible with the ones presented in Section~\ref{sec:sensi_monohbb} of this white paper. The decay channel~$h \to WW$ also offers interesting prospects to search for a mono-Higgs signal in the \hdma model~\cite{GPHeidelberg} but no results from LHC experiments have been presented so far. 

\subsubsection[Mono-$Z$ signature]{Mono-$\bm{Z}$ signature}
%\subsection{Mono-$\bm{Z}$ signature}

\begin{figure}[t!]
\centering
\includegraphics[height=0.45\textwidth]{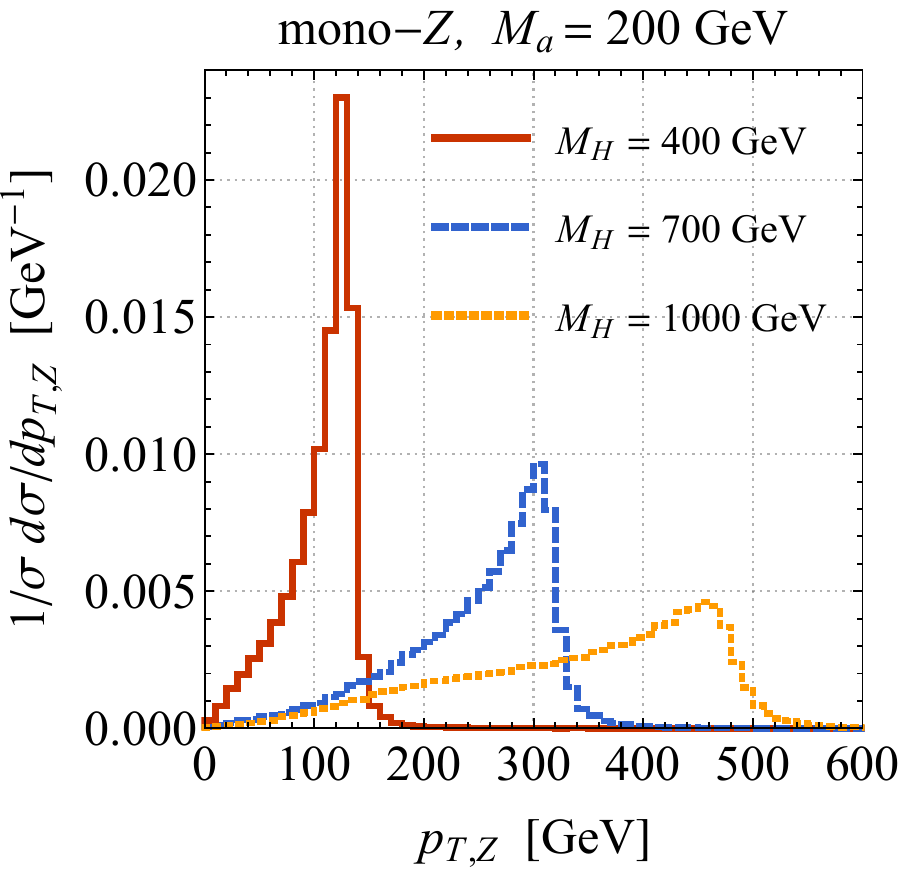}	\qquad 
\includegraphics[height=0.45\textwidth]{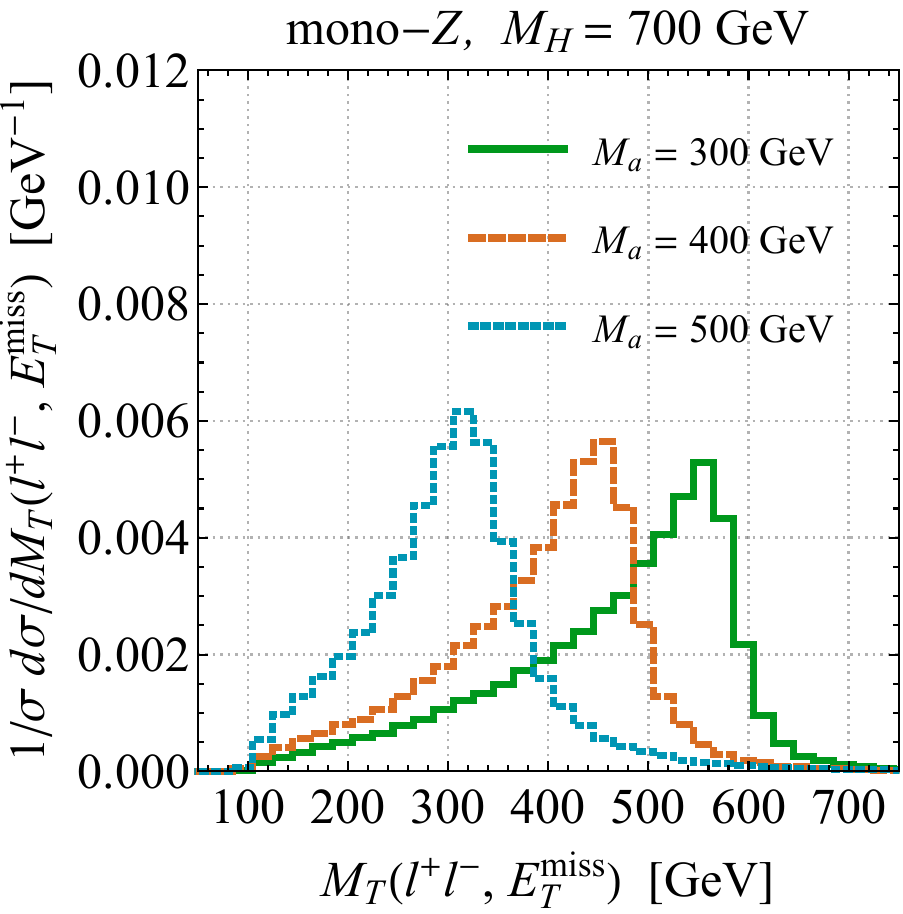}
\vspace{2mm}
\caption{\label{fig:zptmt} Normalised $p_{T,Z}$ (left panel) and $M_T (\ell^+ \ell^-, \MET)$ (right panel) distributions for $Z + \MET$ production followed by $Z \to \ell^+ \ell^-$. The  predictions shown have been obtained for the \hdma benchmark parameter choices  given in~\eqref{eq:benchmark} and employ different values of~$M_H$ and~$M_a$ as indicated in the legends.}
\end{figure}

As for the mono-Higgs signal, an analysis of the shape of the $\MET$ variable in the mono-$Z$ case offers a powerful way to enhance the signal-to-background ratio. The endpoint of the $\MET$ spectrum for the $Z+\MET$ signature can be obtained from~\eqref{eq:monoHMETpeak} by  replacing $M_A \to M_H$ and $M_h \to M_Z$.  Since the four-momenta of the decay products $Z$ and $a$ that enter $H \to Za$ are fixed by $H$ being preferentially on-shell, also the spectrum of the $Z$-boson transverse momentum ($p_{T,Z}$) in mono-$Z$ production will have a characteristic shape if $M_H > M_Z + M_a$.   In fact,  the $p_{T,Z}$ distribution is predicted  to be Jacobian with a cut-off at~\cite{No:2015xqa,Bauer:2017ota}
\begin{equation} \label{eq:monoZpTpeak}
p_{T,Z}^{ {\rm max} } \simeq \frac{\lambda^{1/2}(M_H, M_Z, M_a)}{2 M_H}\,,
\end{equation}
that is smeared by the total decay width $\Gamma_H$ of the heavy Higgs $H$. Ignoring higher-order QED and EW corrections and detector effects the shapes of the $p_{T,Z}$ and $\MET$ spectra are identical. Whether a shape fit to $\MET$ or $p_{T,Z}$ provides a better experimental reach thus depends to first approximation only on which of the two variables can be better measured and the corresponding backgrounds can be controlled.

Another useful observable to study the properties of the mono-$Z$ signal is the transverse mass 
\begin{equation} \label{eq:transversemass}
M_T (\ell^+ \ell^-,\MET) = \sqrt{2 p_{T, \ell^+ \ell^-}  \MET  \left ( 1- \cos \Delta \phi \right )} \,,
\end{equation}
constructed from the $\ell^+ \ell^-$ system and the amount of $\MET$. Here $p_{T, \ell^+ \ell^-}$ denotes the transverse momentum of the lepton pair and $\Delta \phi$ is the azimuthal angle between the $\ell^+ \ell^-$ system and the $\MET$ direction.

Figure~\ref{fig:zptmt} displays  $p_{T,Z}$ and $M_T(\ell^+ \ell^-, \MET)$ distributions for different choices of the masses~$M_H$ and~$M_a$. The parameters not explicitly specified in the plots have been  fixed to the values reported in~\eqref{eq:benchmark}. The differential distributions in $p_{T,Z}$ and $M_T(\ell^+ \ell^-, \MET)$ have Jacobian peaks, a feature that  reflects the resonant production of a~$H$ with the subsequent decay $H \to Z a \to \ell^+ \ell^- \chi \bar \chi$.  Increasing~$\mH$ ($\ma$) again shifts the endpoints of the distributions to higher~(lower) values of $p_{T,Z}$ and $M_T(\ell^+ \ell^-, \MET)$. Like in the mono-Higgs case, for large mass differences $M_H - M_a$, box diagrams lead to a non-negligible mono-$Z$ rate at low values of $p_{T,Z}$ and $M_T(\ell^+ \ell^-, \MET)$. Compared to the $h + \MET$ signature, the interference effects between resonant and non-resonant contributions  are less pronounced  in  the $Z + \MET$  case. 

The existing LHC searches for a mono-$Z$ signal (cf.~\cite{Aaboud:2017bja,Sirunyan:2017qfc} for the most recent results) have focused either on invisible decays of the SM-like Higgs boson or on topologies where the $Z$~boson is produced in the form of ISR.  Since ISR of a $Z$ boson is suppressed by both the coupling of the $Z$ to SM fermions and its mass compared to the radiation of a gluon~\cite{Petriello:2008pu,Bell:2012rg,Carpenter:2012rg}, the mono-$Z$ signal is generically not a discovery channel in models that lead to ISR-like mono-$X$ signatures. In contrast, in the \hdma model the $Z+\MET$ signature is more sensitive than the $j+\MET$ channel.

The above discussion has focused on the leptonic decay of the~$Z$~boson, but searching for a mono-$Z$ signal in the hadronic channel is also possible. In fact, the hadronic and leptonic signatures  are complementary, since hadronic decays of the~$Z$~boson are more frequent than leptonic decays, but suffer from larger backgrounds. An improved background suppression is possible if ``boosted'' event topologies are studied as in~\cite{Aad:2013oja,Sirunyan:2017jix}, making the hadronic mono-$Z$ signature an interesting channel if the \hdma model includes high-mass Higgs states. 

\subsubsection{Single-top signatures}

\begin{figure}[t!]
\centering
\includegraphics[height=0.45\textwidth]{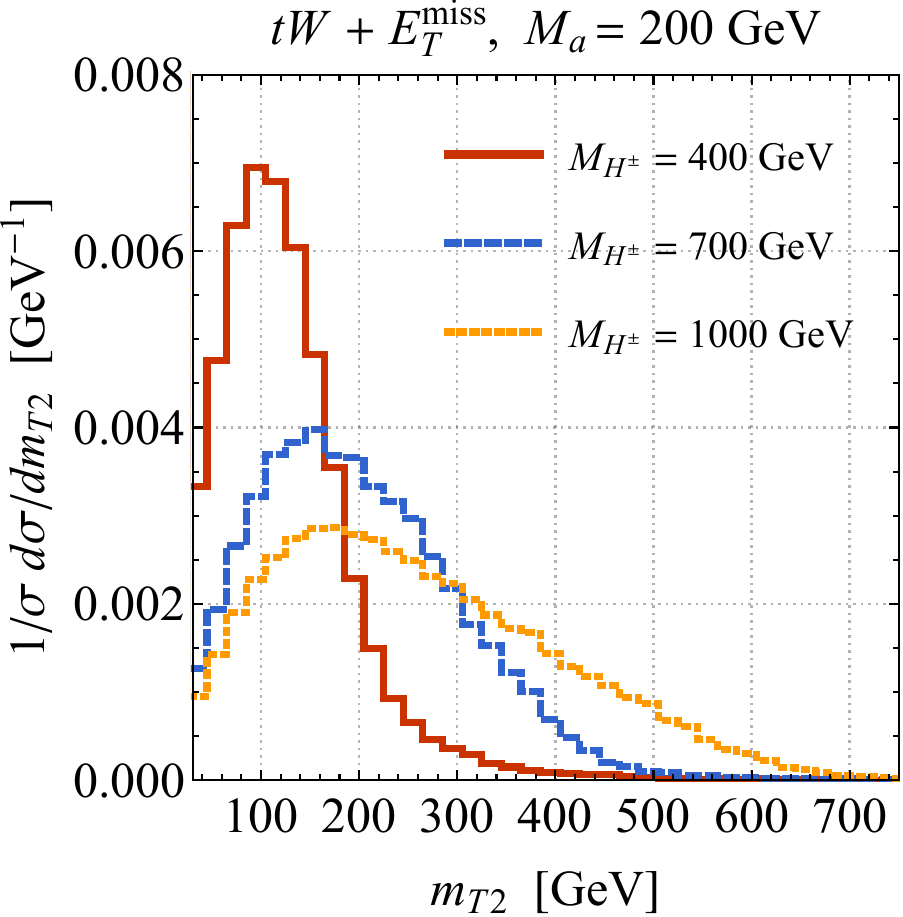} \qquad 
\includegraphics[height=0.45\textwidth]{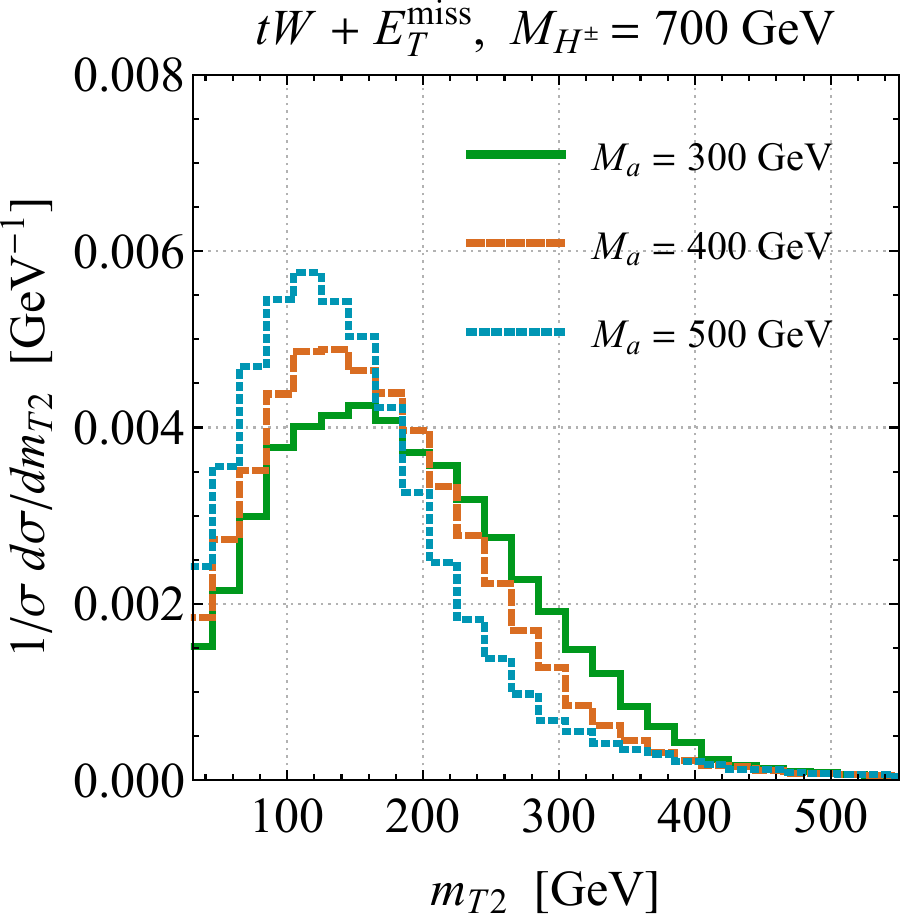}
\vspace{2mm}
\caption{\label{fig:mt2spectra} Normalised $m_{T2}$ distributions for $tW+\MET$ production in the double-lepton channel.  The  results shown correspond to the \hdma benchmark~\eqref{eq:benchmark} and employ different values of~$\mHc$ and~$\ma$ as indicated in the legends.}
\end{figure}

Single-top production in association with $\MET$ is also a promising mono-$X$ channel in the case of spin-0 models~\cite{Pinna:2017tay,Pani:2017qyd,Plehn:2017bys}. The single-top production in the $s$-channel, $t$-channel  or  in association with a $W$ boson can be studied. In the following, we will focus on the $tW + \MET$ channel, which in the context of the \hdma model has been identified as the most interesting mode~\cite{Pani:2017qyd}. Example diagrams leading to a $tW + \MET$ signature are shown in the lower row of~Figure~\ref{fig:resonant}. The $tW + \MET$ signal can be searched for in the single-lepton and double-lepton final state. Analysis strategies for both channels have been developed in~\cite{Pani:2017qyd}. In the former case, $M_T (\ell, \MET)$ and the asymmetric transverse mass $am_{T2}$~\cite{Konar:2009qr,Lester:2014yga} can be used to discriminate between signal and background, while in the latter case the stransverse mass $m_{T2}$~\cite{Lester:1999tx,Barr:2003rg} plays a crucial role in the background suppression.

Examples of normalised $m_{T2}$ distributions obtained in the \hdma model are shown in Figure~\ref{fig:mt2spectra}. The coloured histograms correspond to different masses $\mHc$ and $\ma$. The parameters not indicated in the legends have been set to the values given in~\eqref{eq:benchmark}. The shape of the $m_{T2}$ spectrum is sensitive to the values that are chosen for $\mHc$ and $\ma$. In particular, the maximum of the $m_{T2}$ distribution is shifted to higher values for larger (smaller) values of $\mHc$ ($\ma$). For heavy charged Higgses the $m_{T2}$ spectrum develops a pronounced high-$m_{T2}$ tail.  This feature can be traced back to the resonant contribution $b g \to t H^+  \to  t W^+ a  \to t W^+ \chi \bar \chi$ (see~lower left graph in~Figure~\ref{fig:resonant}).  At present, only a single LHC analysis exists~\cite{CMS-PAS-EXO-18-010}  that considers the $tW + \MET$  or other single-top-like signatures with $\MET$. Performing further studies of these channels would, however, be worthwhile, since enhanced single-top signatures are expected to appear in many DM model that features an extended Higgs sector. 

\subsection[Non-resonant $E_T^{\rm miss}$ signatures]{Non-resonant $\bm{E_T^{\rm miss}}$ signatures}
\label{sec:nonresonant}

\begin{figure}[t!]
\centering
\includegraphics[width=.725\textwidth]{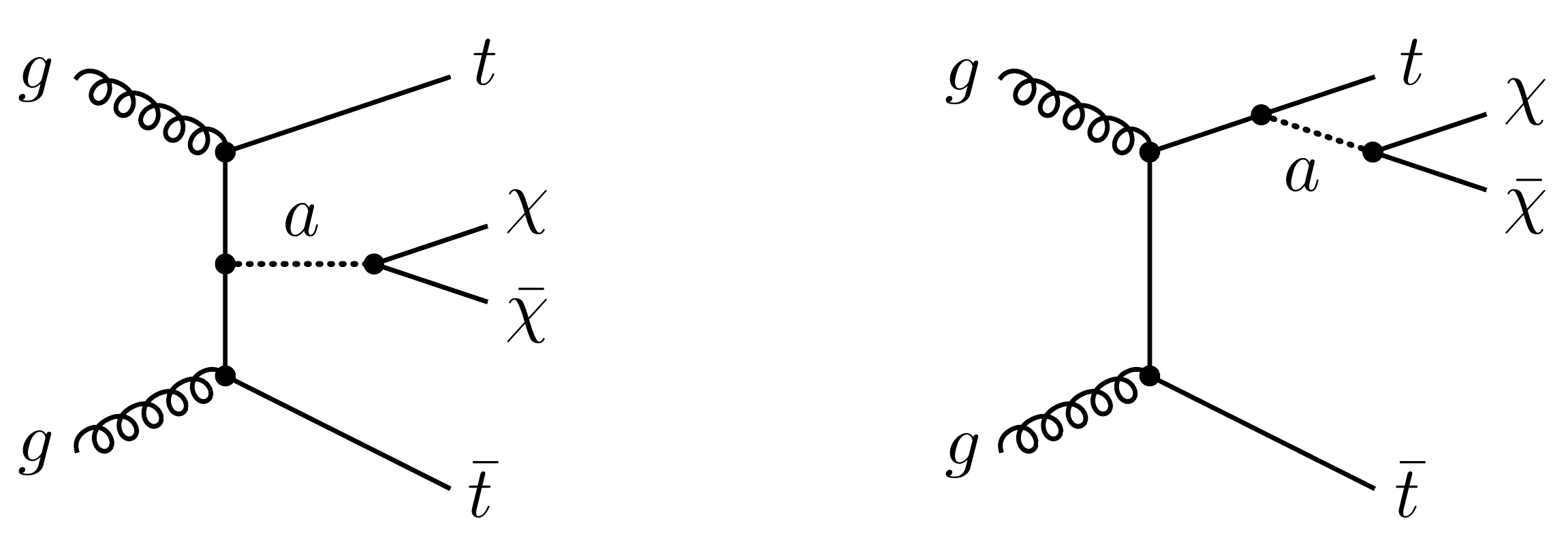}

\vspace{7mm}

\includegraphics[width=.8\textwidth]{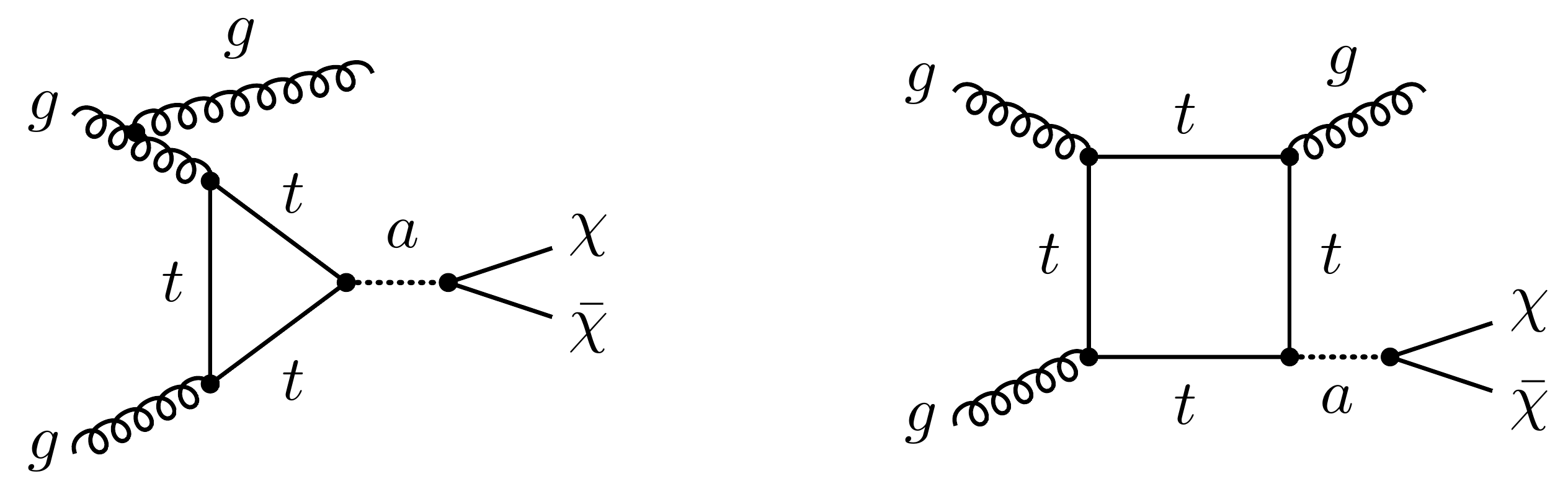}

\vspace{4mm}
\caption{\label{fig:nonresonant} Prototype diagrams that lead  to a $t \bar t+\MET$~(upper row) and $j+\MET$~(lower row) signal in the \hdma model. Graphs involving a heavier pseudoscalar $A$ also contribute to the signals but are not shown explicitly.}
\end{figure}

Besides the resonant $\MET$ signatures discussed in Section~\ref{sec:resonant}, the   \hdma model also predicts to non-resonant mono-$X$ signatures. The most important channels in this class are $t \bar t +\MET$ and $j+\MET$ production. In addition, the $b \bar b +\MET$ mode is interesting because its rate is $\tan \beta$ enhanced if a Yukawa sector of  type-II is realised.  Feynman graphs leading to the first two signatures are depicted in~Figure~\ref{fig:nonresonant}. For $M_A \gg M_a > 2 m_\chi$ the dominant contribution to the  $t \bar t +\MET$ and mono-jet signals arise from diagrams involving the  mediator $a$. In this limit the normalised kinematic distributions of the $t \bar t + \MET$ and $j+\MET$ signals in the \hdma model resemble those obtained in the DMF pseudoscalar   model. Since the contributions  associated to $a$ and $A$ exchange interfere with each other, shape differences can, however, occur if the pseudoscalars are not widely separated in mass~\cite{Bauer:2017ota}.

\subsubsection{Heavy-quark signatures}

Two of the main channels that have been used up to now  to search for spin-0 states with large invisible decay widths at the LHC are $t \bar t + \MET$ and $b \bar b + \MET$. The latest ATLAS and CMS analyses of this type can be found in~\cite{CMS-PAS-EXO-18-010,Aaboud:2017rzf,Sirunyan:2018dub}. These searches have been interpreted in the context of the DMF spin-0   models, and for $M_A \gg M_a$ the obtained cross-section limits can be used  to derive exclusion bounds in the \hdma model by using~\cite{Bauer:2017ota}
\begin{equation} \label{eq:rescaling}
\frac{\sigma \! \left ( pp \to t \bar t + \MET\right )_{\hdma}}{\sigma \! \left ( pp \to t \bar t + \MET\right )_{\rm DMF}} \simeq \left ( \frac{y_\chi \sin \theta}{g_\chi g_q \tan \beta} \right )^2 \, .
\end{equation}
Here $g_\chi$ ($g_q$) denotes the DM-mediator (universal quark-mediator) coupling in the DMF  pseudoscalar  model. An analog formula holds in the case of the $b \bar b + \MET$ signature with $\tan \beta$ replaced by $\cot \beta$ in the type-II \hdma model. 

\begin{figure}[t!]
\centering
\includegraphics[height=0.45\textwidth]{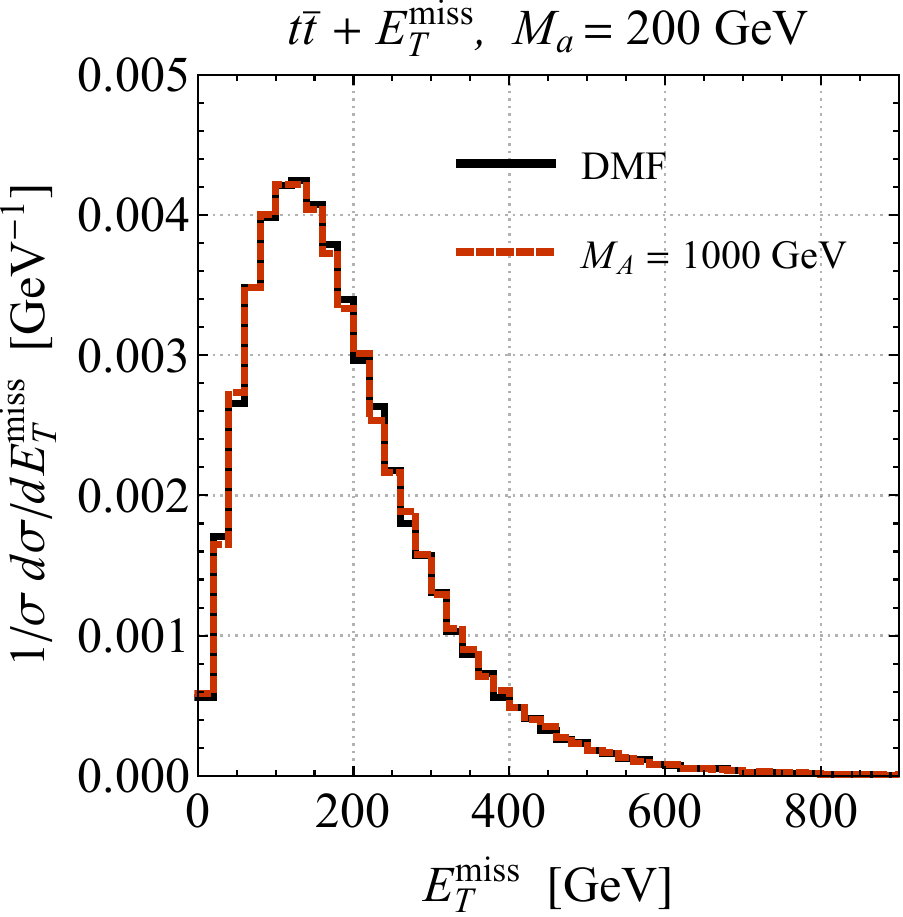} \qquad 
\includegraphics[height=0.45\textwidth]{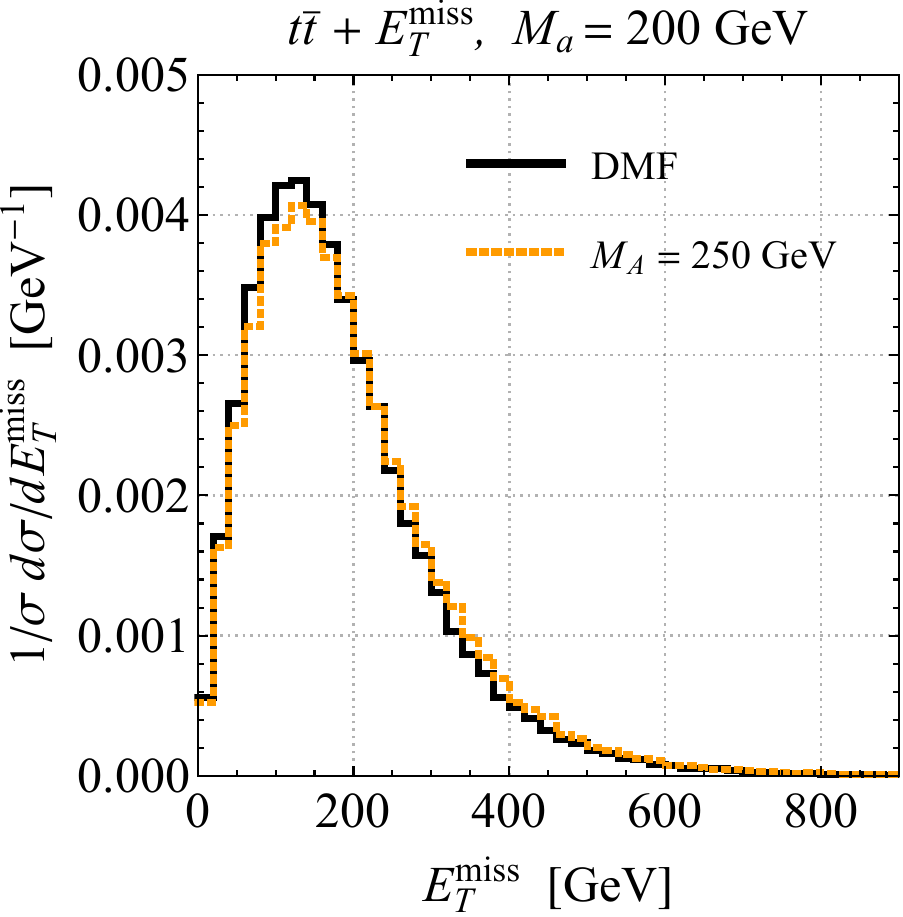}
\vspace{2mm}
\caption{\label{fig:ttmet} Normalised $\MET$ distributions for $t \bar t + \MET$ production. The black curves correspond to the prediction of the DMF pseudoscalar   model, while the coloured predictions illustrate the results in the \hdma benchmark model~\eqref{eq:benchmark} for two different choices of $\mA$ and $\ma$.} 
\end{figure}

In Figure~\ref{fig:ttmet} we compare two normalised $\MET$ spectra obtained in the \hdma model~(coloured histograms) to the prediction of the DMF pseudoscalar   model (black histograms).  The left panel illustrates the case~$M_A \gg M_a$, and one observes that the shape of the \hdma distribution resembles  the one of the DMF model within statistical uncertainties. As shown in the plot on the right-hand side, shape distortions instead arise if~the particle masses $M_A$ and $M_a$ are not widely separated.  Similar findings apply to other variables such as $m_{T2}$ which  plays a crucial role in suppressing the $t \bar t$~background in two-lepton analyses of the $t \bar t + \MET$ signature~\cite{Aaboud:2017rzf,Haisch:2016gry,CMS-PAS-EXO-17-014}. It follows that in order to accurately reproduce the kinematic distributions of the signal in the entire \hdma~parameter space, one should not rely on \eqref{eq:rescaling} but should use a more sophisticated method. A~general approach  that allows to faithfully translate existing limits on DMF spin-0   models into the \hdma~parameter space  is described in~Appendix~\ref{app:recast}. There it is also shown that this rescaling procedure reproduces the results of a direct MC simulation.  In~Appendix~\ref{app:ttMETscalar} we furthermore demonstrate that the same findings apply to the $t \bar t + \MET$ signature in the 2HDM+s model  (see~Section~\ref{sec:2HDMs} for a brief discussion of the model).

\subsubsection{Mono-jet signature}

At the LHC the most studied mono-$X$ signal is the $j +\MET$ channel $\big($the newest analyses have been presented in \cite{Aaboud:2017phn,Sirunyan:2017jix}$\big)$ because this mode typically provides the strongest~$\MET$ constraints on models with ISR-type  signatures. Since only loop diagrams where a mediator couples to a quark (see~the graphs in the lower row in Figure~\ref{fig:nonresonant}) contribute to the mono-jet signature in both the \hdma and the DMF spin-0 models, the normalised kinematic distributions of the $j +\MET$ signal turn out to be very similar in these models.  In the case that the 2HDM pseudoscalar $A$ is decoupled,~i.e.~$\mA \gg \ma$, one can use the right-hand side of the relation~\eqref{eq:rescaling} to translate the existing mono-jet results on the DMF pseudoscalar model into the~\hdma parameter space, while in general one can apply the recasting procedure detailed in~Appendix~\ref{app:recast}.

\subsection{Parameter variations}
\label{sec:parameter_variations}

The  kinematic distributions shown in Sections~\ref{sec:resonant} and \ref{sec:nonresonant}  all employ the parameters~\eqref{eq:benchmark} and consider only variations of the common heavy  spin-0 boson  mass $\mH = \mA = \mHc$ and the mediator mass $\ma$.  In this subsection we study the impact that modifications of the parameters away from the proposed \hdma benchmark scenarios have. The  discussion will thereby focus on the mono-Higgs and mono-$Z$ signatures since the rates and kinematic distributions of these two channels turn out to be  most sensitive to parameter changes. 

\subsubsection[Variations of $\mH$ and $\mA$]{Variations of $\bm{\mH}$ and $\bm{\mA}$}
%\subsubsection{Variations of $\bm{\mH}$ and $\bm{\mA}$}

\begin{figure}[t!]
\centering
\includegraphics[height=0.45\textwidth]{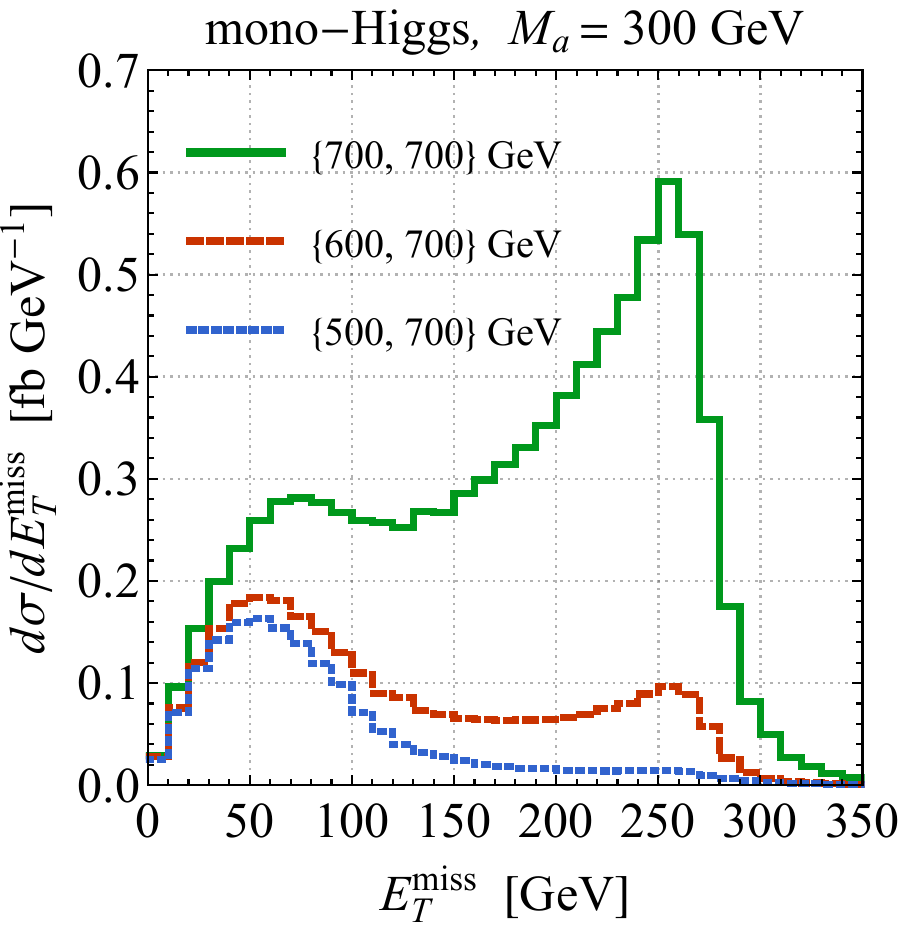} \qquad 
\includegraphics[height=0.45\textwidth]{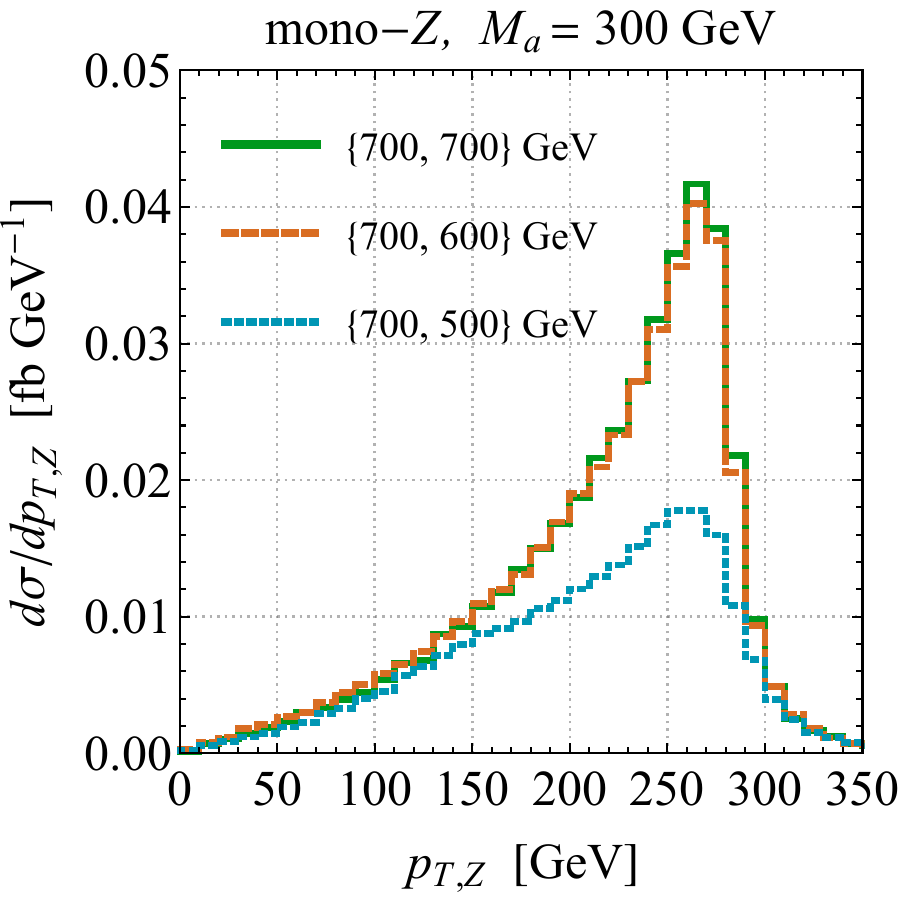}
\vspace{2mm}
\caption{\label{fig:mvar} $\MET$ ($p_{T,Z}$) distributions for mono-Higgs~(mono-$Z$) production at $13 \, {\rm TeV}$ in the \hdma model.  The  predictions shown correspond to different sets $\{M_{H}, M_A\}$ of masses and employ $\mHc = {\rm min} \left ( \mH, \mA \right )$, $M_a = 300 \, {\rm GeV}$  as well as the parameters~\eqref{eq:benchmark}.}
\end{figure}

In Figure~\ref{fig:mvar} we display $\MET$ distributions in $h + \MET$ production~(left panel) and $p_{T,Z}$ distributions in $Z+\MET$ production~(right panel) for different~$\mH$ and~$\mA$ values. As~indicated, the coloured histograms correspond to  different choices of $M_{H,A}$ and $\mHc = {\rm min} \left ( \mH, \mA \right )$, but all employ $M_a = 300 \, {\rm GeV}$. From the figure it is evident that the inclusive mono-Higgs~(mono-$Z$) cross section is reduced compared to the benchmark prediction if~$\mH$~($\mA$) is taken to be smaller than $\mA$~($\mH$).  We furthermore observe that  a change of $\mH$ strongly affects the shape of the $\MET$ distribution in the mono-Higgs channel, while the distortions in the $p_{T,Z}$ distribution of the mono-$Z$ signature under variations of~$\mA$ are much less pronounced. The strong $M_H$-dependence of the~$\MET$ spectrum in $h + \MET$ production can be traced back to the structure of the coupling~$g_{Aha}$. From~(\ref{eq:cubic}) one sees that for smaller~$M_H$ also $g_{Aha}$ is smaller, leading to a reduced $A \to ha$ branching ratio  and in turn to a lower rate of resonant production.  In contrast, the coupling~$g_{HZa} \propto \sin \theta$ that drives resonant production in the case of the mono-$Z$ signal does not depend on the value of~$M_A$.

In order to minimise the constraints from EW precision observables (see the discussion in Section~\ref{sec:EWPO}) we have chosen $\mH = \mHc$ in the  benchmark scenario~\eqref{eq:benchmark}. The further choice of having a common 2HDM  spin-0 boson  mass $\mH = \mA = \mHc$ is then motivated by the observation that in such a case  both the $h + \MET$ and $Z + \MET$ signature are dominated by resonant production. While in our sensitivity studies presented in the next section we will always employ the choice $\mH = \mA = \mHc$, in future \hdma  interpretations of mono-$X$ searches one might, however,  also want to consider cases with~$\mH \neq \mA$. 

\subsubsection[Variation of $\sin \theta$]{Variation of $\bm{\sin \theta}$}
%\subsubsection{Variation of $\bm{\sin \theta}$}

\begin{figure}[t!]
\centering
\includegraphics[height=0.45\textwidth]{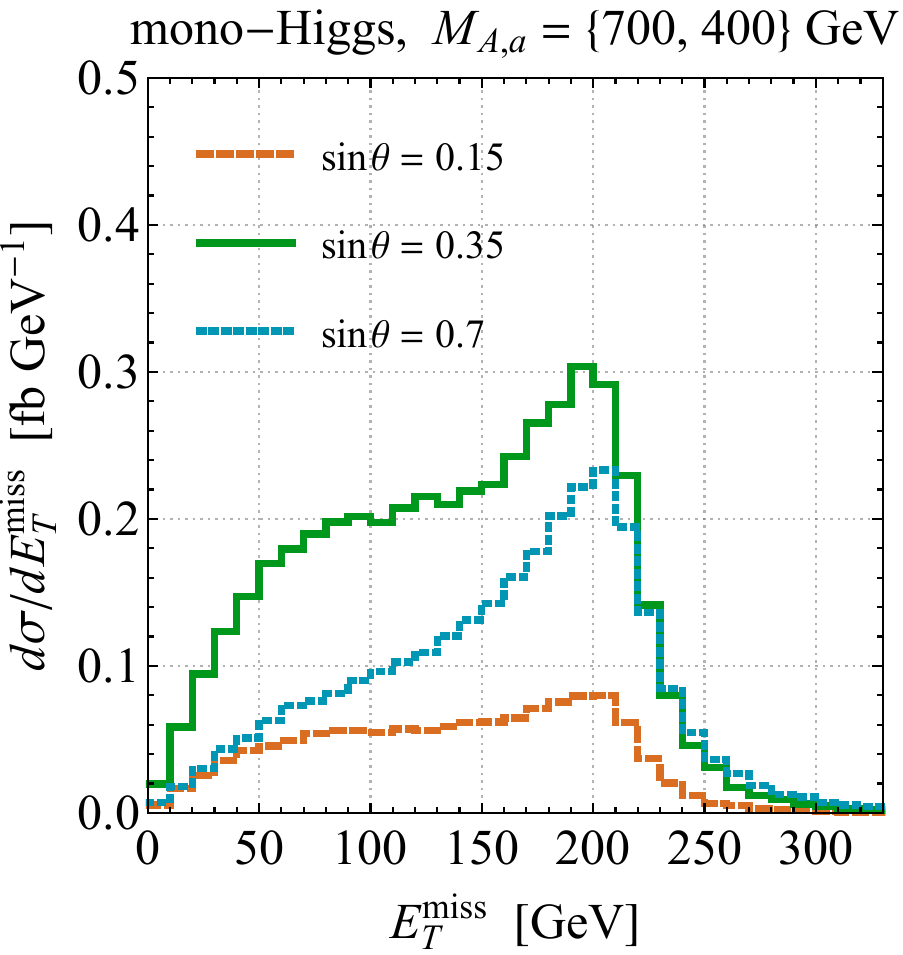} \qquad 
\includegraphics[height=0.45\textwidth]{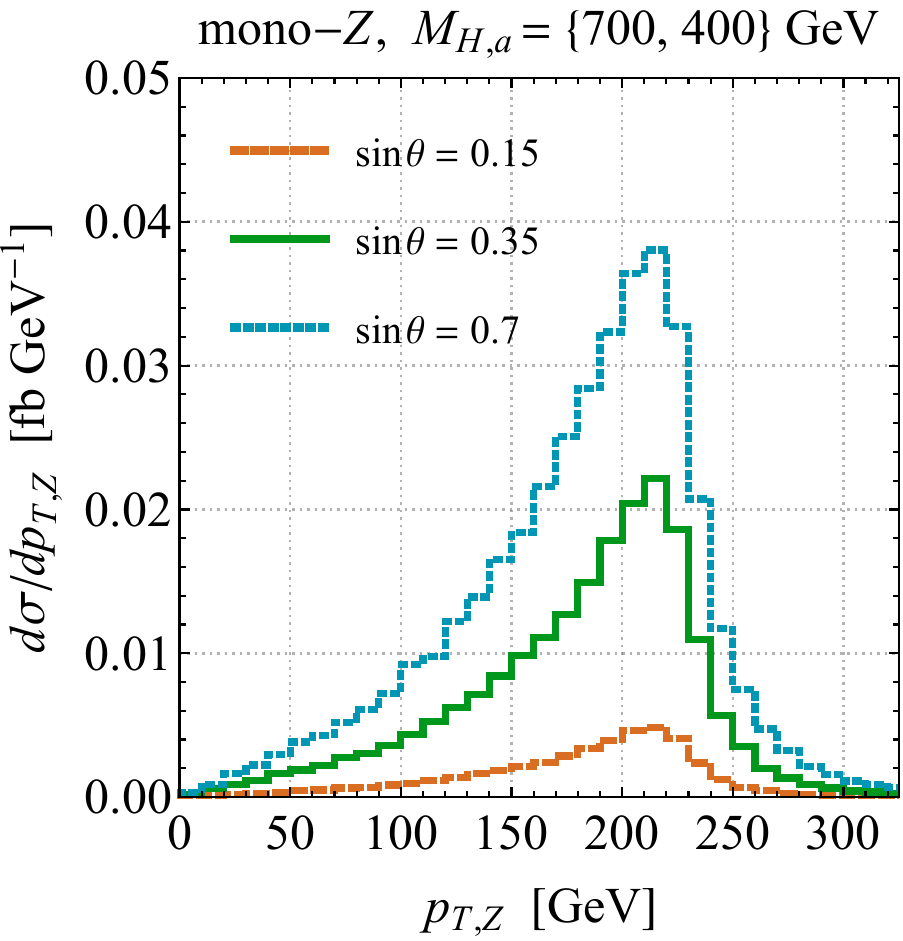}
\vspace{2mm}
\caption{\label{fig:svar} $\MET$ ($p_{T,Z}$) distributions for mono-Higgs~(mono-$Z$) production at $13 \, {\rm TeV}$  in the \hdma model. The displayed results correspond to different choices of $\sin \theta$. The remaining parameters are fixed to~\eqref{eq:benchmark} using $\mH = \mA = \mHc = 700 \, {\rm GeV}$ and $\ma = 400 \, {\rm GeV}$. }
\end{figure}

Figure~\ref{fig:svar} shows $\MET$ distributions in $h + \MET$ production~(left panel) and $p_{T,Z}$ distributions in $Z+\MET$ production~(right panel) for different~values of $\sin \theta$. The spin-0 masses are chosen as $\mH = \mA = \mHc = 700 \, {\rm GeV}$ and $\ma = 400 \, {\rm GeV}$, and the remaining parameters are fixed to~\eqref{eq:benchmark}. It can be observed that the variation of $\sin \theta$ leads to both a rate and shape change in the case of the mono-Higgs signal, while in the case of the mono-$Z$ channel only the total cross section gets rescaled to first approximation (in the peak region the shape changes of the $p_{T,Z}$ distribution amount to at most $\pm 10\%$ for the considered $\sin \theta$ values). The strong sensitivity of the shape of the $\MET$ spectrum in $h + \MET$ production is again a result of the interplay of resonant and non-resonant contributions. While the $gg \to A \to h a \to h \chi \bar \chi$  amplitude scales as $\sin \theta \cos^2 \theta$, the $gg \to h a  \to h \chi \bar \chi$ matrix element shows a $\sin \theta$ dependence. These scalings imply that at moderate  (small and large) $\sin \theta$ the resonant~(non-resonant) amplitudes provide the dominant contribution to the $\MET$ distribution in mono-Higgs production.  In the case of the mono-$Z$ signal the resonant and non-resonant amplitudes both scale as $\sin \theta$ and in consequence  all kinematic distributions are essentially not distorted  under changes of the mixing angle~$\theta$. The latter statement also holds in the case of the $t \bar t +\MET$, $b \bar b + \MET$ and mono-jet signatures.  This can be deduced from~\eqref{eq:rescaling}.

From the above discussion it follows that the choice $\sin \theta = 0.35$ made in~\eqref{eq:benchmark} leads to an enhanced sensitivity of the mono-Higgs signal to the \hdma parameter space. To perform parameter scans in scenarios with larger mixing angles like $\sin \theta = 0.7$ would, however, also be worthwhile because such a choice would lead to an improved coverage via the mono-$Z$ channel. We finally note that in scenarios with $\sin \theta >0.35$ the maximal allowed size of mass splitting $|M_H - M_a|$ can, depending on the choice of $\lambda_3$, be severely constrained by vacuum stability arguments. This can be seen from~\eqref{eq:BFB2}. 

\subsubsection[Variation of $\tan \beta$]{Variation of $\bm{\tan \beta}$}
\label{sec:variationtanb}

\begin{figure}[t!]
\centering
\includegraphics[height=0.45\textwidth]{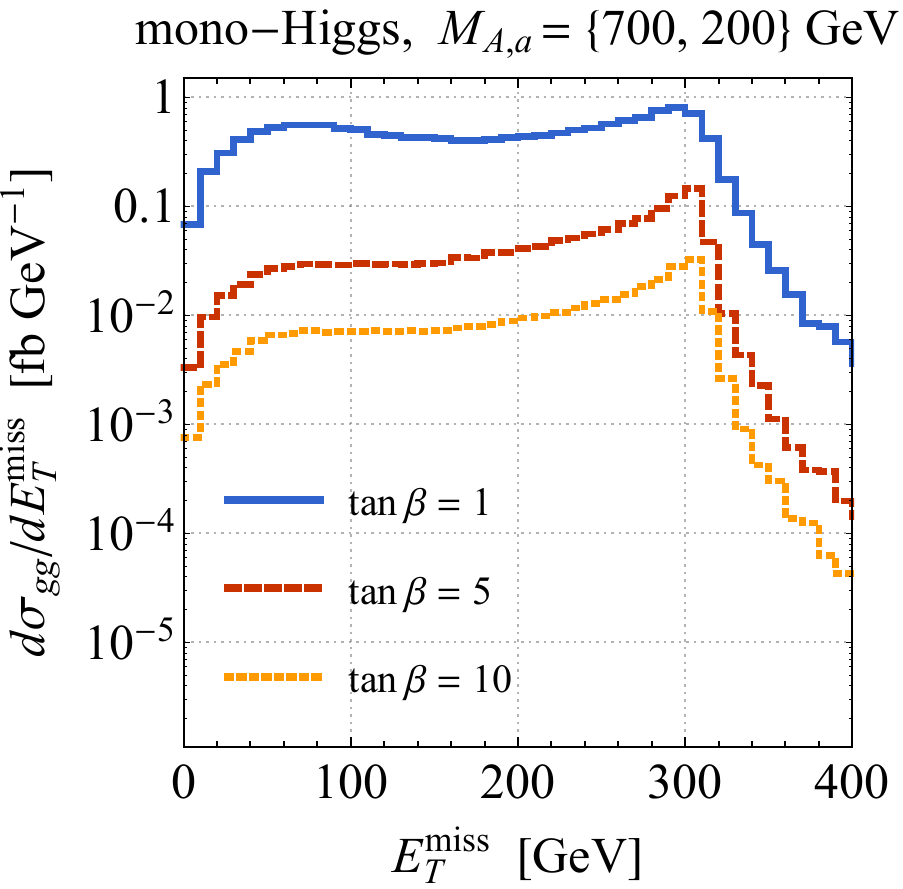} \qquad 
\includegraphics[height=0.45\textwidth]{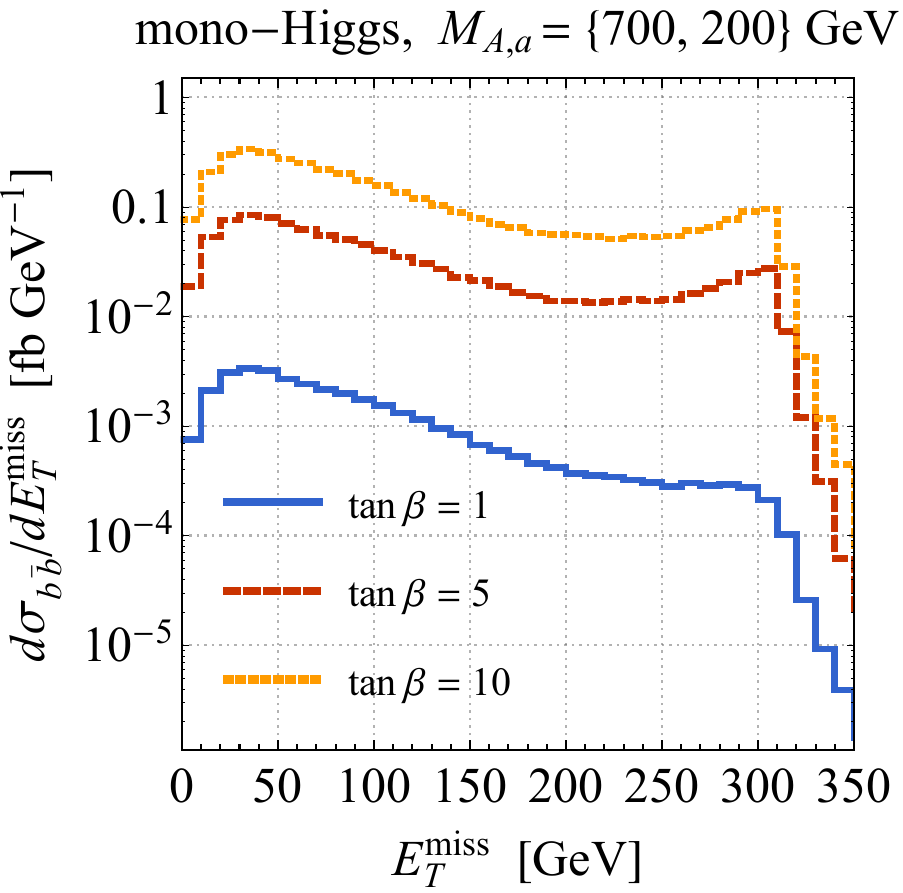}
\vspace{2mm}
\caption{\label{fig:tbvar1} 
$\MET$ distributions for mono-Higgs production in $gg$-fusion (left panel) and $b \bar b$-fusion (right panel) in the \hdma model. The displayed results correspond to $pp$ collisions at  $13 \, {\rm TeV}$ and different choices of $\tan \beta$. The parameters not detailed in the plots are set to~\eqref{eq:benchmark} using $\mH = \mA = \mHc = 700 \, {\rm GeV}$ and $\ma = 200 \, {\rm GeV}$.}
\end{figure}

In Figure~\ref{fig:tbvar1} we display $\MET$ distributions in mono-Higgs production for different choices of $\tan \beta$. The left (right) panel illustrates the contributions from the $gg \to h + \MET$ ($b \bar b \to h + \MET$) channel. The  results shown employ~\eqref{eq:benchmark} with $\mH = \mA = \mHc = 700 \, {\rm GeV}$ and $\ma = 200 \, {\rm GeV}$. The total production cross section in $gg$-fusion strongly decreases with increasing $\tan \beta$, while in the case of $b \bar b$-fusion the opposite behaviour is observed. The strong reduction/enhancement of the production rates originates from the fact that in  the type-II~\hdma model considered in this white paper the couplings of $H,A,a$ to top quarks are proportional to $\cot \beta$, while the corresponding couplings to bottom quarks are proportional to $\tan \beta$. Numerically, we find that at the inclusive level the $gg$-fusion and $b\bar b$-fusion contributions to mono-Higgs production are comparable in size for $\tan \beta \simeq 5$. This means that for $\tan \beta \gtrsim 5$ both channels have to be included to obtain accurate predictions. From the two panels it is furthermore apparent that variations of $\tan \beta$ do not only change the overall signal strength, but also have a pronounced impact on the shapes of the $\MET$ distributions. In particular, changes in $\tan \beta$ influence the importance of resonant versus non-resonant contributions. 

\begin{figure}[t!]
\centering
\includegraphics[height=0.45\textwidth]{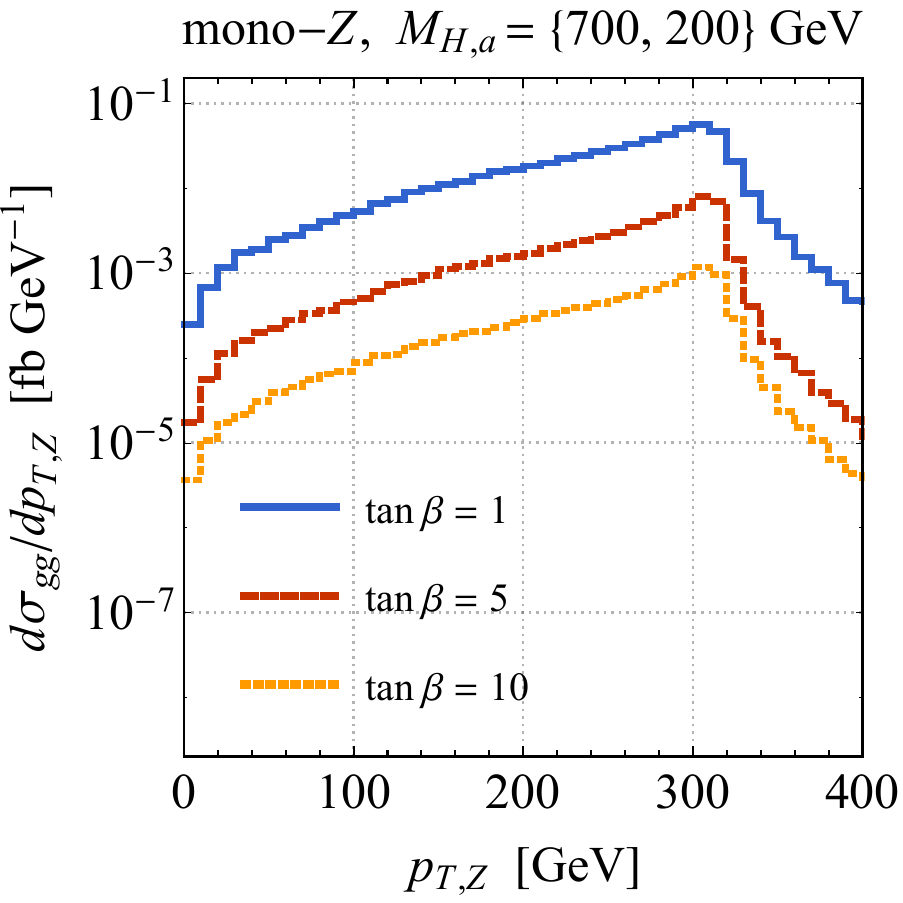} \qquad 
\includegraphics[height=0.45\textwidth]{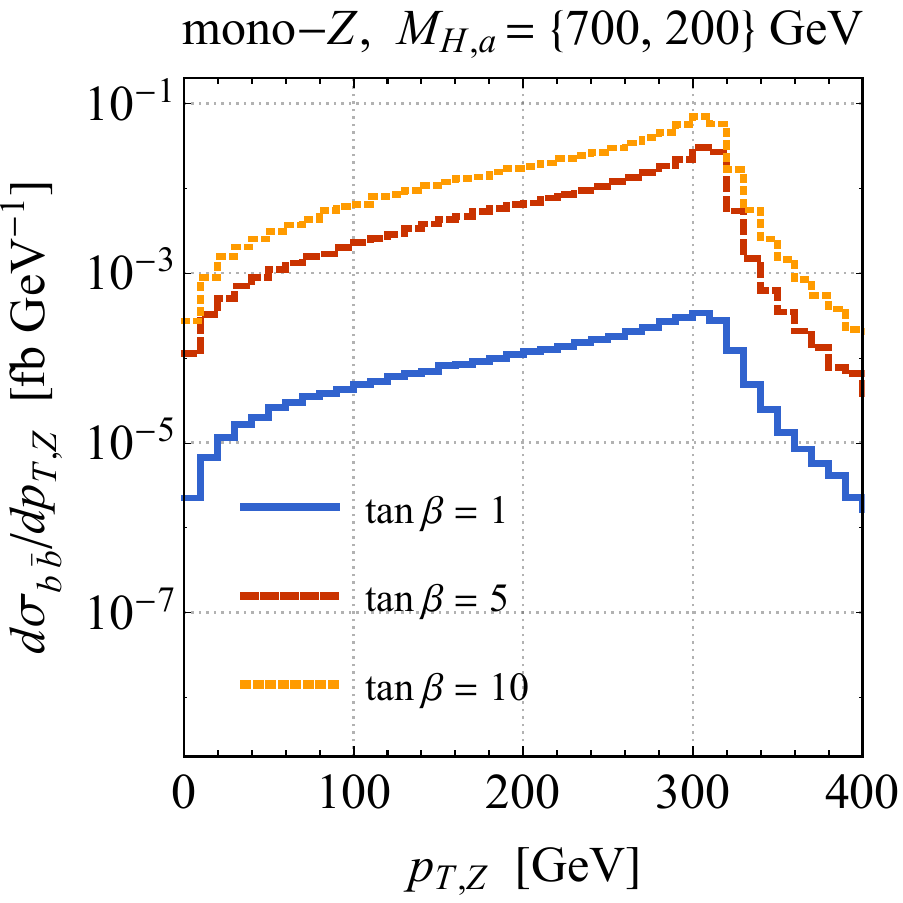}
\vspace{2mm}
\vspace{2mm}
\caption{\label{fig:tbvar2} 
$p_{T,Z}$ distributions for mono-$Z$ production in $gg$-fusion (left panel) and $b \bar b$-fusion~(right panel) in the \hdma model. The  predictions shown correspond to $pp$ collisions at $13 \, {\rm TeV}$ and the same choices of parameters as in Figure~\ref{fig:tbvar1} are employed.}
\end{figure}

 Similar to the mono-Higgs channel, $b \bar b$-initiated production can also be relevant for the mono-$Z$ signal,  if $\tan \beta$ is sufficiently large~\cite{Bauer:2017ota}. Figure~\ref{fig:tbvar2} displays $p_{T,Z}$ spectra in mono-$Z$ production for different choices of $\tan \beta$ in both the $gg$-fusion~(left panel) and $b \bar b$-fusion~(right panel) channel.  From the plots one sees that for the considered parameters $\mH = \mA = \mHc = 700 \, {\rm GeV}$ and $\ma = 200 \, {\rm GeV}$,  production in $b \bar b$-fusion dominates over $gg$-fusion already for the choice $\tan \beta = 5$. In the mono-$Z$ case   the shapes of the differential distributions are less distorted under changes of $\tan \beta$ than the mono-Higgs spectra. We furthermore  add that the modifications in the kinematic distributions of $t \bar t + \MET$ and $j +\MET$ production under changes of $\tan \beta$ are, like in the mono-$Z$ case, not very pronounced.
 
Our  scans in the $M_{a}\hspace{0.5mm}$--$\hspace{0.5mm} M_H$ plane are based on the choice $\tan \beta = 1$, since for this value the existing mono-Higgs and mono-$Z$ searches already provide sensitivity to/exclude large regions in the mass planes.  These  scans are complemented by sensitivity studies in the $\ma \hspace{0.5mm}$--$\hspace{0.5mm} \tan \beta$ (cf.~Section~\ref{sec:sensitivitystudies} and~\cite{No:2015xqa,Bauer:2017ota,Pani:2017qyd}).  We add that, if $\tan \beta$ is scanned, special attention has to be given to the fact that in the large-$\tan \beta$ limit the total decay widths of some of the Higgs states can become very large, potentially invalidating the narrow width assumption.   To~give an example, for the choice made in~\eqref{eq:matbscan} one has $\Gamma_{H}/M_H \gtrsim 30\%$ for $\tan \beta \gtrsim 10$ and $M_a \lesssim 300 \, {\rm GeV}$. 

\subsubsection[Variation of $m_\chi$]{Variation of $\bm{m_\chi}$}
%\subsubsection{Variation of $\bm{m_\chi}$}

\begin{figure}[t!]
\centering
\includegraphics[height=0.45\textwidth]{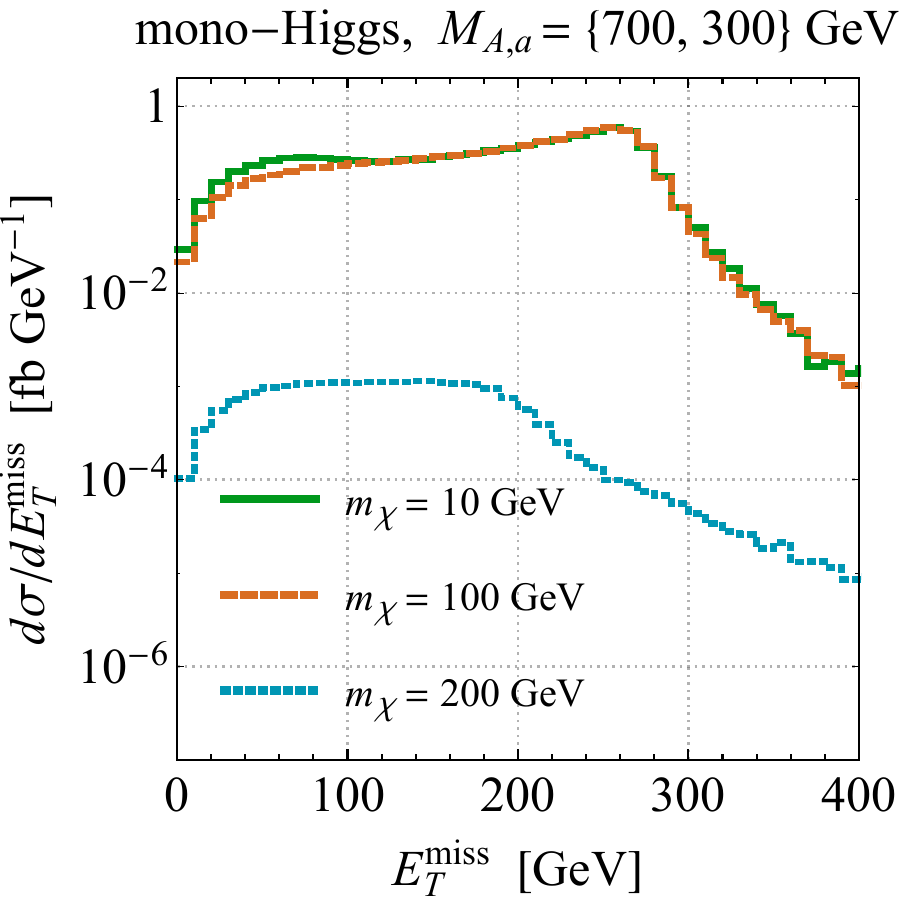} \qquad 
\includegraphics[height=0.45\textwidth]{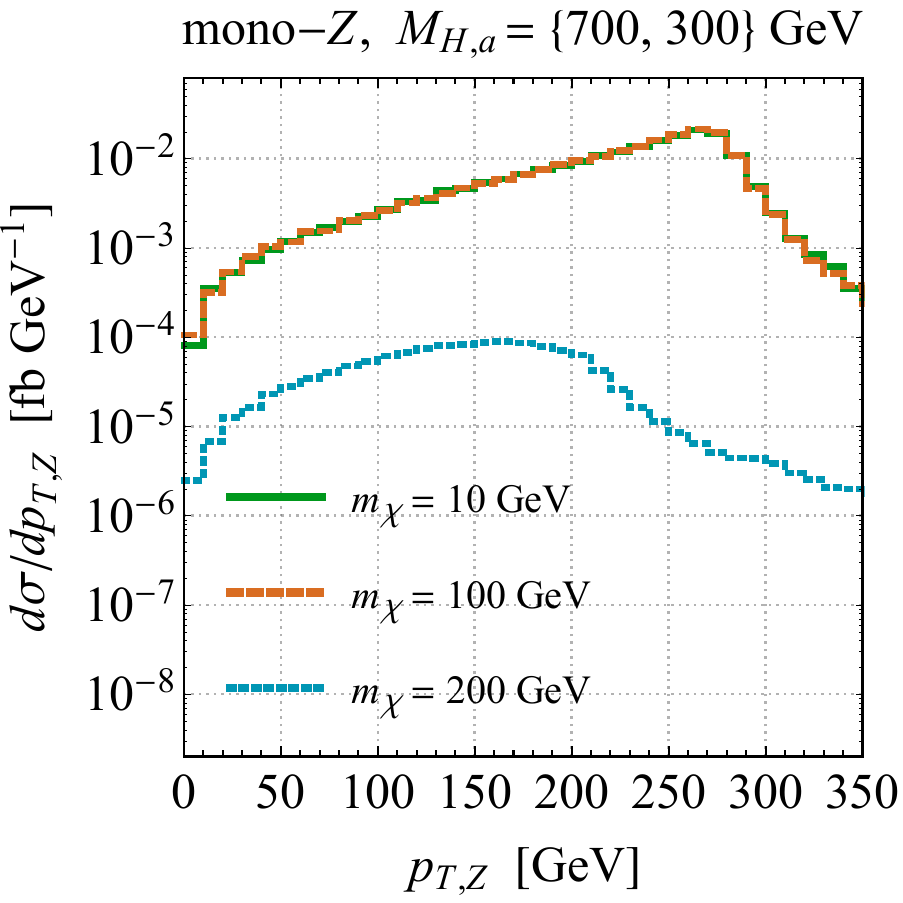}
\vspace{2mm}
\caption{\label{fig:mdmvar} $\MET$ ($p_{T,Z}$) distributions for mono-Higgs~(mono-$Z$) production at $13 \, {\rm TeV}$. The presented results correspond to different values of the DM mass $m_\chi$. The other \hdma parameters are set to~\eqref{eq:benchmark} using $\mH = \mA = \mHc = 700 \, {\rm GeV}$ and $\ma = 300 \, {\rm GeV}$. }
\end{figure}

The modifications of the $\MET$~($p_{T,Z}$) spectrum in $h+\MET$~($Z+\MET$) production under a variation of the DM mass $m_\chi$ are illustrated in the two panels of Figure~\ref{fig:mdmvar}. The given predictions correspond to $pp$ collision at $13 \, {\rm TeV}$ and employ the benchmark parameters~\eqref{eq:benchmark} with $\mH = \mA = \mHc = 700 \, {\rm GeV}$ and $\ma = 300 \, {\rm GeV}$. The depicted scenarios with $\ma  > 2 m_\chi$ (green and orange histograms) lead to almost identical rates, $\MET$ and $p_{T,Z}$ spectra, while the choice $\ma  < 2 m_\chi$ (blue histograms) largely reduces the total rates and also modifies the shapes of the shown distributions. This feature is expected since for $\ma > 2 m_\chi$ the decay channel $a \to \chi \bar \chi$ is kinematically allowed, while for $\ma < 2 m_\chi$ it is forbidden. In order to have detectable mono-$X$ signals even for light mediators~$a$, we have chosen a value of  $m_\chi = 10 \, {\rm GeV}$ as the baseline for the following sensitivity studies.  We will discuss the role that the DM mass $m_\chi$ plays in the context of DD, ID and the DM relic density in Section~\ref{sec:DMdetection}  and Section~\ref{sec:relic}, respectively. 

%%%%%%%%%%%%%%%%%%%%%%%%%%%%%%%%%%%%%%%%%%%%%%%%%%%%%%%%%%%%%%%%%%%%
%%%%%%%%%%%%%%%%%%%%%%%%%%%%%%%%%%%%%%%%%%%%%%%%%%%%%%%%%%%%%%%%%%%%
%%%%%%%%%%%%%%%%%%%%%%%%%%%%%%%%%%%%%%%%%%%%%%%%%%%%%%%%%%%%%%%%%%%%

\section{Non-$\bm{E_T^{\rm miss}}$ collider signatures  in the \hdma model}
\label{sec:nonMET}

In this section we will discuss the most important non-$\MET$ signals that can be used to explore the parameter space of the \hdma model at the LHC. Most of the discussion will be centred around final states containing top quarks since these channels provide the best sensitivity to model realisations with low $\tan \beta$ such as our benchmark parameter choice~\eqref{eq:benchmark}. Final states that give access to the \hdma parameter space with large $\tan \beta$ such as di-tau searches will, however, also be discussed briefly. 

\subsection{Di-top  searches}
\label{sec:ttbarresonances}

 In all 2HDM models, the spin-0 bosons $H,A$  decay dominantly to top-quark pairs if these states have masses above the top threshold,~i.e.~$M_{H, A} > 2 m_t$, and if $\cos (\beta - \alpha) \simeq 0$ and $\tan \beta = {\cal O}(1)$. New-physics scenarios of this kind can thus be tested by studying the~$t \bar t$ invariant mass spectrum $m_{t \bar t}$.  Interference effects between the signal process and the SM  $t \bar t$ background, however,  distort the $m_{t \bar t}$ signal shape from a single peak to a peak-dip structure~\cite{Gaemers:1984sj,Dicus:1994bm,Bernreuther:1997gs,Frederix:2007gi,Hespel:2016qaf,BuarqueFranzosi:2017jrj}, a feature that represents a serious obstacle to probe 2HDM models with $M_{H,A} > 350 \, {\rm GeV}$ and small $\tan \beta$ values~\cite{Craig:2015jba,Hajer:2015gka,Gori:2016zto,Carena:2016npr}. 

\begin{figure}
\centering
\includegraphics[width=.475\textwidth]{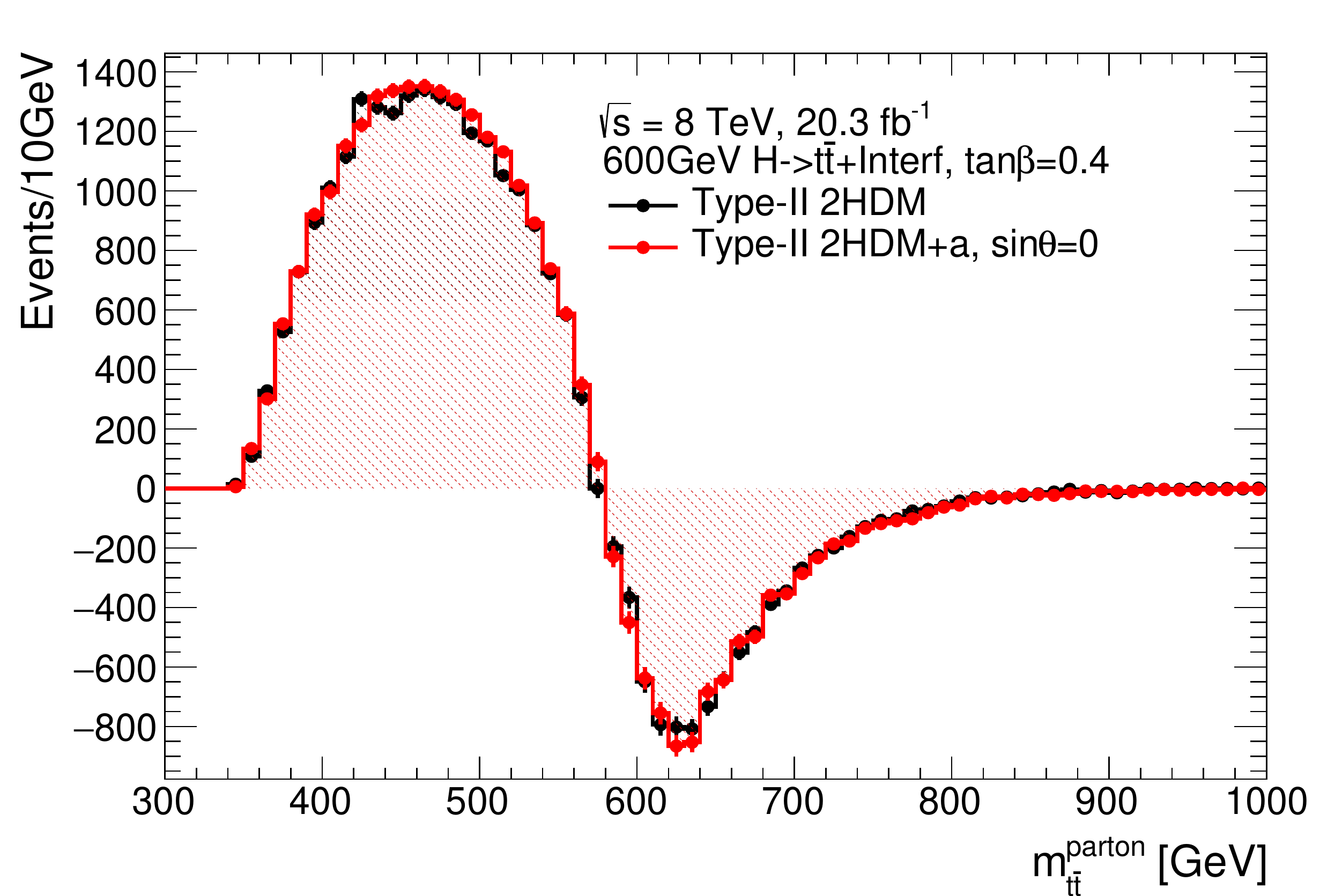} \quad 
\includegraphics[width=.475\textwidth]{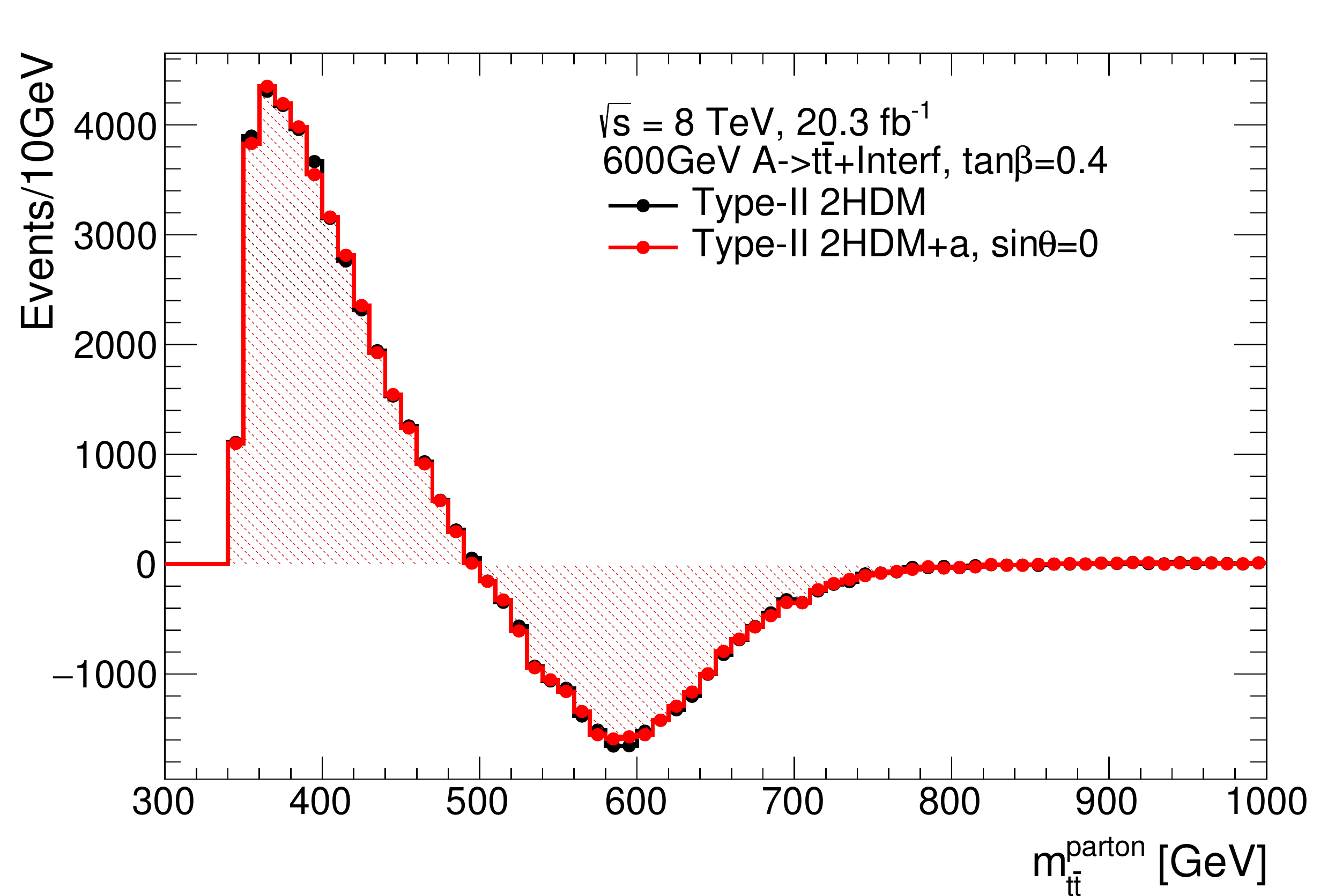}
\vspace{4mm}
\caption{$m_{t \bar t}$ spectra for $gg \to H \to t \bar t$~(left) and  $gg \to A \to t \bar t$~(right). The black (red) predictions correspond to the type-II 2HDM (\hdma) model.  The  results shown employ $\mH=\mA=\mHc = 600 \, {\rm GeV}$,  $\ma=100 \, {\rm GeV}$, $\tan \beta =0.4$ and $\sin \theta = 0$, and correspond to $20.3 \, {\rm fb}^{-1}$ of 8 TeV data.  The parameters not explicitly specified are chosen as in~\eqref{eq:benchmark}.}
\label{fig:ttres_2HDMvs2HDMa}
\end{figure}

The first LHC analysis that takes into account interference effects between the signal process $gg \to H/A \to t \bar t$ and the SM background $gg \to t \bar t$ is the ATLAS search~\cite{Aaboud:2017hnm}. This search  is based on an integrated luminosity of $20.3 \, {\rm fb}^{-1}$ collected at $8 \, {\rm TeV}$. The results are interpreted in the alignment limit of the usual type-II 2HDM. The obtained  exclusion limits in the $M_{H,A}\hspace{0.5mm}$--$\hspace{0.5mm} \tan \beta$ plane turn out to be significantly stronger than previously published LHC constraints on the 2HDM parameter space with low $\tan \beta$ and $M_{H,A} \simeq [500, 650] \, {\rm GeV}$. For instance,  for $M_{H,A} = 500 \, {\rm GeV}$ values of $\tan \beta < 1$ are excluded at 95\%~CL. 

\begin{figure}
\centering
\includegraphics[width=0.475\textwidth]{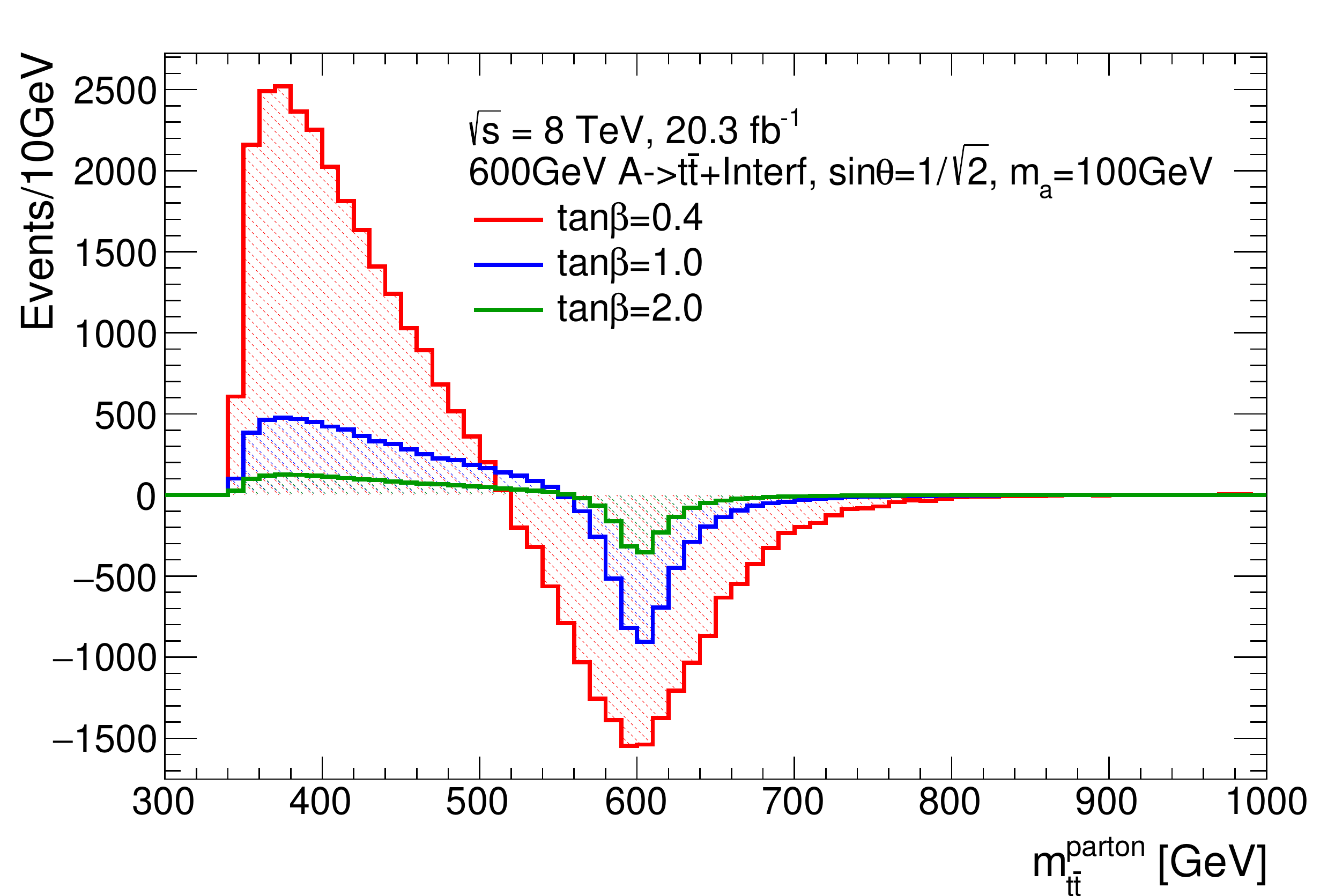} \quad 
\includegraphics[width=0.475\textwidth]{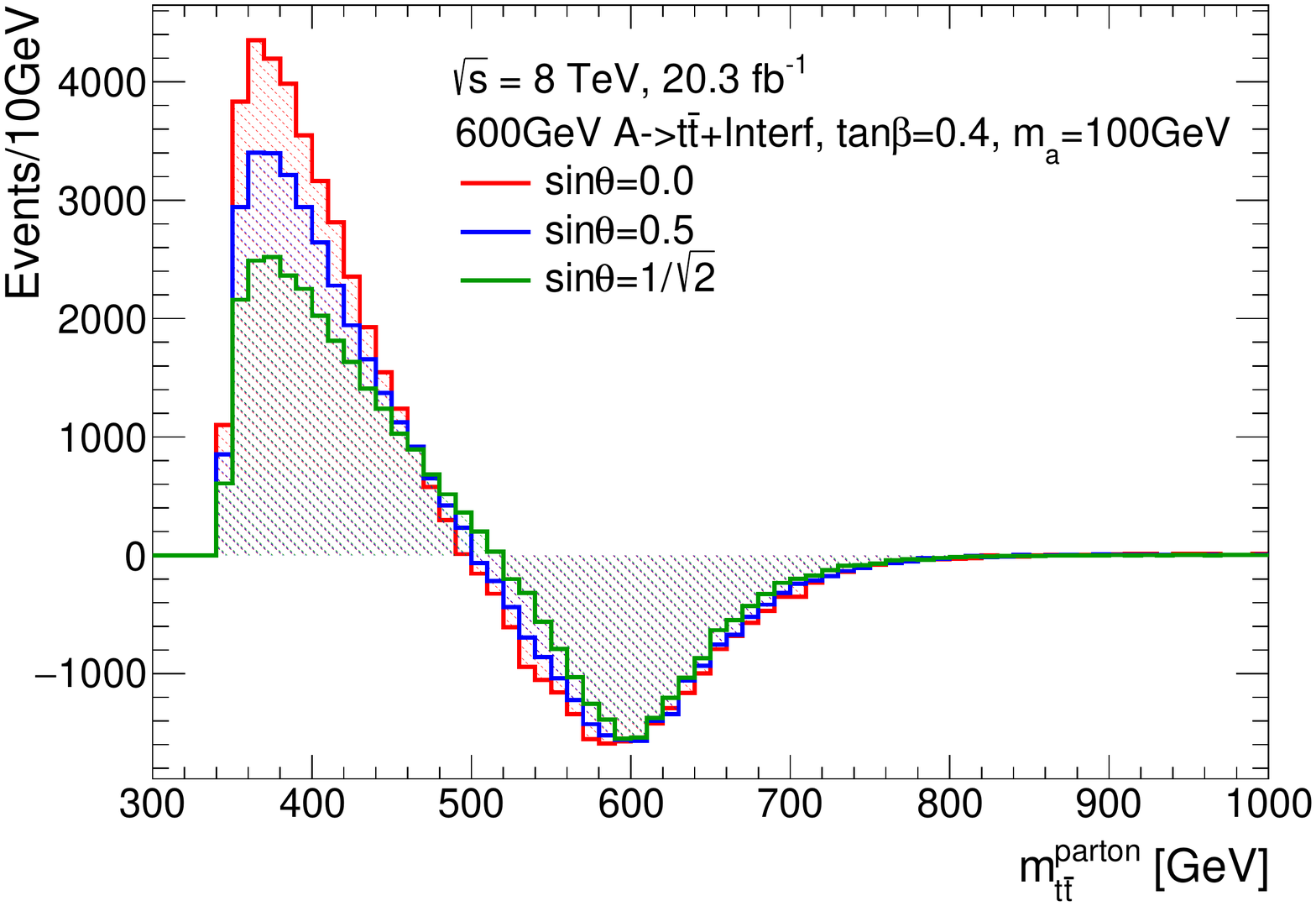}
\vspace{4mm}
\caption{Left: $\tan \beta$ dependency of $m_{t \bar t}$ spectrum for fixed $\sin \theta = 1/\sqrt{2}$. Right:  $\sin \theta$ dependency of $m_{t \bar t}$ spectrum for fixed $\tan \beta = 0.4$. The chosen \hdma parameters are $\mH=\mA=\mHc =600 \, {\rm GeV}$ and $\ma=100 \, {\rm GeV}$ and the depicted distributions correspond to $20.3 \, {\rm fb}^{-1}$ of integrated luminosity collected at 8 TeV. Parameters not explicitly specified are set to~\eqref{eq:benchmark}.}
\label{fig:ttres_2HDM_A}
\end{figure}

Di-top invariant mass spectra for various $\tan \beta$ and $\sin \theta$ scenarios and 2HDM models are shown in Figures~\ref{fig:ttres_2HDMvs2HDMa} and \ref{fig:ttres_2HDM_A}. The signal process has been obtained by treating the loop contributions from top and bottom quarks as form factors~\cite{FranzosiZhang}. In this way the interference between the signal and the tree-level SM background from $gg \to t \bar t$ can be calculated at leading order in QCD. In Figure~\ref{fig:ttres_2HDMvs2HDMa}, we show predictions for the $m_{t \bar t}$ spectra in $gg \to H \to t \bar t$~(left panel) and $gg \to A \to t \bar t$~(right panel). The black histograms illustrate the 2HDM predictions~\cite{Aaboud:2017hnm}, while the red curves represent the corresponding \hdma predictions for the choice $\sin \theta = 0$, which effectively decouples the mediator $a$ from the 2HDM Higgs sector. The agreement between the black and red predictions serves as a validation of our form factor implementation in the \hdma model. 

As examples of the parameter dependencies of the $t \bar t$ predictions in the \hdma model, we display in  Figure~\ref{fig:ttres_2HDM_A} several $m_{t \bar t}$ spectra in $pp \to A \to t \bar t$, either fixing $\sin \theta$ and varying $\tan \beta$ (left panel) or vice versa (right panel). The  spin-0 boson  masses are chosen $\mH=\mA=\mHc =600 \, {\rm GeV}$ and $\ma=100 \, {\rm GeV}$, which implies that only the  decays $H/A \to t \bar t$  are kinematically possible but not $a \to t \bar t$.  From the left panel one sees that increasing $\tan \beta$ leads  to a reduction of the signal strength.  Likewise, larger values of $\sin \theta$ also lead to lower $t \bar t$ cross sections as illustrated on the right-hand side of the latter figure. These are expected features because the $g g \to A$ amplitude  scales as $\cot \beta \cos \theta$.  Additionally, the interference between the $t \bar t$ signal and the corresponding background, and thus the shape of the $m_{t \bar t}$ spectrum, depends on the precise choice of $\tan \beta$  and $\sin \theta$.  Before moving on, let us add that the results of~\cite{Aaboud:2017hnm} have already been recasted to the \hdma model in~\cite{Bauer:2017ota}. For the parameter benchmarks studied in the latter paper it turns out that only values $\tan \beta < {\cal O} ( 0.5)$ can be excluded based on the 8 TeV ATLAS search, making the resulting $t \bar t$ constraints weaker than those arising at present from flavour physics~(see~Section~\ref{sec:flavour}). 

\subsection{Four-top searches}
\label{sec:fourtopmain}

The four-top final state is a rare, yet increasingly important signature (see for instance~\cite{ATLAS-CONF-2016-104,Sirunyan:2017roi,Aaboud:2018xuw,Hajer:2015gka,Gori:2016zto,Alvarez:2016nrz}).   In fact, in the work~\cite{ATLAS-CONF-2016-104} the results of a search for the four-top final state based on $13.2 \, {\rm fb}^{-1}$ of 13 TeV LHC data has already been interpreted in the context of the standard 2HDMs. The comparison to the predictions for a type-II 2HDM in the alignment limit allows the exclusion at the 95\%~CL of $\tan \beta$  below 0.17 (0.11) for $\mH = 400 \, {\rm GeV}$ ($\mH = 1 \, {\rm TeV}$). While these limits are weaker than those that can be obtained from $t \bar t$ production in the $M_{H} \simeq [500, 650] \, {\rm GeV}$ range~\cite{Aaboud:2017hnm}, in the long run, four-top searches can be expected to have a better sensitivity than $t \bar t$ searches for mediators with  masses either close to the top threshold or  in the ballpark of $1 \, {\rm TeV}$. 

\begin{figure}[!t]
\centering
\includegraphics[width=0.475\textwidth]{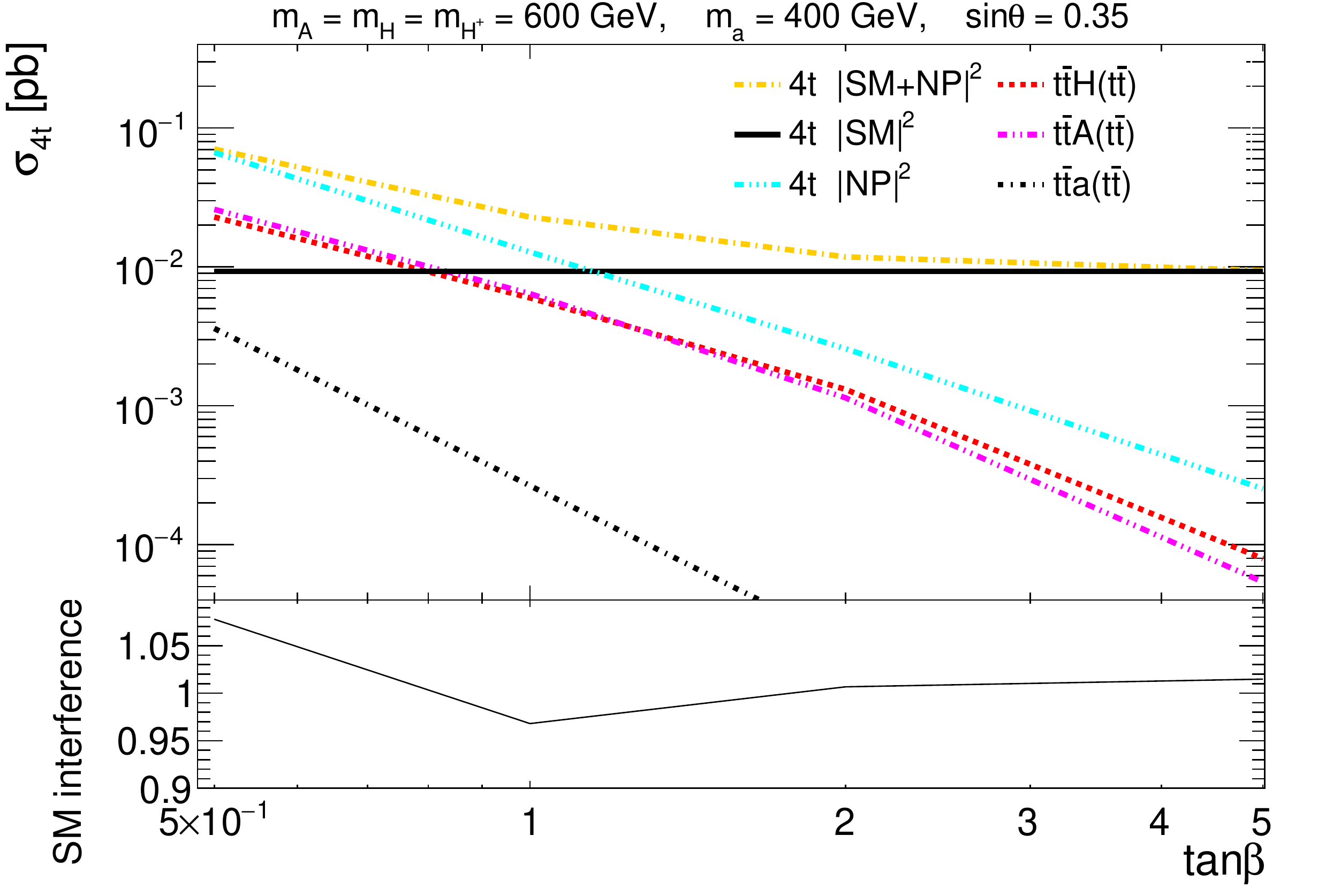} \quad 
\includegraphics[width=0.475\textwidth]{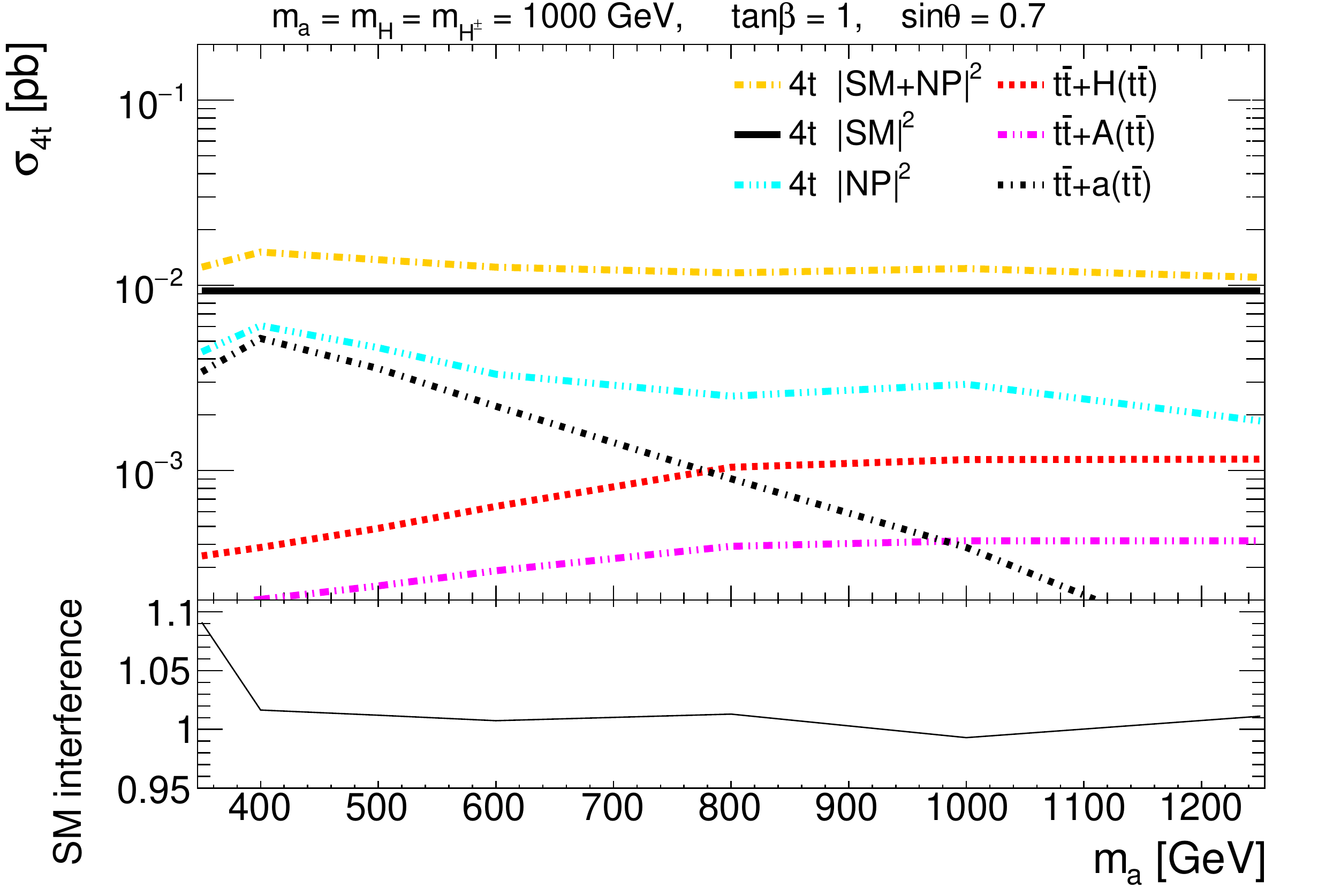}
\vspace{4mm}
\caption{\label{fig:4top} Four-top cross sections as function of $\tan \beta$ (left) and $\ma$ (right) for $pp$~collisions at $13 \, {\rm TeV}$.   In the left panel $\mH =\mA=\mHc = 600 \, {\rm GeV}$ and  $\ma = 400 \, {\rm GeV}$ have been used, while in the right panel $\mH =\mA=\mHc = 1 \, {\rm TeV}$ and  $\sin \theta = 0.7$ have been employed. Parameters not explicitly specified are set to~\eqref{eq:benchmark}.  The SM and the different new-physics contributions are indicated by the black and coloured lines. See text for further explanations.}
\end{figure}

 In this white paper we perform a first characterisation of the four-top signature in the \hdma context by studying the predicted cross section for different parameter choices. Predictions for the four-top cross section ($\sigma_{4t}$) as a function of $\tan \beta$~(left panel) and $\ma$~(right panel) are presented in~Figure~\ref{fig:4top}. The total four-top production cross section in the SM ($|{\rm SM}|^2$) is indicated by a black line in both panels, while the new-physics~(NP) contributions ($|{\rm NP}|^2$) are represented by the blue curves. The predictions that account for both the SM and the \hdma contribution as well as  their interference ($|{\rm SM} + {\rm NP}|^2$) are coloured yellow. The contributions from associated $t \bar t$ production of an on-shell $H, A, a$ with the subsequent decay $H/A/a \to t \bar t$ are also given.  A brief description of how the different channels have been separated in our MC simulations  is given in Appendix~\ref{app:mcgeneration}. From the left panel one can see that for the chosen parameters on-shell production of $H$ and $A$ provides the dominant contribution to inclusive cross section. Interference effects turn out to be small as they modify the results  by only ${\cal O} (5\%)$ at the inclusive level. This feature is illustrated in the lower part of the left plot.
 
On the right-hand side in Figure~\ref{fig:4top} we instead  study the $\ma$ dependence of the cross section. For the chosen parameters the $|{\rm NP}|^2$ contribution is rather flat in $\ma$. The breakdown of the on-shell contributions furthermore shows that for $\ma \lesssim 800 \, {\rm GeV}$ the contribution from $t \bar t a$ production dominates, while for $\ma \gtrsim 800 \, {\rm GeV}$ the $t \bar t H/A$ channels are numerically more important. The small bump at $1 \, {\rm TeV}$ is due to interference effects between the three Higgs states.  As for the previous benchmark,  the impact of the signal-background interference on the inclusive cross section is found to be small (i.e.~below 2\%), except for $\ma$ values close to the top threshold. 

\begin{figure}[t!]
\centering
\includegraphics[width=.625\textwidth]{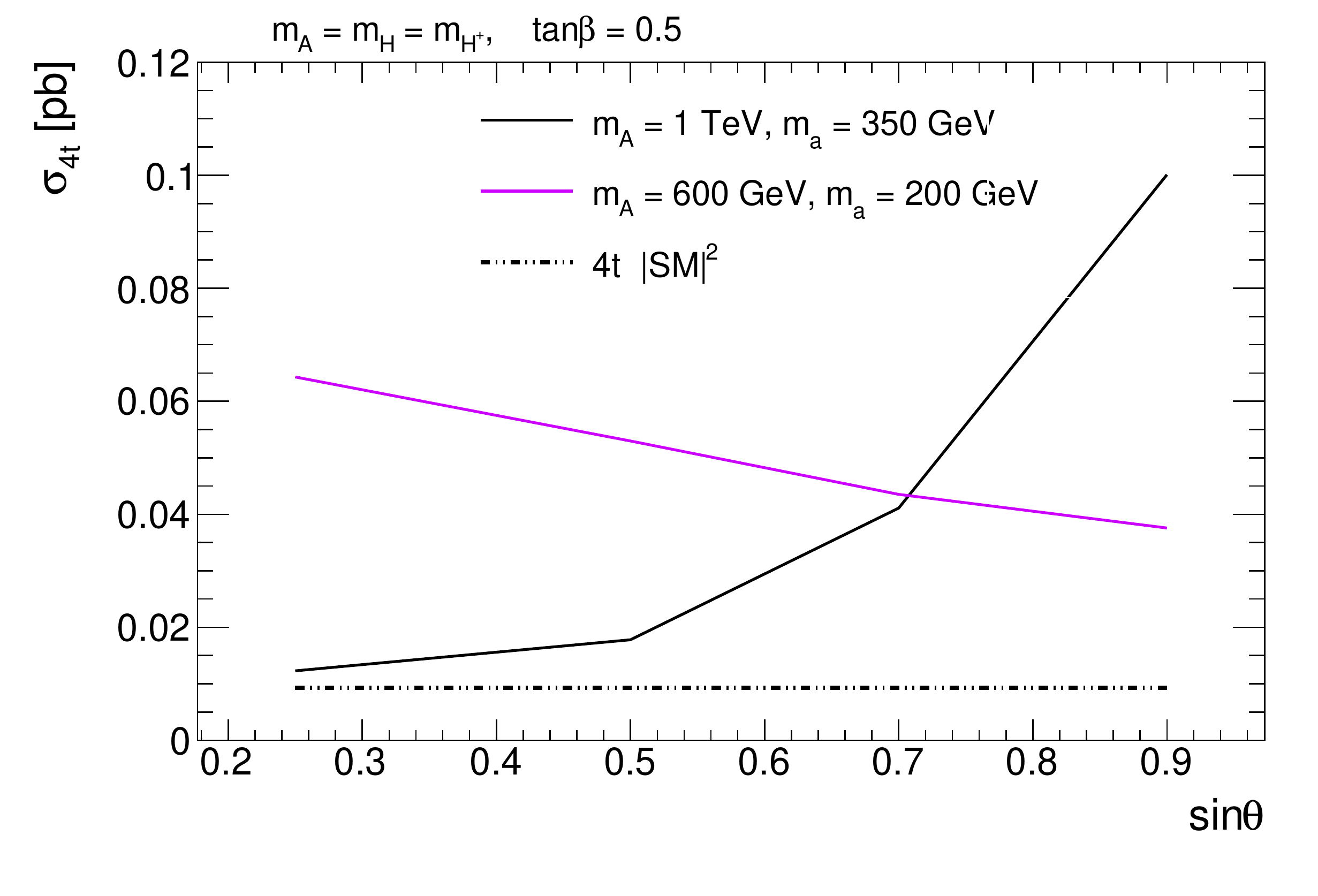}
\vspace{1mm}
\caption{\label{DMHF-4top-scan3} Four-top cross sections as a function of $\sin \theta$ at the $13 \, {\rm TeV}$ LHC. The black dashed-dotted corresponds to the SM prediction, while the solid black (magenta) line employs  $\mH =\mA=\mHc = 1 \, {\rm TeV}$ and $\ma = 350  \, {\rm GeV}$ ($\mH =\mA=\mHc = 600 \, {\rm GeV}$ and $\ma = 200 \, {\rm GeV}$). Both $|{\rm NP}|^2$ curves are based on $\tan \beta = 0.5$ and all parameters not explicitly specified in the legend are set to~\eqref{eq:benchmark}. }
\end{figure}

In Figure~\ref{DMHF-4top-scan3} we finally plot the  $\sin \theta$ dependence of the new-physics contribution~$|{\rm NP}|^2$ to the cross section of four-top production for the two benchmarks studied before. 
In the case of $\mH =\mA=\mHc = 1 \, {\rm TeV}$ and $\ma = 350 \, {\rm GeV}$ (black curve) the cross section  increases for increasing $\sin \theta$. This is expected because the dominant contribution to the signal arises from $t \bar t  a$ production followed by $a \to t \bar t$ and the coupling of the $a$ to top quarks scales as $\sin \theta$. In the case of $\mH =\mA=\mHc = 600 \, {\rm GeV}$ and $\ma = 200 \, {\rm GeV}$~(magenta curve) the cross section instead decreases with increasing $\sin \theta$.  In this case  the $H \to t \bar t$ decay gives the largest  contribution, since $a \to t \bar t$ is kinematically closed. The observed $\sin \theta$ dependence then arises from the interplay between $\Gamma (H \to t \bar t)$ which does not depend on $\sin \theta$ and $\Gamma (H \to aa)$ as well as $\Gamma (H \to Za)$ which are both proportional to $\sin^2 \theta$.
 
\subsection{Other final states}
\label{sec:others}

The $\tau^+ \tau^-$ final state is one of the most common channels that experiments have considered to search for additional neutral Higgs bosons (see~\cite{Aaboud:2017sjh,Sirunyan:2018zut} for the latest LHC results).   The sensitivity of the $\tau^+ \tau^-$ searches to the \hdma parameter space has been studied in~\cite{Bauer:2017ota} and found to be weak. The limited sensitivity of the $\tau^+ \tau^-$ channel arises because the rates in  $A/a \to \tau^+ \tau^-$  are predicted to be generically small if the $A/a \to \chi \bar \chi$ decays are open. In fact, the decay rate $\Gamma ( a \to \chi \bar \chi )$ dominates over  $\Gamma ( a \to \tau^+ \tau^- )$ for all parameter choices that fulfill $M_a > 2 m_\chi$ and  $y_\chi^2  \cot^2 \beta \cot^2 \theta > m_\tau^2/v^2 \simeq 5.2 \cdot 10^{-5}$ in the type-II~\hdma~model. The latter inequality implies that it will be very difficult to test the benchmark models~\eqref{eq:matbscan} through  $\tau^+ \tau^-$ searches. Future  $\tau^+ \tau^-$ analyses may however be able to exclude scenarios like~\eqref{eq:benchmark} for $\mH = \mA = \mHc = {\cal O} (300 \, {\rm GeV})$ and $M_H \lesssim 2M_a$. Since such realisations are not easy to test otherwise, interpreting the results of forthcoming~$\tau^+ \tau^-$ searches in the \hdma context seems to be worthwhile.

If $\mH > \ma + M_Z$ and the mediator $a$ is sufficiently heavy,~i.e.~$\ma > 2 m_t$, another channel that offers sensitivity to the  $\hdma$ parameter space is $pp \to aZ$ with $a \to t \bar t$ instead of $a \to \chi \bar \chi$~\cite{GPHeidelberg}. The corresponding $t \bar t Z$ final state has been recently studied~\cite{Haisch:2018djm} in the context of the standard 2HDMs and shown to lead to a robust coverage of the 2HDM parameter space with $M_{H,A} > 350 \, {\rm GeV}$, $|M_H - M_A| > M_Z$ and $\tan \beta = {\cal O} (1)$ at future LHC runs.  The analysis strategy detailed in~\cite{Haisch:2018djm} can be directly applied to the \hdma case, and should provide sensitivity to realisations that feature $a$ mediators with masses above the top threshold in the high-luminosity phase of the LHC. Such scenarios are generically difficult to explore via a mono-$Z$ search (see Section~\ref{sec:sensi_monozll}).

The ATLAS and CMS Collaborations have set limits on the production of charged Higgses in both the $\tau \nu$~\cite{Aaboud:2016dig,CMS-PAS-HIG-16-031} and the $tb$~\cite{Aad:2015typ,Khachatryan:2015qxa,ATLAS:2016qiq} final state. The limits given in~\cite{ATLAS:2016qiq} have been used in~\cite{Pani:2017qyd} to derive constraints on the \hdma model. It turns out that  the constraints on the \hdma parameter space are generically weaker than those obtained in the 2HDM context, because in the \hdma model  the $H^\pm \to tb$ branching ratio tends to be reduced compared to the 2HDM since the partial decay width $H^\pm \to aW^\pm$ is generically non-vanishing.  However, compared to the $tW+\MET$ signature, $tb$ searches can still provide complementary information, because the non-$\MET$  search can test $\mHc$ values below around $350 \, {\rm GeV}$ which are not easily accessible with the corresponding $\MET$ signature~\cite{Pani:2017qyd}. Another signal that can be used to search for charged Higgses is the $tbW$ final state~\cite{Haisch:2018djm}. This channel has, however, not yet been explored in the \hdma context. 

%%%%%%%%%%%%%%%%%%%%%%%%%%%%%%%%%%%%%%%%%%%%%%%%%%%%%%%%%%%%%%%%%%%%
%%%%%%%%%%%%%%%%%%%%%%%%%%%%%%%%%%%%%%%%%%%%%%%%%%%%%%%%%%%%%%%%%%%%
%%%%%%%%%%%%%%%%%%%%%%%%%%%%%%%%%%%%%%%%%%%%%%%%%%%%%%%%%%%%%%%%%%%%

\section{Sensitivity studies}
\label{sec:sensitivitystudies}

In this section we present sensitivity estimates for two of the main $\MET$ signatures in the \hdma model, namely the $h +\MET$  and the $Z+\MET$ channels. Specifically, we will consider the mono-Higgs (mono-$Z$) signal in the $b \bar b$ ($\ell^+ \ell^-$) channel.  Our studies are based on reinterpretation of existing results that use $36 \, {\rm fb}^{-1}$ of LHC data taken at $\sqrt{s} = 13 \, {\rm TeV}$. These results contain different amounts of public information. In the mono-Higgs case model-independent limits presented in~\cite{Aaboud:2017yqz} are used for the reinterpretation, while in the mono-$Z$ case  the sensitivity is estimated using information on the signal together with published background estimates~\cite{Aaboud:2017bja}.  The sensitivities that other mono-$X$ searches provide are also briefly discussed below.  A concise description of how the  mono-$X$ signals considered in our sensitivity study have been generated  can be found in Appendix~\ref{app:mcgeneration}.

\subsection{Mono-Higgs study}
\label{sec:sensi_monohbb}

\begin{figure}[t!]
\centering
\includegraphics[width=0.75\textwidth]{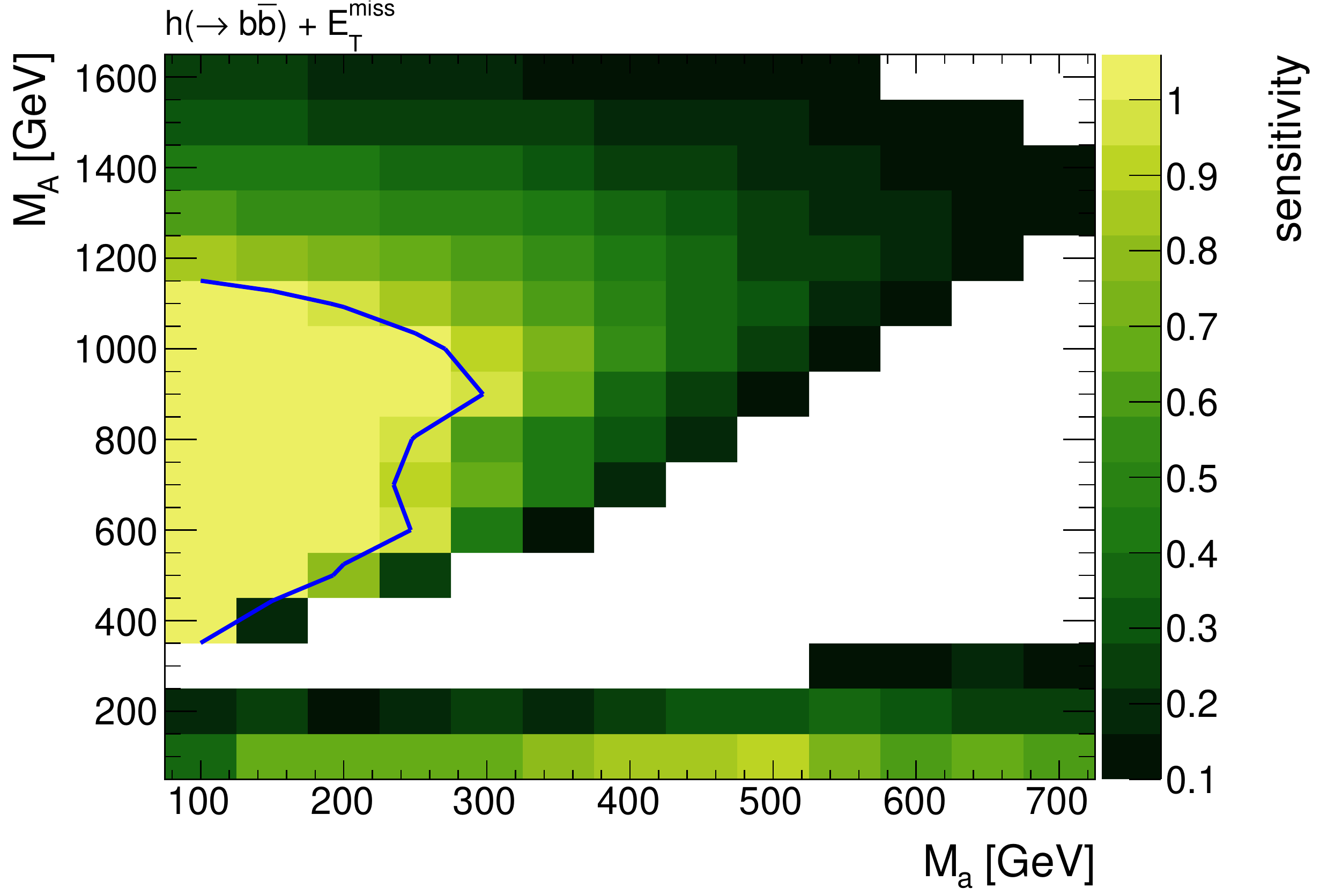}

\vspace{2mm}

\includegraphics[width=0.75\textwidth]{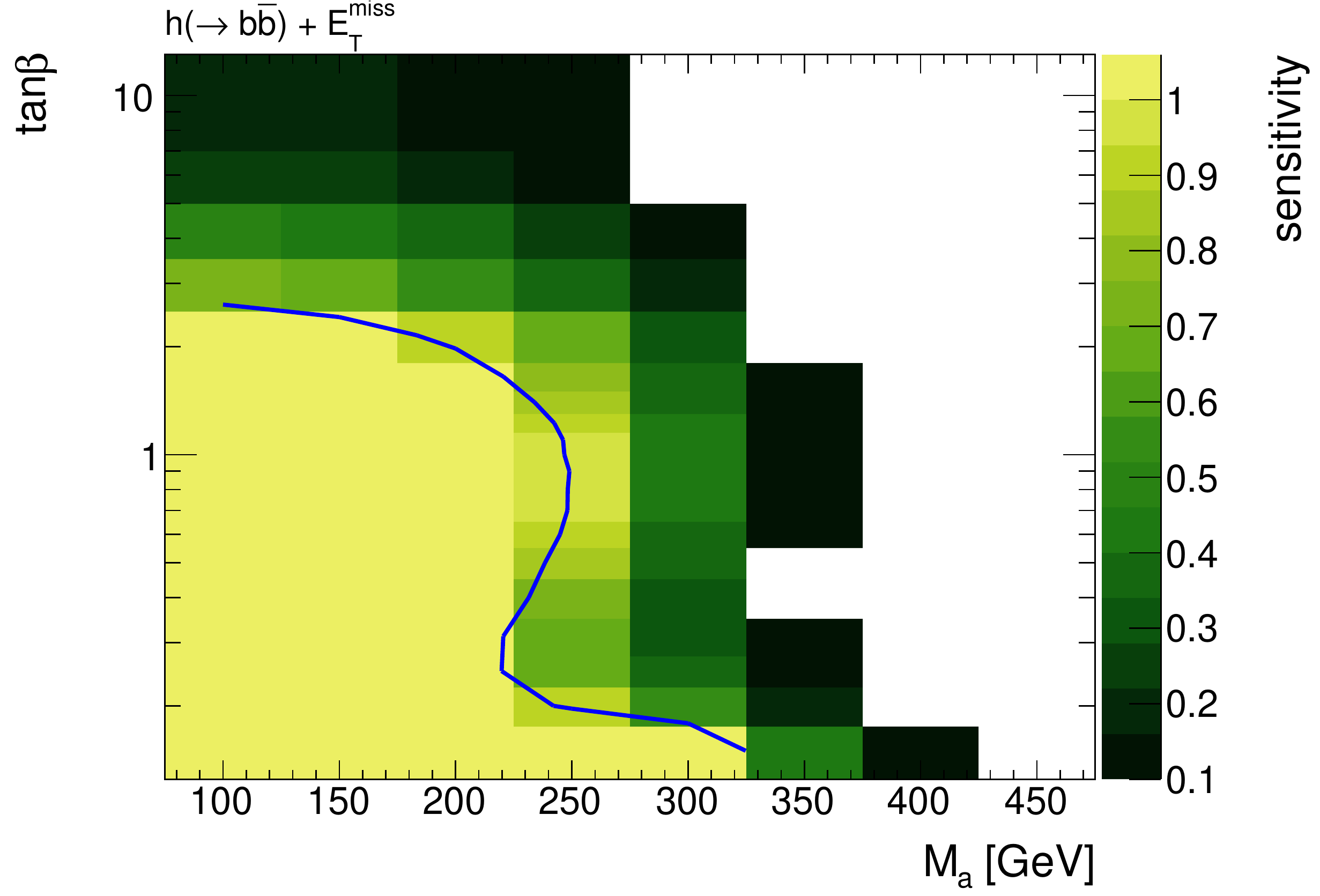}
\vspace{2mm}
\caption{Estimated sensitivities with the $h+\MET$ signature in the $h \to b \bar b$ channel. The upper (lower) panel shows our results in the $\ma\hspace{0.25mm}$--$\hspace{0.25mm}\mA$ ($\ma\hspace{0.25mm}$--$\hspace{0.25mm}\tan \beta$) plane.  The remaining parameters are set to \eqref{eq:benchmark} in the upper panel, while in the lower panel $\tan \beta$ is left to vary but the common 2HDM  spin-0 boson  mass is fixed to~\eqref{eq:matbscan}. The blue contours correspond to $S =1$ and bins with no content have a negligible sensitivity $S<0.1$ (see text for further explanations). The grid generated is evenly spaced in \mA and \ma, each bin corresponding to one grid point.}
\label{fig:monoHbb_sensi}
\end{figure}

The sensitivity estimates of the ATLAS and CMS mono-Higgs searches in the $b \bar b$ channel to the \hdma model are based on the model-independent limits on the anomalous production of the SM Higgs boson in association with \met derived in~\cite{Aaboud:2017yqz}.  As these limits are set in terms of the observed production cross section of non-SM events with large $\MET$ and a Higgs boson, they can be compared directly to the cross sections obtained in the \hdma model after taking into account the kinematic acceptance $\mathcal{A}$ of the event selection and the detection efficiency $\varepsilon$. The variables of interest for the sensitivity study of the $h \, (b \bar b) + \MET$ searches are
\begin{equation}
\label{eq:monoHbb_sensi}
S_i = \frac{\sigma_{i} \left ( p p \to h + \MET \right)_{\hdma} \cdot {\rm BR} \left (h \to b \bar b \right )_{\rm SM} \cdot \left ( \mathcal{A} \cdot \varepsilon \right)_i}
{\sigma_{i} \left ( p p \to h + \MET \to b \bar b + \MET \right)_{\rm obs}}\,,
\end{equation}
where $\sigma_{i} \left ( p p \to h + \MET \right)_{\hdma}$ is the partonic cross section of the \hdma signal,  the branching ratio of the SM Higgs boson is denoted by ${\rm BR} \left (h \to b \bar b \right )_{\rm SM} \simeq 58\%$  and $\sigma_{i} \left ( p p \to h + \MET \to b \bar b + \MET \right)_{\rm obs}$ represents the  observed upper cross-section limit on $h + \MET$ production with $h \to b \bar b$. In our mono-Higgs sensitivity study we include $gg$-fusion as well as $b \bar b$-initiated production.  The cross sections as well as the product $ \mathcal{A} \cdot \varepsilon$ depend on the considered $\MET$ bin as indicated by the index $i$.   A particular point in parameter space is expected to be excluded if the sum $S = \sum_i S_i$ of the individual sensitivities is larger than $1$.

The results of our sensitivity study for the mono-Higgs signal in the $b \bar b$ decay channel are shown in Figure~\ref{fig:monoHbb_sensi}. The upper panel in the figure displays $S$  as a function of $\ma$ and $\mA$. The existing mono-Higgs searches allow us to probe/exclude \hdma scenarios with  $\mA > M_h + \ma$  and sufficiently small $\ma$ values, while they are only weakly  sensitive to models where the mass hierarchy between $A$ and $a$ is reversed,~i.e.~$\ma > M_h + \mA$. Numerically, we find that  for  a light $a$ with $\ma \simeq 100 \, {\rm GeV}$ one has $S > 1$ for all values $\mA \simeq [350, 1150] \, {\rm GeV}$. In the parameter region  $\mA > M_h + \ma$ the strong sensitivity of the search arises because the mono-Higgs signature is resonantly produced via $pp \to A \to ha \to h \chi \bar \chi$ --- see the discussion in Section~\ref{sec:resonant}. The sensitivity of the search decreases for increasing (decreasing) $M_A$ because the production rate of $pp \to A$  decreases $\big($the Jacobian peak~\eqref{eq:monoHMETpeak} is shifted to lower $\MET$ values$\big)$. In the region $\ma > M_h + \mA$, the largest contribution to the $h + \MET$ cross section again originates from resonant production, namely $pp \to a \to hA \to h \chi \bar \chi$.  The resulting sensitivities are, however, much smaller compared to the case discussed before, because  $\sigma \left (p p \to a \right )/\sigma \left (pp \to A \right ) = \sin^2 \theta/\cos^2 \theta \simeq 1/7$, ${\rm BR} \left ( a \to A h \right )/{\rm BR} \left (  A \to a h  \right ) < 1$ and ${\rm BR} \left ( A \to \chi \bar \chi \right )/{\rm BR} \left (  a \to \chi \bar \chi \right ) \ll~\!\!1$ for the parameter choices made in \eqref{eq:benchmark}. Notice that in the parameter region with $M_A \gtrsim 1250 \, {\rm GeV}$ the BFB condition~\eqref{eq:BFB2} is not satisfied (see the right panel in Figure~\ref{fig:EWVAC}) for the choice of parameters employed in the upper panel of Figure~\ref{fig:monoHbb_sensi}.

The lower panel in Figure~\ref{fig:monoHbb_sensi} shows the sensitivity $S$ in the $\ma\hspace{0.25mm}$--$\hspace{0.25mm}\tan \beta$ plane fixing  $\mH, \mA$ and  $\mHc$ to~\eqref{eq:matbscan}. The existing mono-Higgs searches allow to exclude $\tan \beta \lesssim 2.5$ for $M_a \simeq 100 \, {\rm GeV}$ and $\tan \beta \lesssim 1$ for $M_a \lesssim 240 \, {\rm GeV}$. From~Figure~\ref{fig:tbvar1} it is apparent that for such small values of $\tan \beta$, the $h + \MET$ signal is dominantly produced through top-quark loops in $gg$-fusion. The corresponding production rate scales as $\sigma \left ( gg \to A \right ) \propto \cot^2 \beta$, and as a result the sensitivity rapidly decreases for $\tan \beta > 1$. The decrease is to some extent counteracted by the fact that the Jacobian peak becomes more pronounced when  $\tan \beta$ is increased (cf.~the left panel in Figure~\ref{fig:tbvar1}). For $\tan \beta \gtrsim 10$ the sensitivity of the mono-Higgs search  starts to increase again, because the $b \bar b$-initiated production cross section behaves like $\sigma \hspace{0.5mm} ( b \bar b \to A  ) \propto \tan^2 \beta$. Further plots of our mono-Higgs sensitivity study can be found in Appendix~\ref{app:extramonoh}.  We add that our $h \, (b \bar b) + \MET$ results are compatible with those provided very recently by the CMS~Collaboration in~\cite{CMS-PAS-EXO-16-050}. More restrictive experimental analyses such as~\cite{Sirunyan:2018qob} are expected to have an even higher sensitivity than the case  studied here. 

\subsection[Mono-$Z$ study]{Mono-$\bm{Z}$ study}
\label{sec:sensi_monozll}

\begin{figure}[t!]
\centering
\includegraphics[width=0.75\textwidth]{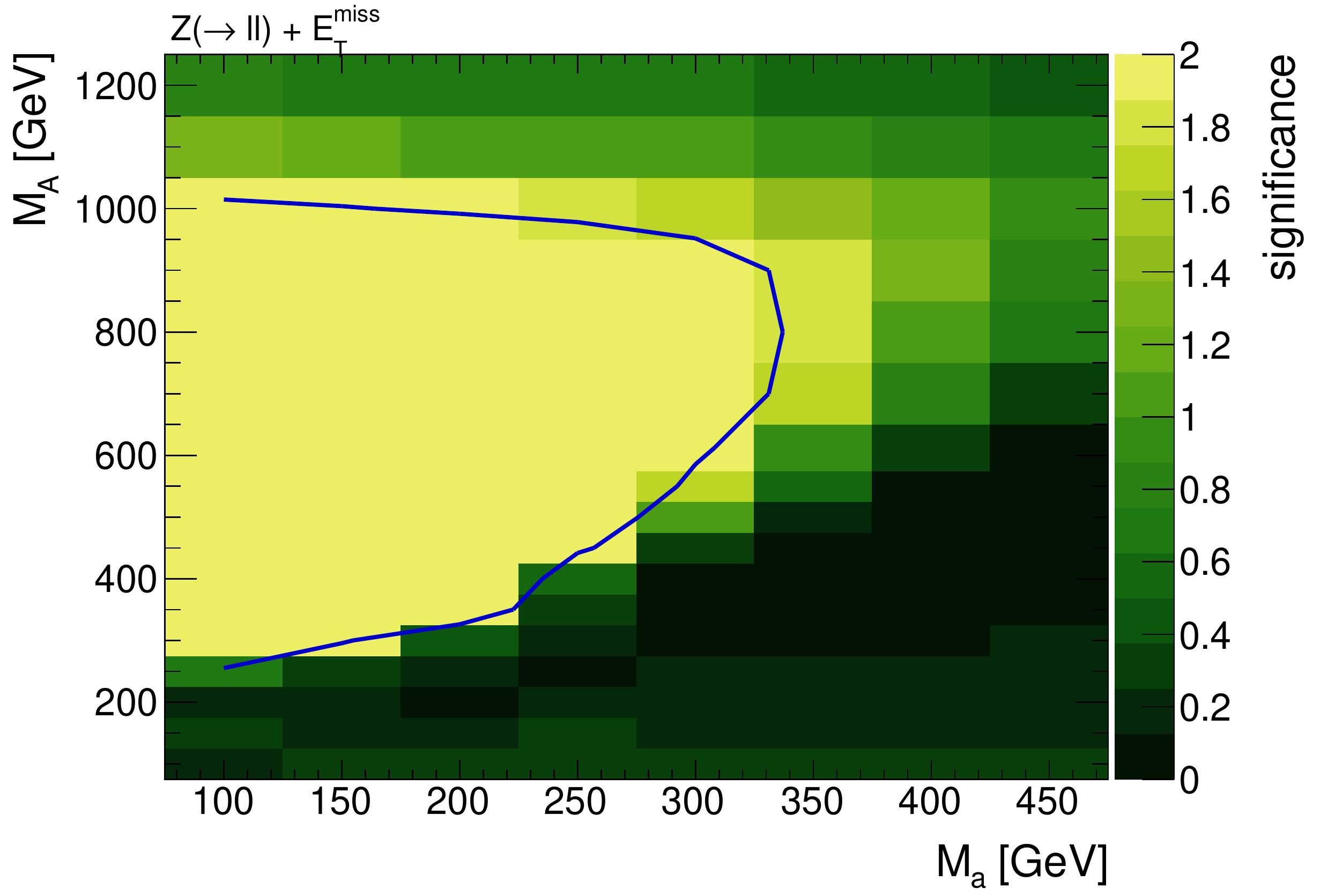} 

\vspace{2mm}

\includegraphics[width=0.75\textwidth]{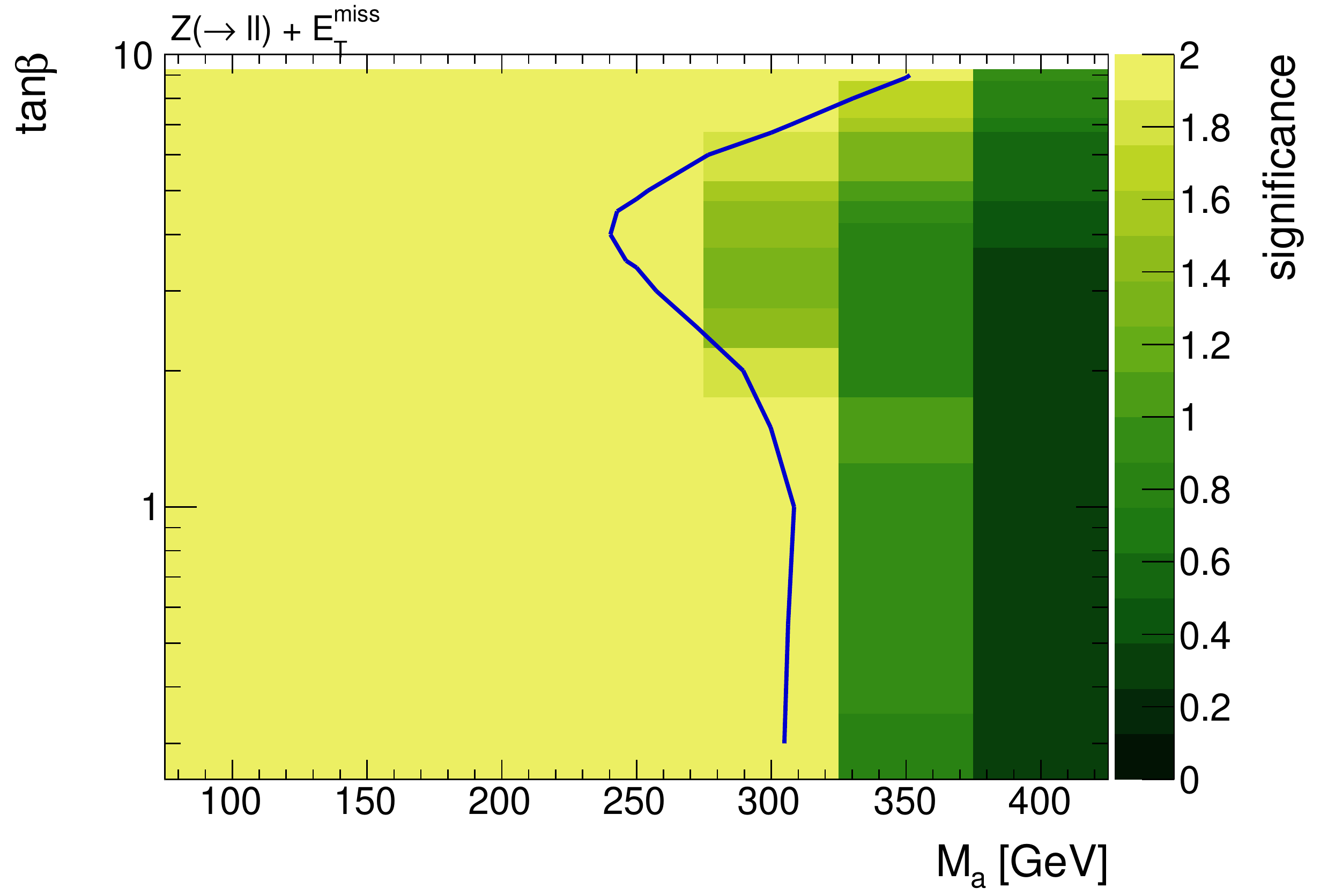} 
\vspace{2mm}
\caption{Estimated significance of the $Z+\MET$ signature in the $Z \to \ell^+ \ell^-$ channel. The upper (lower) panel shows our results in the $\ma\hspace{0.25mm}$--$\hspace{0.25mm}\mA$ ($\ma\hspace{0.25mm}$--$\hspace{0.25mm}\tan \beta$) plane. The choice of parameters is identical to those made in Figure~\ref{fig:monoHbb_sensi}. The blue contours correspond to $Z_A = 2$ and the grid generated is evenly spaced in \mA and \ma, each bin corresponding to one grid point. Further details can be found in the text. }
\label{fig:monoZll_sensi}
\end{figure}

 The expected sensitivity of the mono-$Z$ search to the \hdma model is estimated by comparing the number of  signal events to the number of expected background events. Published background predictions for $Z+\MET$ production followed by $Z \to \ell^+ \ell^-$~\cite{Aaboud:2017bja} are used which correspond to $36 \, {\rm fb}^{-1}$ of $13 \, {\rm TeV}$ data. The selection requirements and \MET binnings that are applied to the signal events resemble those employed in the ATLAS analysis~\cite{Aaboud:2017bja}. A typical reconstruction efficiency of 75\% is assumed for signal events~\cite{Sirunyan:2017qfc},  and a conservative background systematic uncertainty of 20\% (10\%) is taken for events with $\MET < 120  \, {\rm GeV}$ ($\MET > 120  \, {\rm GeV}$). Following the Asimov approximation~\cite{Cowan:2010js}, the significance $Z_{A,i}$ for individual bins $i$ is calculated as a Poisson ratio of likelihoods modified to incorporate systematic uncertainties on the background. Explicitly one has \cite{Cowan:2012}
\begin{equation}
\label{eq:significance_wsyst}
Z_{A, i} = \sqrt{ 2 \left ( \left ( s + b \right ) \ln \left [ \frac{\left (s+b \right ) \left (b + \sigma_b^2 \right )}{b^2 + \left ( s +b \right ) \sigma_b^2 } \right ]  - \frac{b^2}{\sigma_b^2} \ln \left [ 1 + \frac{\sigma_b^2 s}{b \left ( b + \sigma_b^2 \right )} \right ] \right ) } \,, 
\end{equation}
where $s \hspace{0.5mm} \mathrm{(}b\mathrm{)}$ represents the expected number of signal (background) events and $\sigma_b$ denotes the standard deviation that characterises the systematic uncertainties of the background. The total significance $Z_A$ is then defined by adding the individual $Z_{A, i}$ in quadrature.   In this approximation, one expects to exclude regions with total significances of $Z_A > 2$. 

The results of our sensitivity study for the mono-$Z$ signature in the $\ell^+ \ell^-$ channel are presented in Figure~\ref{fig:monoZll_sensi}. The upper (lower) panel displays the total significance $Z_A$  in the $\ma\hspace{0.25mm}$--$\hspace{0.25mm}\mA$ ($\ma\hspace{0.25mm}$--$\hspace{0.25mm}\tan \beta$) plane. Comparing the obtained results to those depicted in Figure~\ref{fig:monoHbb_sensi}, one observes that for the parameter choices~\eqref{eq:benchmark} the mono-$Z$ and mono-Higgs searches allow to test quite similar parameter regions in the $\ma\hspace{0.25mm}$--$\hspace{0.25mm}\mA$ plane. Numerically, we find  that for $\ma \simeq 100 \, {\rm GeV}$  the existing $Z + \MET$ searches are sensitive to 2HDM pseudoscalar masses  in the range of $\mA \simeq [250, 1000] \, {\rm GeV}$. The mono-$Z$ sensitivity to lower values of $M_A$ is slightly better than the one found in the mono-Higgs case. This enhanced sensitivity arises because for fixed $M_A$ and $M_a$ and given that $M_Z < M_h$ the endpoint $E_{T, \rm max}^{\rm miss}$ of the $\MET$ distribution in $Z+ \MET$ production is always higher  than that in  the $h+ \MET$ channel. In contrast, in the parameter region with $M_a > M_Z + M_A$ the sensitivity of the mono-$Z$ signature is weaker than that of the mono-Higgs  signal. This feature is readily understood by noticing that the $pp \to a \to Z H$ channel does not lead to an $\MET$ signature, since the scalar $H$ does not decay invisibly in the \hdma model. For $M_a > M_Z + M_A$ hence only non-resonant diagrams contribute to the $Z+\MET$ signature, and the sensitivity to such model realisations is consequently very weak. 

 In the lower panel of Figure~\ref{fig:monoZll_sensi} we show the significance $Z_A$ in the $\ma\hspace{0.25mm}$--$\hspace{0.25mm}\tan \beta$ plane for the choice~\eqref{eq:matbscan}.  We see that present mono-$Z$ searches are expected to exclude all $\tan \beta$ values for $M_a \lesssim 240 \, {\rm GeV}$ and for $M_a \simeq 300 \, {\rm GeV}$ the ranges $\tan \beta \lesssim 1.5$ and $\tan \beta \gtrsim 6.5$. The drop in sensitivity for $\tan \beta \simeq 4$ is a result of the interplay between $gg$- and $b \bar b$-fusion production cross sections $\sigma ( g g \to H ) \propto \cot^2 \beta$ and $\sigma ( b \bar b  \to H ) \propto \tan^2 \beta$. The existing $Z \, (\ell^+ \ell^-) + \MET$ searches thus have sensitivity to  values $\tan \beta \gtrsim 2.5$, which is presently not the case for the $h \, (b \bar b) + \MET$ searches (cf.~Figure~\ref{fig:monoHbb_sensi}). Both features can be understood from the discussion presented in Section~\ref{sec:variationtanb}.
 
\subsection[Sensitivity of other mono-$X$ channels]{Sensitivity of other mono-$\bm{X}$ channels}
\label{sec:sensi_others}

The sensitivities of the LHC to the associated production of DM with a single top have been  studied in the framework of the \hdma model in~\cite{Pani:2017qyd}. This analysis assumes $300 \, {\rm fb}^{-1}$ of data and finds that the $t X + \MET$ signatures complement the parameter space coverage of the mono-Higgs and  mono-$Z$ signals considered by us in detail.   In fact, repeating  the analysis of~\cite{Pani:2017qyd} using only   $36 \, {\rm fb}^{-1}$ of integrated luminosity, one finds that a combination of the single-lepton and double-lepton channel allows to exclude values of $M_H = M_A = \mHc$ in the range of around $[400, 1000 ] \, {\rm GeV}$ for $\ma = 150 \, {\rm GeV}$, $\tan \beta < 1$ and $\sin \theta = 1/\sqrt{2}$. For $M_H = M_A = \mHc = 700 \, {\rm GeV}$ even a bound of $\tan \beta < 2$ can be set at 95\% CL. While a direct comparison with the limits obtained in the mono-Higgs and mono-$Z$ case is not possible due to the different value of $\sin \theta$ used in Sections~\ref{sec:sensi_monohbb} and~\ref{sec:sensi_monozll},  we note that the $\tan \beta$ values probed by all three searches  lie in the same ballpark. Another feature that is worth recalling is that the $h+\MET$, $Z+\MET$ and $tW+\MET$ signature can be resonantly enhanced through $A$, $H$ and $H^\pm$ exchange in the \hdma model (see Figure~\ref{fig:resonant}). Observing correlated deviations in all three channels might hence allow to determine the complete non-SM Higgs spectrum. 

Sensitivity studies of the $t \bar t + \MET$ and $j + \MET$ channels in the \hdma have been performed in~\cite{Bauer:2017ota}. The results presented in that work imply that for the benchmark parameter choices~\eqref{eq:benchmark}, the latest  $t \bar t + \MET$ and  mono-jet searches that are based on $36 \, {\rm fb}^{-1}$ of 13 TeV data have  only a very weak sensitivity to the parameter space shown in Figures~\ref{fig:monoHbb_sensi} and \ref{fig:monoZll_sensi}. Given the limited sensitivity of the $t \bar t + \MET$ and $j + \MET$  modes, we leave detailed sensitivity studies for these channels for future work. The $b \bar b + \MET$ channel is also not considered here due to the same reason. Notice, however, that a reinterpretation of existing $t \bar t + \MET$, $b \bar b + \MET$ and $j + \MET$ results is straightforward by using the general rescaling strategy discussed in Appendix~\ref{app:recast}. 

%%%%%%%%%%%%%%%%%%%%%%%%%%%%%%%%%%%%%%%%%%%%%%%%%%%%%%%%%%%%%%%%%%%%
%%%%%%%%%%%%%%%%%%%%%%%%%%%%%%%%%%%%%%%%%%%%%%%%%%%%%%%%%%%%%%%%%%%%
%%%%%%%%%%%%%%%%%%%%%%%%%%%%%%%%%%%%%%%%%%%%%%%%%%%%%%%%%%%%%%%%%%%%

\begin{figure}[t!]
    \centering
    \includegraphics[width=0.3\textwidth]{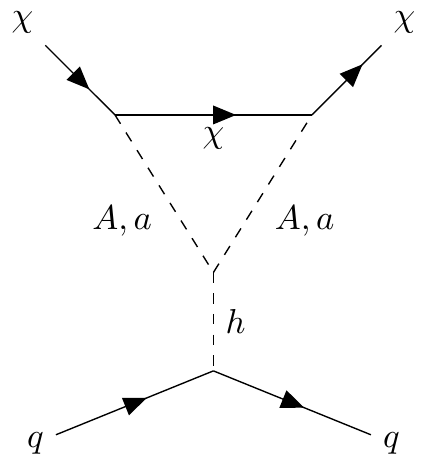} \quad 
    \includegraphics[width=0.3\textwidth]{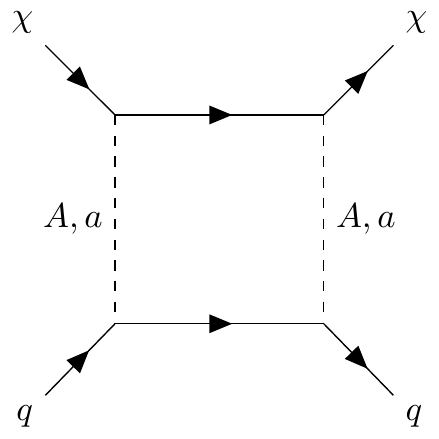} \quad 
    \includegraphics[width=0.3\textwidth]{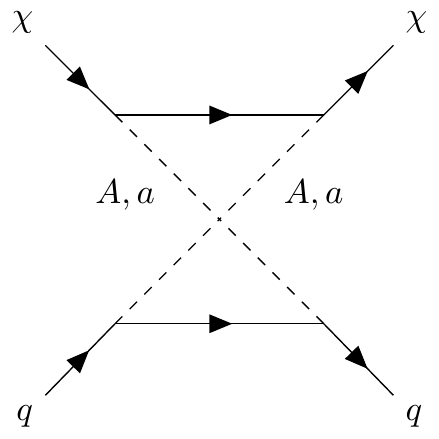} 
    \vspace{4mm}
    \caption{One-loop diagrams that lead to a SI DM-nucleon scattering cross section in the \hdma model.  Both triangle diagrams (left) as well as box graphs (middle and right) contribute in general. For further details consult the text. }
    \label{fig:feynDDPS}
\end{figure}

\section{Constraints from other DM experiments}
\label{sec:DMdetection}

In this section we  discuss the constraints that DD and ID experiments set on the parameter space of the \hdma model. We will illustrate both the existing constraints as well as show future projections. 

\subsection{DD experiments}

The constraints from DD for pseudoscalar mediators are generally suppressed at tree level, so that the dominant contributions arise from one-loop Feynman diagrams~\cite{Ipek:2014gua,Arcadi:2017wqi,Sanderson:2018lmj,Li:2018qip}. In the case of the \hdma model a spin-independent (SI) DM-nucleon  scattering cross section is generated by the graphs shown in Figure~\ref{fig:feynDDPS}. Notice that the triangle diagram shown on the left-hand side  is proportional to a single power of the Yukawa coupling~$y_q$, while the box diagrams that are displayed in the middle and on the right of the figure  are proportional to $y_q^3$. It follows that the triangle graph generically provides the dominant contribution to the SI DM-nucleon  scattering cross section. The only exceptions are models that feature a Yukawa sector with  $\tan \beta$-enhanced down-type Yukawa couplings such as type-II models, where the box diagrams can be numerically important if $\tan\beta\gtrsim50$. This has first been pointed out in \cite{Ipek:2014gua}. Unlike the box graphs the triangle diagram does not depend on the Yukawa sector of the \hdma model~\cite{Arcadi:2017wqi,Sanderson:2018lmj}.

\begin{figure}[t!]
\begin{center}
\includegraphics[width=0.475\textwidth]{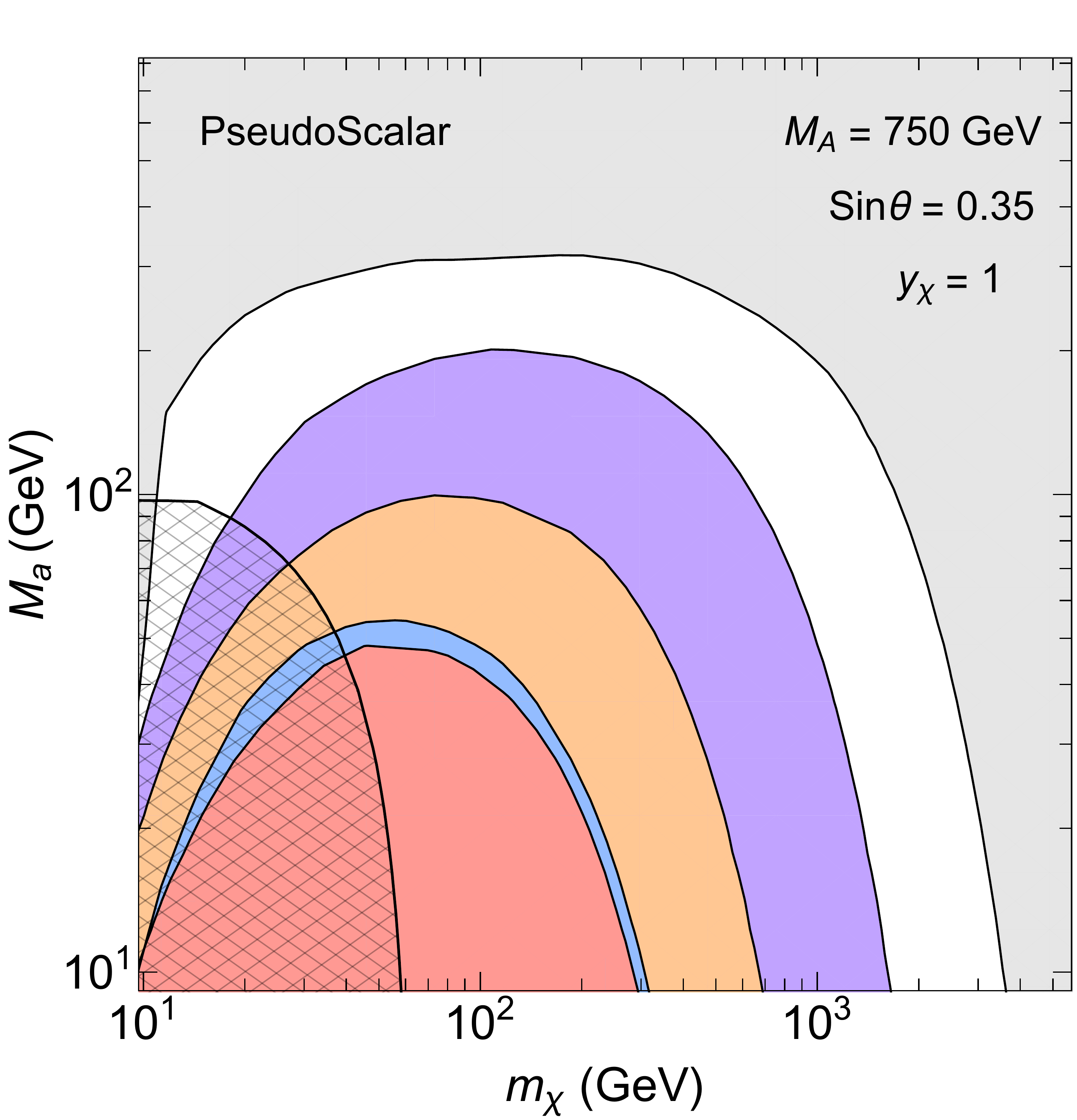} \quad 
\includegraphics[width=0.475\textwidth]{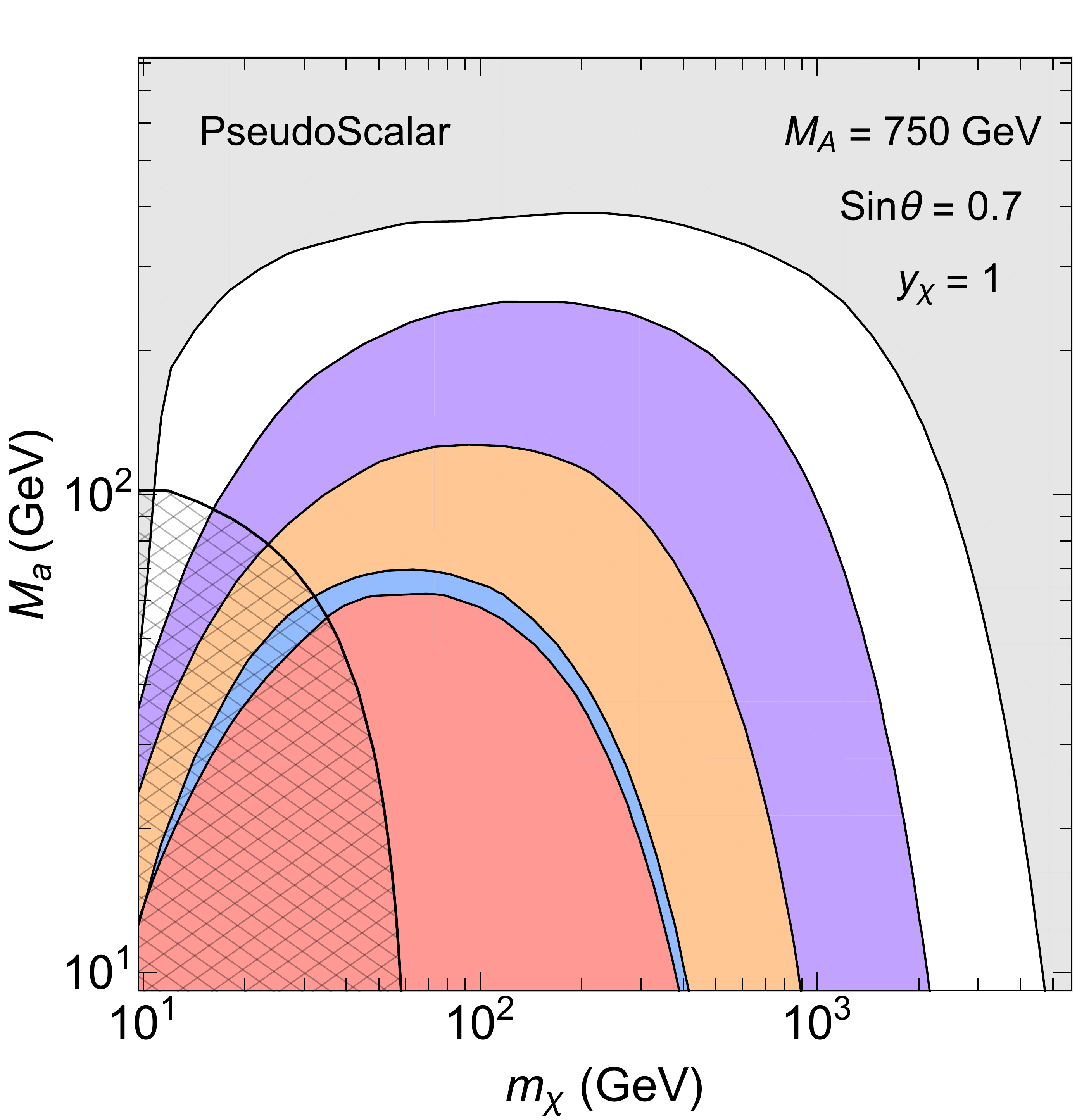}
\vspace{4mm}
\caption{DD exclusions in the \hdma model as function of $m_\chi$ and $M_a$.  The constraints from LUX~2016~(red)~\cite{Akerib:2016vxi}, XENON1T~2017~(blue)~\cite{Aprile:2017iyp} and the projections from XENON1T~2ty~(orange) and  XENONnT~20ty~(purple) \cite{Aprile:2015uzo} are shown. The grey shaded area is not accessible to ordinary DD experiments due to the presence of the neutrino background~\cite{Billard:2013qya}, while the   black hatched regions are excluded by the LHC bounds on invisible Higgs decays. In the left (right) panel the parameters $\sin \theta = 0.35$ ($\sin \theta = 0.7$), $M_A = 750 \, {\rm GeV}$ and $y_\chi = 1$ are employed.} 
\label{fig:PSDD}
\end{center}
\end{figure} 

The bounds that DD experiments can or may set on the  \hdma model are presented in Figure~\ref{fig:PSDD}. In the left (right) panel the choices $\sin \theta = 0.35$ ($\sin \theta = 0.7$), $M_A = 750 \, {\rm GeV}$ and $y_\chi = 1$ are employed. For~$\sin \theta = 0.35$,  current limits from LUX~2016~(red)~\cite{Akerib:2016vxi} and  XENON1T~2017~(blue)~\cite{Aprile:2017iyp} are able to exclude the portion of parameter space with $m_\chi \simeq [10, 300 ]\, {\rm GeV}$ and $M_a \lesssim 50 \, {\rm GeV}$. Projected limits from XENON1T~2ty~(orange) and XENONnT~20ty~(purple)~\cite{Aprile:2015uzo} are expected to expand the exclusions to $m_\chi \lesssim 1700 \, {\rm GeV}$ and $M_a \lesssim 200 \, {\rm GeV}$. In the case $\sin \theta = 0.7$, the obtained limits are slightly better because of the larger mixing angle.  The XENON1T~1ty constraints~\cite{Aprile:2018dbl} are not explicitly shown  in Figure~\ref{fig:PSDD}. They would fall between the  XENON1T~2017 and the XENON1T~2ty exclusions. For~comparison also the regions in the $m_\chi \hspace{0.5mm}$--$\hspace{0.5mm} M_a$ plane excluded by the present  LHC bounds on invisible Higgs decays --- see Section~\ref{sec:invisiblehiggs} and \cite{Bauer:2017ota} --- are shown as black hatched regions.  The results displayed in Figure~\ref{fig:PSDD} imply that present and future DD experiments cannot probe benchmarks like~\eqref{eq:benchmark} since these employ $m_\chi = 10 \, {\rm GeV}$. In fact, the sensitivity of DD is complementary to that of the mono-$X$ searches because the former constraints are strongest for $M_a < 2 m_\chi$ while the latter searches provide the best exclusions for~$M_a > 2 m_\chi$. 

The loop calculations of  $\sigma_{\rm SI}$ performed in~\cite{Ipek:2014gua,Arcadi:2017wqi,Sanderson:2018lmj,Li:2018qip} have been recently revisited and improved in \cite{Abe:2018emu}.  In~fact, the latter article has presented the first complete leading order calculation of the SI DM-nucleon scattering cross section in the  \hdma context. It includes the full set of two-loop diagrams that induce an effective interaction between DM and gluons and takes into account all terms in \eqref{eq:VHP}. In contrast, in the works~\cite{Ipek:2014gua,Arcadi:2017wqi,Sanderson:2018lmj,Li:2018qip}  as well as in this white paper, two-loop effects have merely been included in an approximate fashion and only the term $P  \hspace{0.5mm} ( i  b_P   H_1^\dagger H_2 + {\rm h.c.} )$ has been considered. Depending on the specific choice of parameters, the additional contributions calculated in \cite{Abe:2018emu} can lead to both an enhancement and a reduction of the SI DM-nucleon scattering cross section in the \hdma model. For parameters where $\sigma_{\rm SI}$ is enhanced, the predicted SI DM-nucleon scattering cross sections, however, still turn out to be smaller than the current upper bounds from DD experiments, if $y_\chi$ is fixed so as for the  thermal relic abundance to coincide with the observed value of $\Omega h^{2}$. The main conclusion drawn before that DD experiments have only a limited sensitivity to benchmarks like~\eqref{eq:benchmark} thus remains valid.  

\subsection{ID experiments}

\begin{figure}[t!]
\centering
\includegraphics[width=0.35\textwidth]{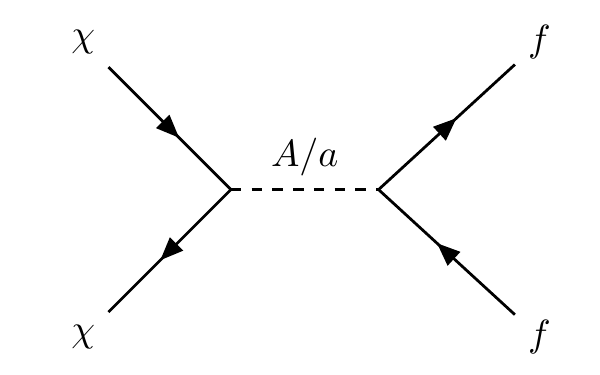} \qquad 
\includegraphics[width=0.35\textwidth]{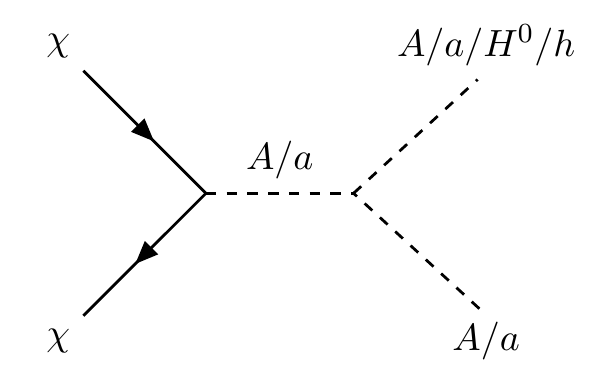} 

\vspace{5mm}

\includegraphics[width=0.35\textwidth]{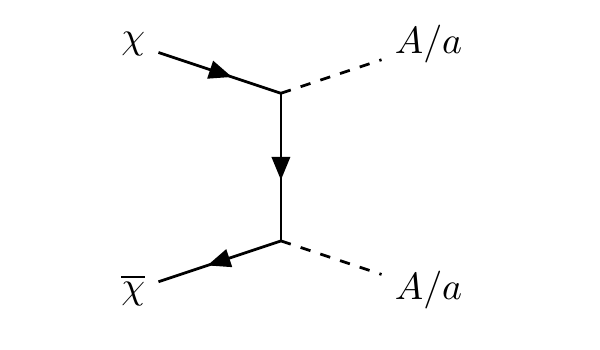} \qquad 
\includegraphics[width=0.35\textwidth]{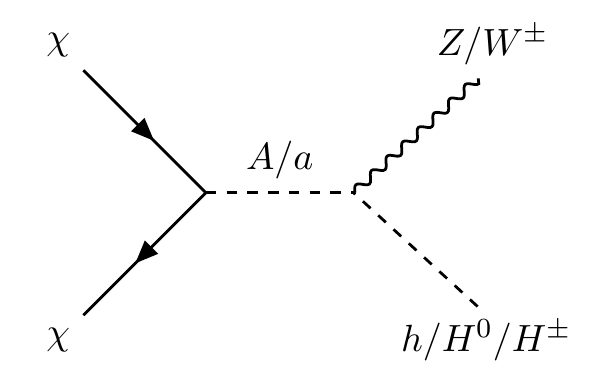}
\vspace{4mm}
\caption{Tree-level annihilation diagrams of DM in the \hdma model. Annihilation into pairs of SM fermions $(f)$, spin-0 states ($h, H,A,a$) and a spin-0 particle and a EW gauge boson~($HZ$ and $H^\pm W^\pm$) are possible in the alignment limit.}
\label{fig:feyn_annihilation}
\end{figure}

Due to the large number of couplings that the $A, a$ have with SM or 2HDM states, the ID signals in the \hdma model are  complex. In fact, for~$\cos (\beta - \alpha) = 0$, the possible annihiliation channels of DM are $ f \bar f$, $hA$, $HA$, $HZ$,  $H^\pm W^\mp$, $ha$, $Ha$, $AA$, $aa$ and~$Aa$.  Here~$f$ denotes all SM fermions that are kinematically accessible for a given DM mass,~i.e.~those fermions with $m_f < m_\chi$. Relevant diagrams are shown in Figure~\ref{fig:feyn_annihilation}. Since the SM gauge bosons and the Higgs states decay further into pairs of SM fermions, the final states resulting from the $\chi \bar \chi$ annihilation can contain either two or four SM particles. 

\begin{figure}[t!]
\centering
\includegraphics[height=0.4\textwidth]{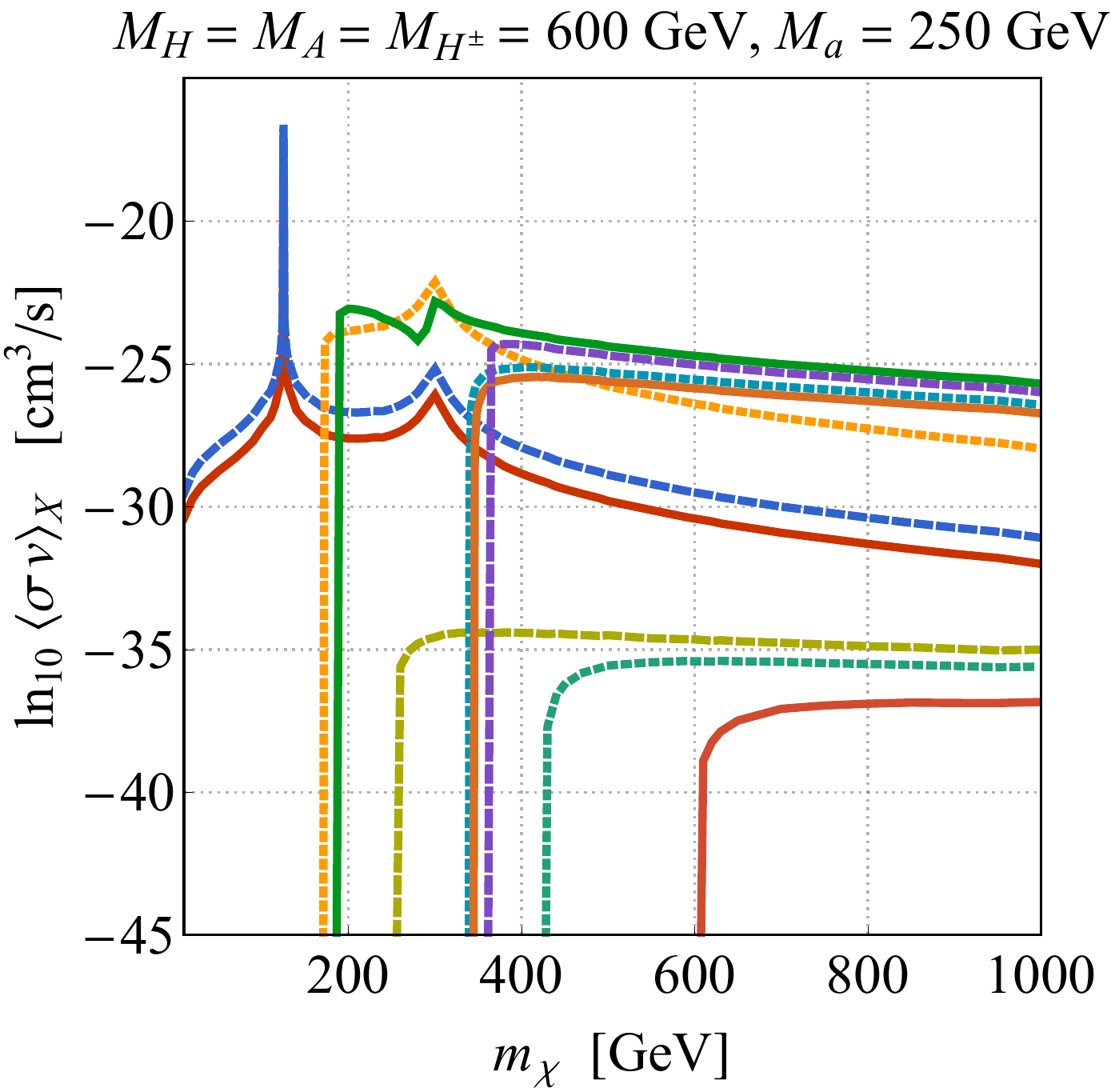} \quad 
\includegraphics[height=0.4\textwidth]{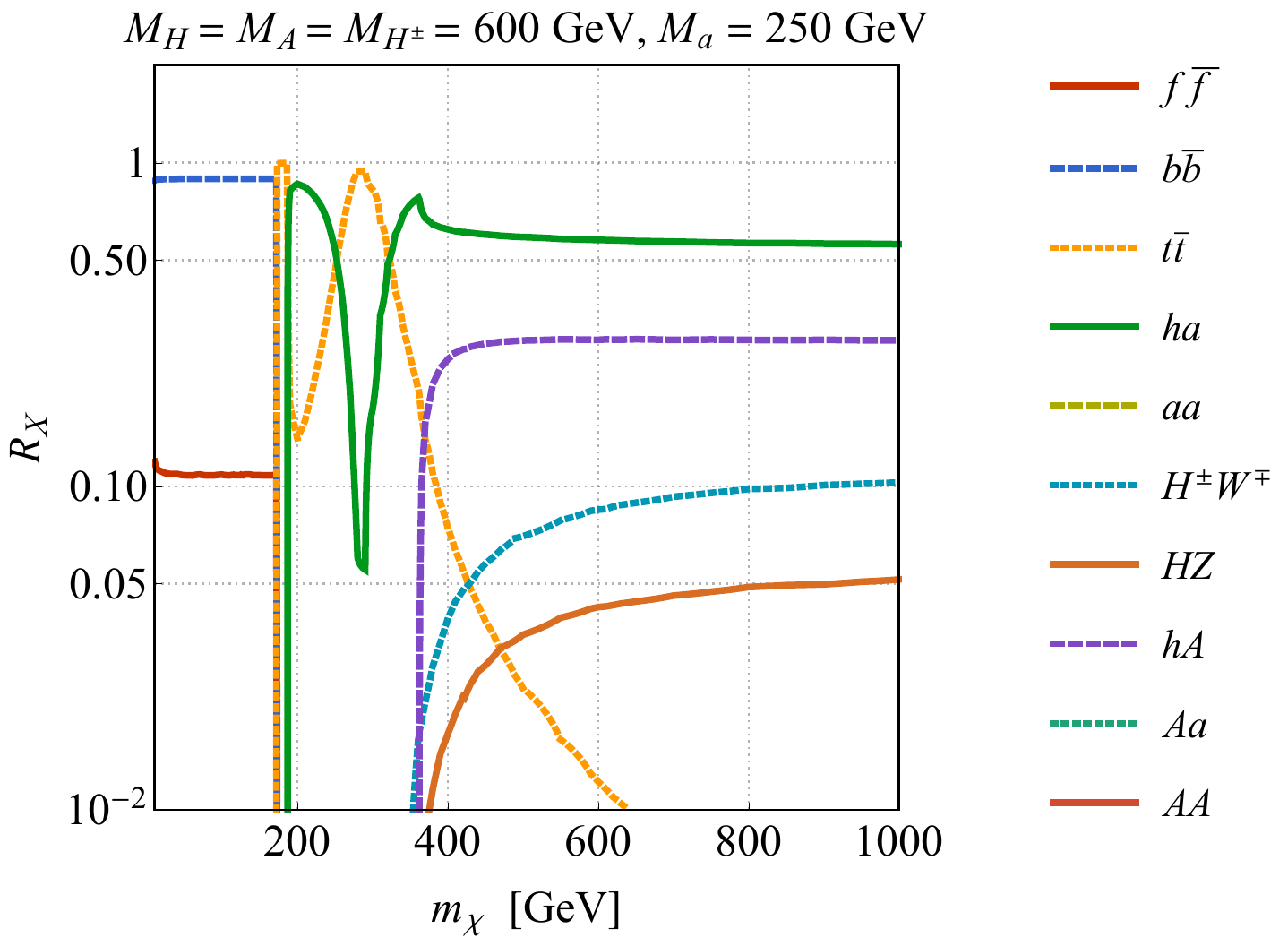}
\vspace{4mm}
\caption{The velocity-averaged DM annihilation cross sections  (left panel) and the corresponding relative rates (right panel) in the \hdma model. The shown results correspond to $M_H = M_A = M_{H^\pm} = 600 \, {\rm GeV}$, $M_a = 250 \, {\rm GeV}$ and the benchmark choices made in~\eqref{eq:benchmark}. See text for further details. }
\label{fig:IDbenchmark}
\end{figure}

In  Figure~\ref{fig:IDbenchmark} we display  an example of the various velocity-averaged DM annihilation cross sections (left) and the corresponding relative rates $R_X = \langle \sigma \, v \rangle_X/\sum_Y \langle \sigma \, v \rangle_Y$~(right) predicted  in the \hdma model. Here $Y =  f \bar f$, $b \bar b$, $t \bar t$,  $hA$, $HA$, $HZ$,  $H^\pm W^\mp$, $ha$, $Ha$, $AA$, $aa$, $Aa$. The employed input parameters are given by $M_H = M_A = M_{H^\pm} = 600 \, {\rm GeV}$, $M_a = 250 \, {\rm GeV}$ and~\eqref{eq:benchmark}. The numerical results for~$\langle \sigma v \rangle$ have been obtained with {\tt MadGraph5\_aMC@NLO}  using the latest  {\tt MadDM}~\cite{Ambrogi:2018jqj} plugin.  The average velocity of DM is taken to be $2 \cdot 10^{-5} \, c$, which is a typical velocity for Milky Way dwarf spheroidal satellite galaxies (see~e.g.~\cite{Simon:2007dq,Walker:2008ax}). Focusing on the region of DM masses below the top threshold, one sees that in this case only the annihilation cross sections for $\chi \bar \chi \to f \bar f$ with $f = e, \mu, \tau, u, d, s, c$ and $\chi \bar \chi \to b \bar b$ are non-zero.   Notice that both cross sections are resonantly enhanced at $m_\chi \simeq M_a/2$ due to $\chi \bar \chi \to a \to f \bar f$, leading to narrow peaks in the spectra. For $m_\chi > m_t$ the process $\chi \bar \chi \to t \bar t$ is also possible, representing the dominant fermionic annihilation channel for DM masses above the top threshold. One furthermore notices that all fermionic channels are enhanced for $m_\chi \simeq M_A/2$. This is again a resonance effect driven  by $\chi \bar \chi \to A \to f \bar f$ with an on-shell~$A$. The remaining annihilation processes $\chi \bar \chi \to A B$ with $A,B$ either  two spin-0 bosons or a spin-0 and a EW gauge boson turn on whenever the relevant threshold is reached,~i.e.~$m_\chi > (m_A + m_B)/2$. The largest channel of this type is $\chi \bar \chi \to ha$ which for the chosen parameters dominantly leads to a $b \bar b b \bar b$ final state. Also DM annihilation to  $H^\pm W^\mp$, $HZ$ and $hA$ is relevant for $m_\chi \gtrsim (M_h + M_A)/2$, while the remaining channels involving two pseudoscalars,~i.e.~$aa$, $Aa$ and $AA$, are all numerically negligible. We also add that the annihilation cross section corresponding to $\chi \bar \chi \to HA$ is exactly zero due to our parameter choices  $\tan \beta = 1$ and~$\lambda_{P1} = \lambda_{P2}$.  

Figure~\ref{fig:IDbenchmark} also shows that the total DM annihilation cross section is lowest for light~DM and in this mass region fully dominated by the annihilation into the $b \bar b$ final state. For our benchmark choice $m_\chi = 10 \, {\rm GeV}$, we obtain  for instance a velocity-averaged $\chi \bar \chi \to b \bar b$ annihilation cross section   of  $\langle \sigma v \rangle_{b \bar b} = 3.0 \cdot 10^{-30} \, {\rm cm}^3/{\rm s}$. The corresponding Fermi-LAT bound is compared to this weaker by more than three orders of magnitude as it amounts to $4.8 \cdot 10^{-27} \, {\rm cm}^3/{\rm s}$~\cite{Fermi-LAT:2016uux}.  In fact, for the parameters chosen to obtain the results depicted in the latter figure, we find that DM masses in the range of $m_\chi \simeq [110, 115] \, {\rm GeV}$ and $m_\chi \simeq [190, 405] \, {\rm GeV}$ are excluded at 95\%~CL by the Fermi-LAT constraints on $\chi \bar \chi \to b \bar b$ and $\chi \bar \chi \to t \bar t$, respectively. Notice that for $m_\chi > (m_h + m_a)/2$ also the $\chi \bar \chi \to A B \to 4f$ channels may lead to constraints when confronted with the Fermi-LAT data. These additional annihilation contributions have however not been considered when quoting the excluded~$m_\chi$ ranges $\big($a full treatment of ID bounds would require to calculate the sum of the photon energy ($E_\gamma$) spectra  from all contributing channels and to construct a joint  likelihood across all $E_\gamma$ bins of the photon flux to determine if a specific point in parameter space is ruled out; cf.~\cite{Carpenter:2015xaa,Carpenter:2016thc}~for instance$\big)$. The different dependence on~$m_\chi$ makes ID experiments and LHC mono-$X$ searches    complementary in constraining the \hdma parameter space. 

%%%%%%%%%%%%%%%%%%%%%%%%%%%%%%%%%%%%%%%%%%%%%%%%%%%%%%%%%%%%%%%%%%%%
%%%%%%%%%%%%%%%%%%%%%%%%%%%%%%%%%%%%%%%%%%%%%%%%%%%%%%%%%%%%%%%%%%%%
%%%%%%%%%%%%%%%%%%%%%%%%%%%%%%%%%%%%%%%%%%%%%%%%%%%%%%%%%%%%%%%%%%%%

\section{DM relic density}
\label{sec:relic}

In this section, we check the consistency of the \hdma model as a function of the parameters chosen for the scans with the measured DM relic density, according to the standard thermal relic ``freeze-out'' scenario. This exercise requires the following assumptions, already detailed in~\cite{Albert:2017onk}. First,  the DM annihilation cross section receives only contributions from the interactions of the \hdma model, while possible additional degrees of freedom and couplings not included in the model are ignored. Second, the DM number density in the Universe today is entirely determined by the DM annihilation cross section predicted by the \hdma model. In particular, no additional mechanisms exist that enhance or deplete the DM relic density. It is important to realise that if one or both of these assumptions are violated there is no strict correlation between the relic density and the strength of mono-$X$ signals. For instance, if DM is overproduced, the relic density can be reduced if the DM has large annihilation cross sections to new hidden sector states. These states might however not be directly accessible at LHC energies. Conversely, the correct DM relic density can still be obtained if the DM is underproduced. For instance, if the hidden sector carries a particle-antiparticle asymmetry (similar to the baryon asymmetry) then this necessarily leads to a larger relic density compared to the conventional freeze-out picture.

\begin{figure}[t!]
\centering
\includegraphics[width=0.475\textwidth]{{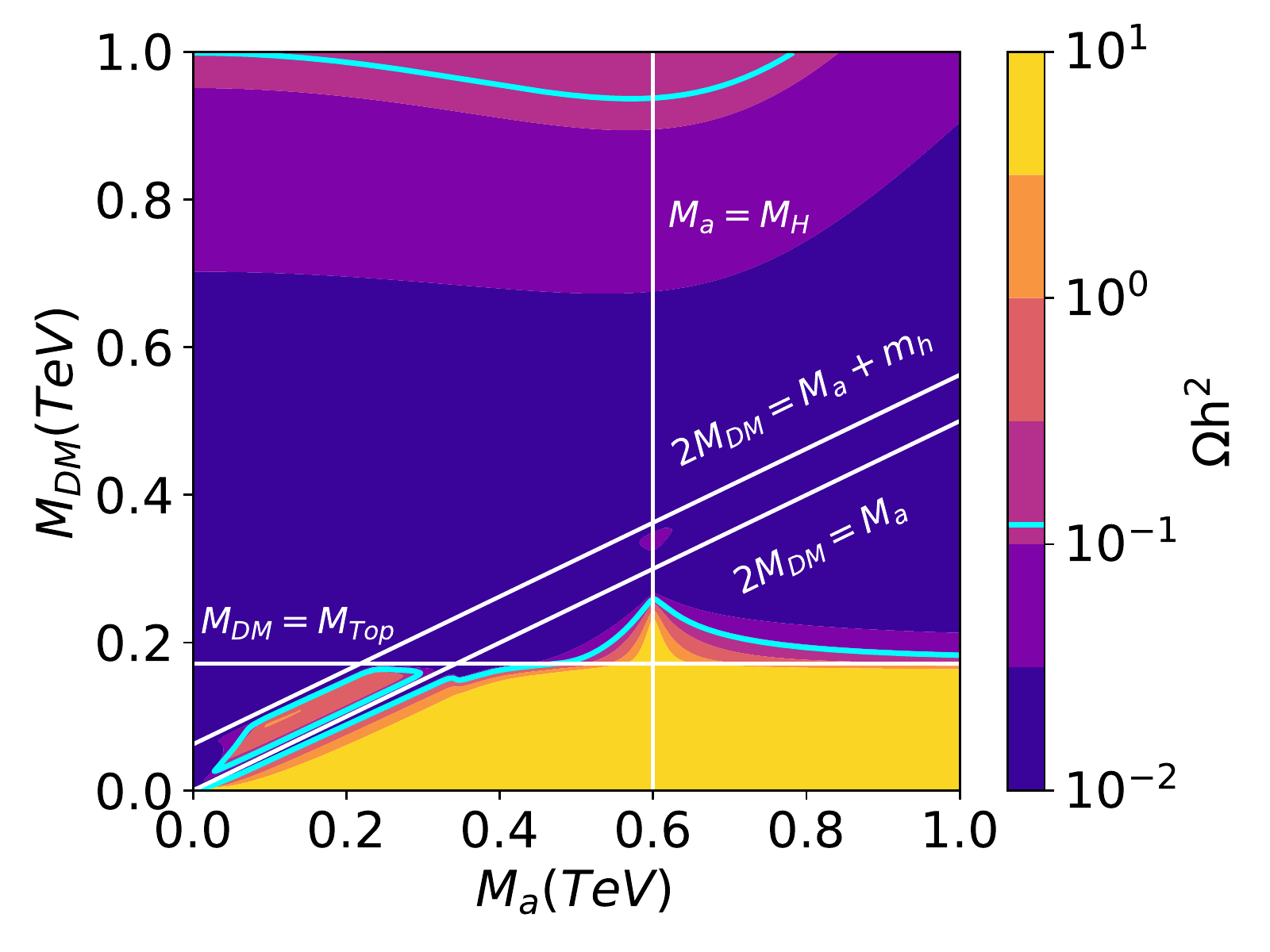}} \quad 
\includegraphics[width=0.475\textwidth]{{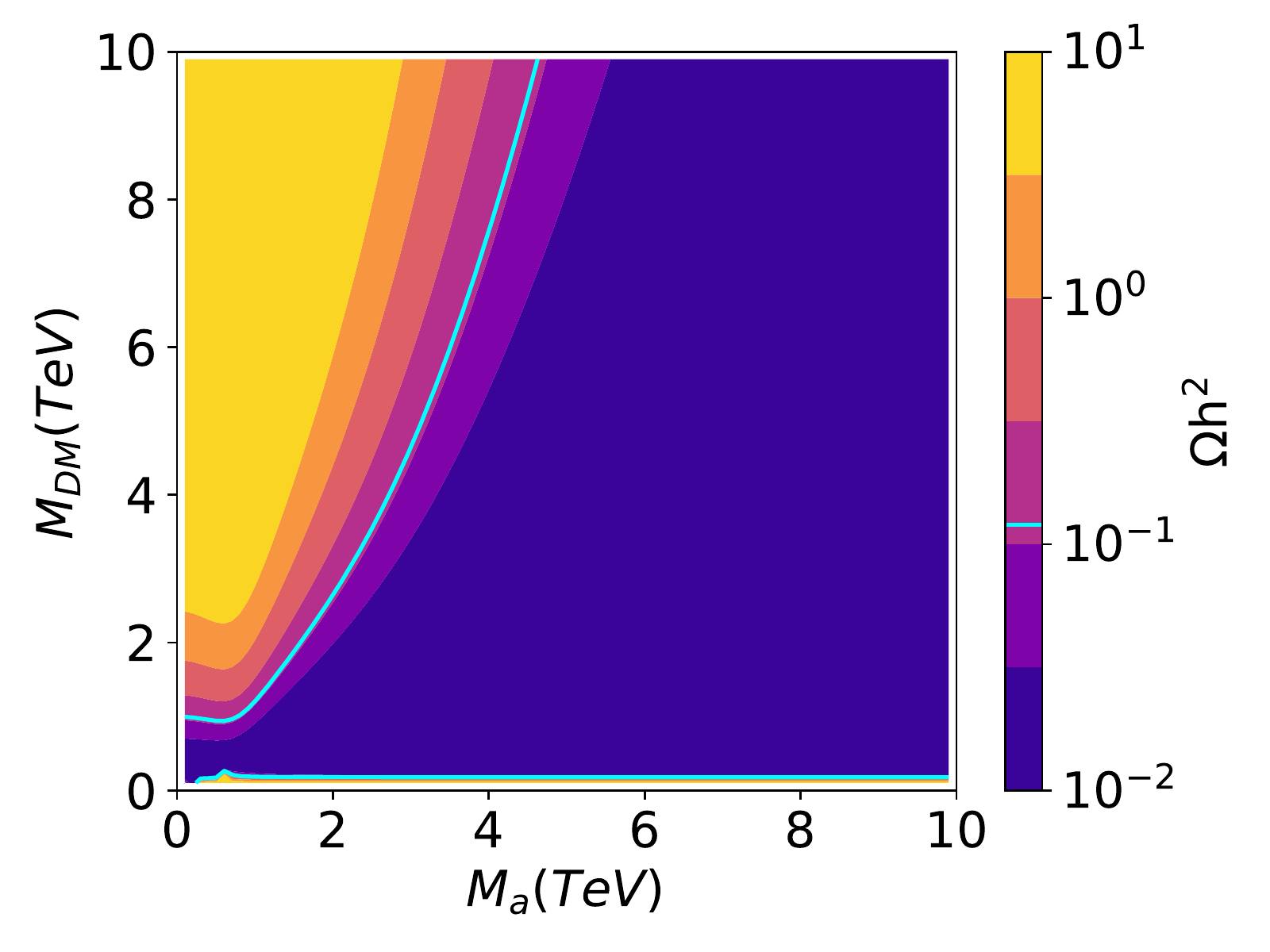}}
\vspace{2mm}
\caption{Predicted DM relic density for a two-dimensional scan of $\ma$ and $m_\chi$. The other parameters remain fixed at  $\mH=\mA=\mHc= 600 \, {\rm GeV}$ and $\tan \beta=1$, as well as the benchmark choices given in \eqref{eq:benchmark}. The color scale indicates the DM relic density, the cyan solid line shows the observed value of $\Omega h^{2} = 0.12$. The color scale is truncated at its ends,~i.e.~values larger than the maximum or smaller than the minimum are shown in the same color as the maximum/minimum. While the left panel focuses on the mass region relevant to collider searches, the right panel shows the development of the DM relic density for a larger mass region.}
\label{fig:relic_scan_mxd_ma}
\end{figure}

\subsection{Calculation}

The Feynman diagrams of the annihilation processes taken into account in the calculation of the DM relic density in the \hdma model are shown in Figure~\ref{fig:feyn_annihilation}. Generally, the annihilation proceeds via single (upper and lower right graphs) or double exchange (lower left graph) of the pseudoscalars  $A$ and~$a$ with subsequent decays. The {\tt MadDM}~\cite{Backovic:2013dpa,Backovic:2015cra} plugin for {\tt MadGraph5\_aMC@NLO} is used to calculate the present-day DM relic density.  Since {\tt MadDM} uses only $2 \to 2$ scattering diagrams, contributions from off-shell pseudoscalars can only be taken into account for the case of single mediation with direct decay of the pseudoscalars to SM fermions. If the pseudoscalars instead decay to other bosons or if the annihilation proceeds through double-exchange diagrams, the outgoing bosons are taken to be on-shell and their decays are not simulated. All tree-level annihilation processes are considered, and the Yukawa couplings of all fermions are taken to be non-zero.

\subsection{Scan results}

 If not stated otherwise, the results shown in this section use the benchmark values from~\eqref{eq:benchmark}. The DM relic density is displayed in the $\ma \hspace{0.5mm}$--$\hspace{0.5mm} m_\chi$ plane in the two panels of Figure~\ref{fig:relic_scan_mxd_ma}. The parameters not indicated in the plots are  fixed to  $\mH=\mA=\mHc= 600 \, {\rm GeV}$ and $\tan \beta=1$. For values of $m_\chi$ below the mass of the top quark, DM is mostly overabundant. In this regime, annihilation to quarks is suppressed by the small Yukawa couplings of the light fermions. The observed DM relic density can only be achieved for $m_\chi \simeq \ma/2$, where annihilation is resonantly enhanced, or for $m_\chi \simeq (\ma+\mh)/2$, close to the threshold for the $\chi\chi \to h a$ process.  Above the top threshold, annihilation into fermions becomes very efficient and DM is typically underabundant. Exceptions are regions in parameter space where $M_a \gtrsim m_t$ and $m_\chi \simeq m_t$ in which the observed DM relic density can be achieved.  As $m_\chi$ increases further, annihilation via single-exchange diagrams is more and more suppressed and the observed DM relic density can again be reproduced.  At low values of $\ma$ this happens for $m_\chi \simeq 1 \, {\rm TeV}$. The right panel of Figure~\ref{fig:relic_scan_mxd_ma}   shows in addition the two branches of solutions for masses up to $10 \, {\rm TeV}$. 

\begin{figure}[t!]
\centering
\includegraphics[width=0.475\textwidth]{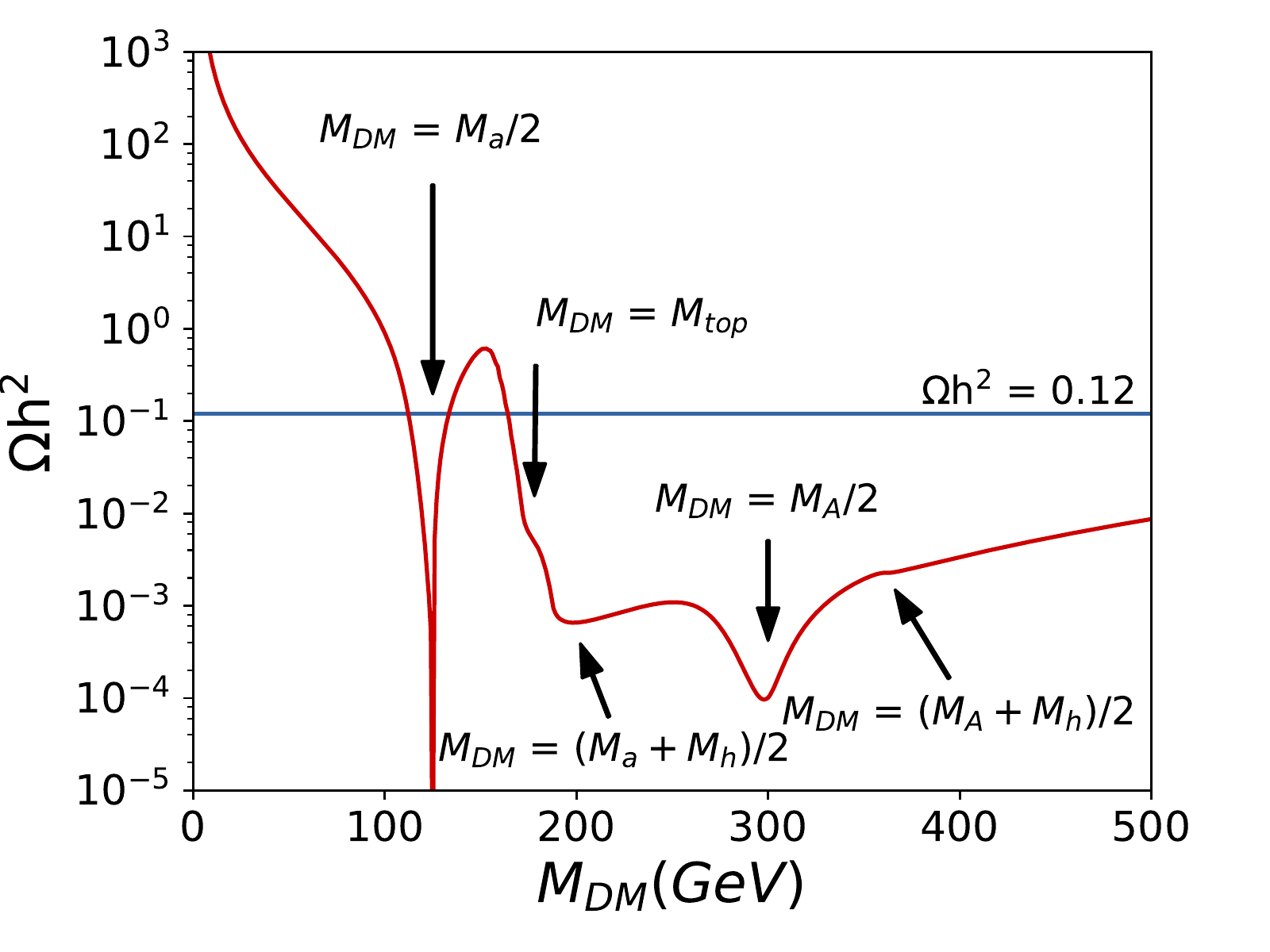} \quad 
\includegraphics[width=0.475\textwidth]{{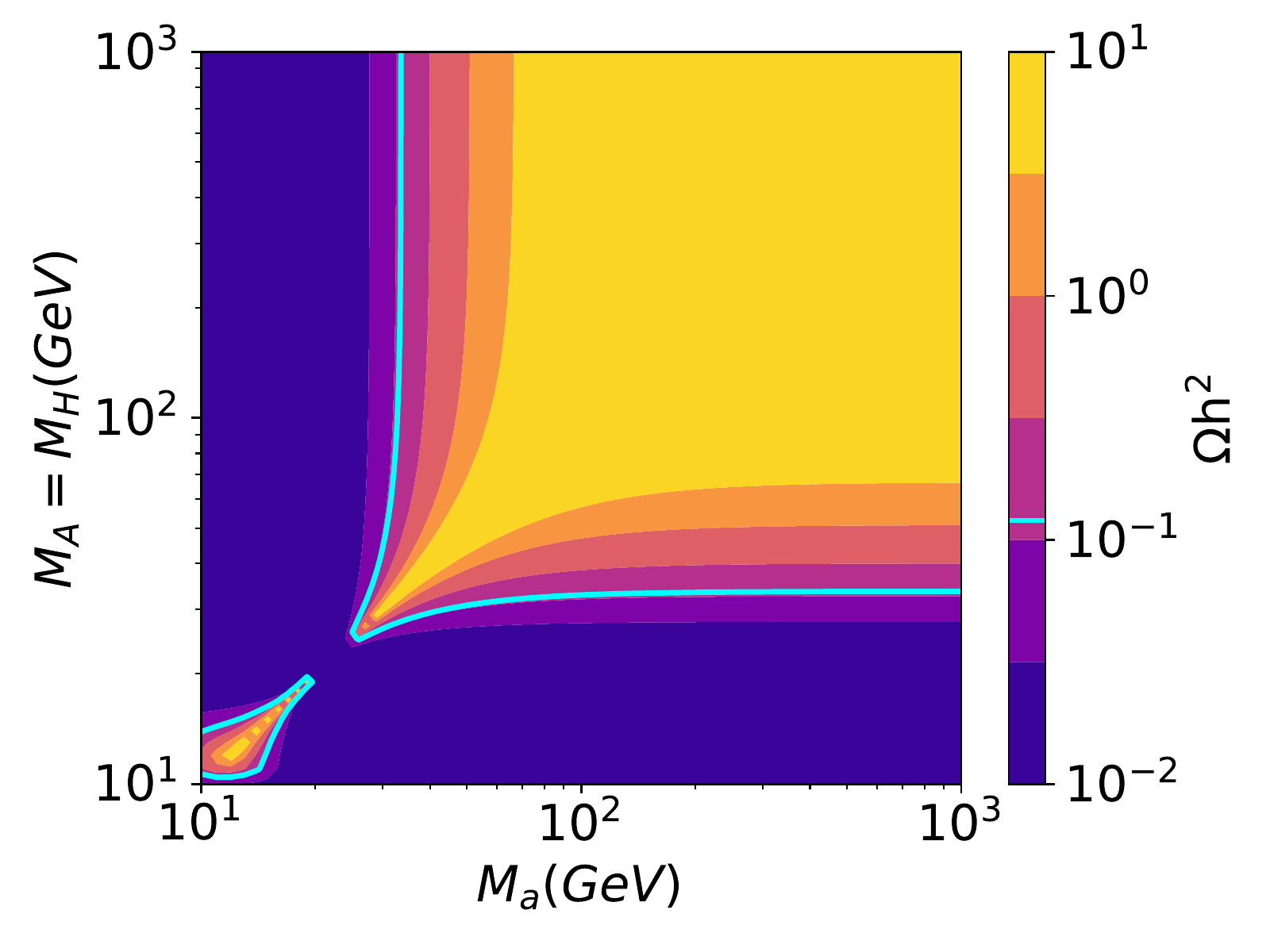}}
\vspace{2mm}
\caption{Left: DM relic density in the \hdma model as a function of $m_\chi$. The  predictions shown are obtained for $\mH=\mA=\mHc= 600 \, {\rm GeV}$, $\ma =250 \, {\rm GeV}$ and $\tan \beta=1$. See text for further details. Right: Predicted DM relic density for the \hdma model in the $\ma \hspace{0.5mm}$--$\hspace{0.5mm} \mA$ plane.  A common mass $\mH = \mA = \mHc$ is used. The colour coding resembles that of Figure~\ref{fig:relic_scan_mxd_ma}.}
\label{fig:relic_scan_mxd}
\end{figure}
 
The dependence of the DM relic density on the choice of $m_\chi$ is further explored in Figure~\ref{fig:relic_scan_mxd} (left). The red curve represents the choices $\mH=\mA=\mHc= 600 \, {\rm GeV}$, $\ma =250 \, {\rm GeV}$ and $\tan \beta=1$. The  result shown confirms the presence of the previously discussed regions of resonant enhancement and  kinematic boundaries. Overall, the behaviour is dominated by the low-$m_\chi$ suppression of the annihilation cross section, the resonant enhancement at $m_\chi = \ma/2$ and the  top thresholds. Other effects, such as the resonant enhancement of $\chi\chi \to A$ annihilation are present, but are small. 

 The DM relic density values for the $\ma$--$\mA$ scan  are shown in the right panel of Figure~\ref{fig:relic_scan_mxd}. The regions where the \hdma model predicts a DM relic density compatible with the measured value $\Omega h^{2} = 0.12$ are located either at $\ma < 30 \, {\rm GeV}$ or at $\mA=\mH=\mHc < 30 \, {\rm GeV}$. As explained in Section~\ref{sec:invisiblehiggs} the first option is excluded by the LHC bounds on invisible Higgs decays, while the second possibility is ruled out directly by LEP and LHC searches for charged Higgses and indirectly by flavour physics. This means that the benchmark~\eqref{eq:benchmark} employed in this white paper cannot give rise to the correct DM relic density as it generically predicts $\Omega h^{2} \gg 0.12$.  Since the cosmological production of DM is largely driven by the choice of $m_\chi$ it is however possible to tune the DM mass such that the correct DM relic density is obtained in scenarios~\eqref{eq:benchmark} with $m_\chi \neq 10 \, {\rm GeV}$.  For instance, by choosing the DM mass to be slightly below the $a$ threshold,~i.e.~$m_\chi = \ma/2$, one can obtain $\Omega h^{2} \lesssim 0.12$ (see the left panel in Figure~\ref{fig:relic_scan_mxd}). Given that both the total cross sections and the kinematic distributions of the mono-$X$ signatures are largely insensitive to the precise choice of $m_\chi$ as long as $m_\chi < \ma/2$ (cf.~Figure~\ref{fig:mdmvar}), our sensitivity studies performed in Section~\ref{sec:sensitivitystudies} apply to first approximation also to scenarios like~\eqref{eq:benchmark} where the measured DM relic density is obtained by tuning $ m_\chi \simeq \ma/2$. From the collider perspective another interesting parameter region is $M_a \gtrsim  2 m_t$ and $m_\chi \simeq m_t$ since it can be probed by LHC searches and can lead to the observed DM relic density (see~the left-hand side of Figure~\ref{fig:relic_scan_mxd_ma}).  

\begin{figure}[t!]
\centering
\includegraphics[width=0.45\textwidth]{{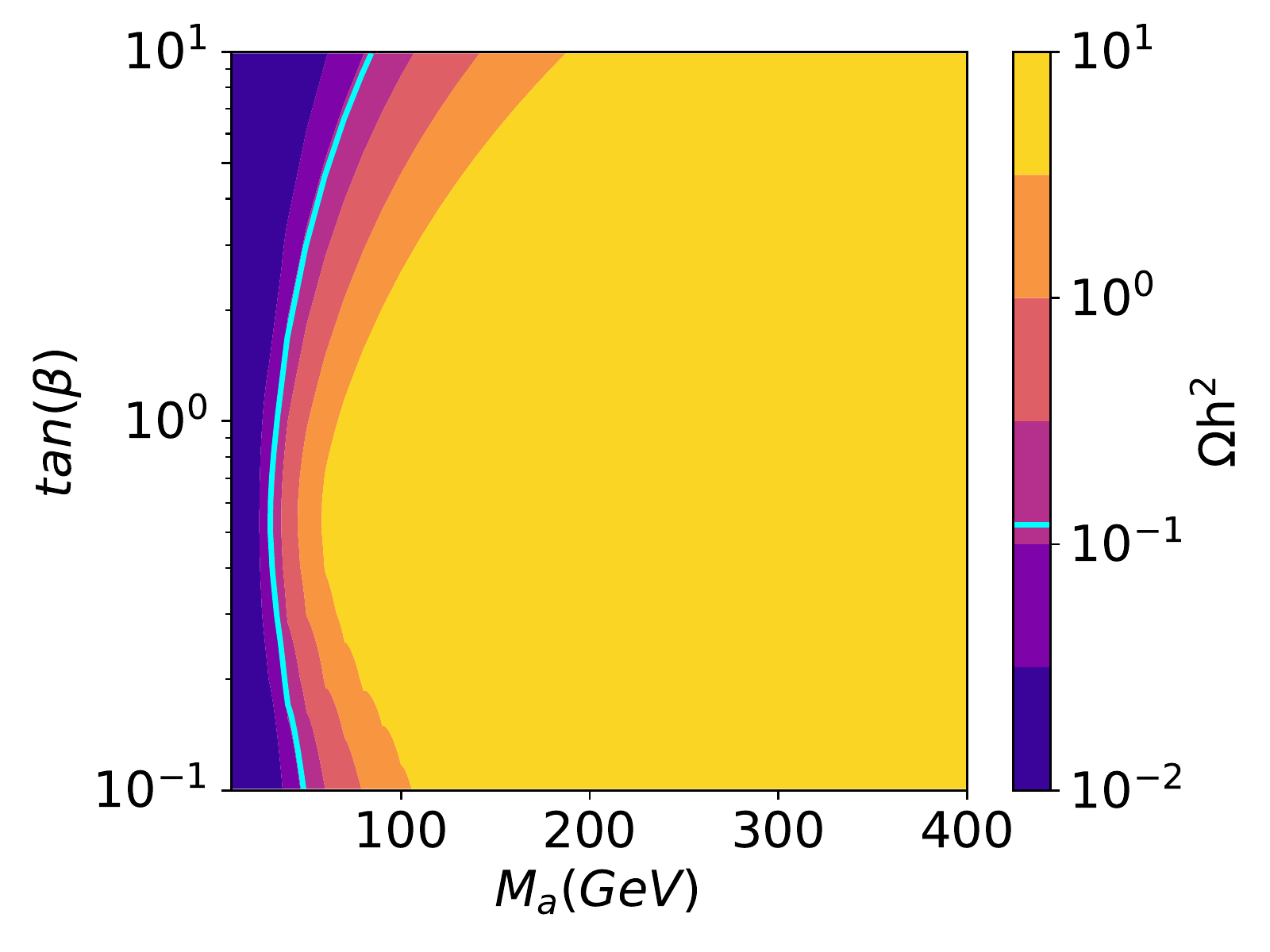}} \qquad 
\includegraphics[width=0.45\textwidth]{{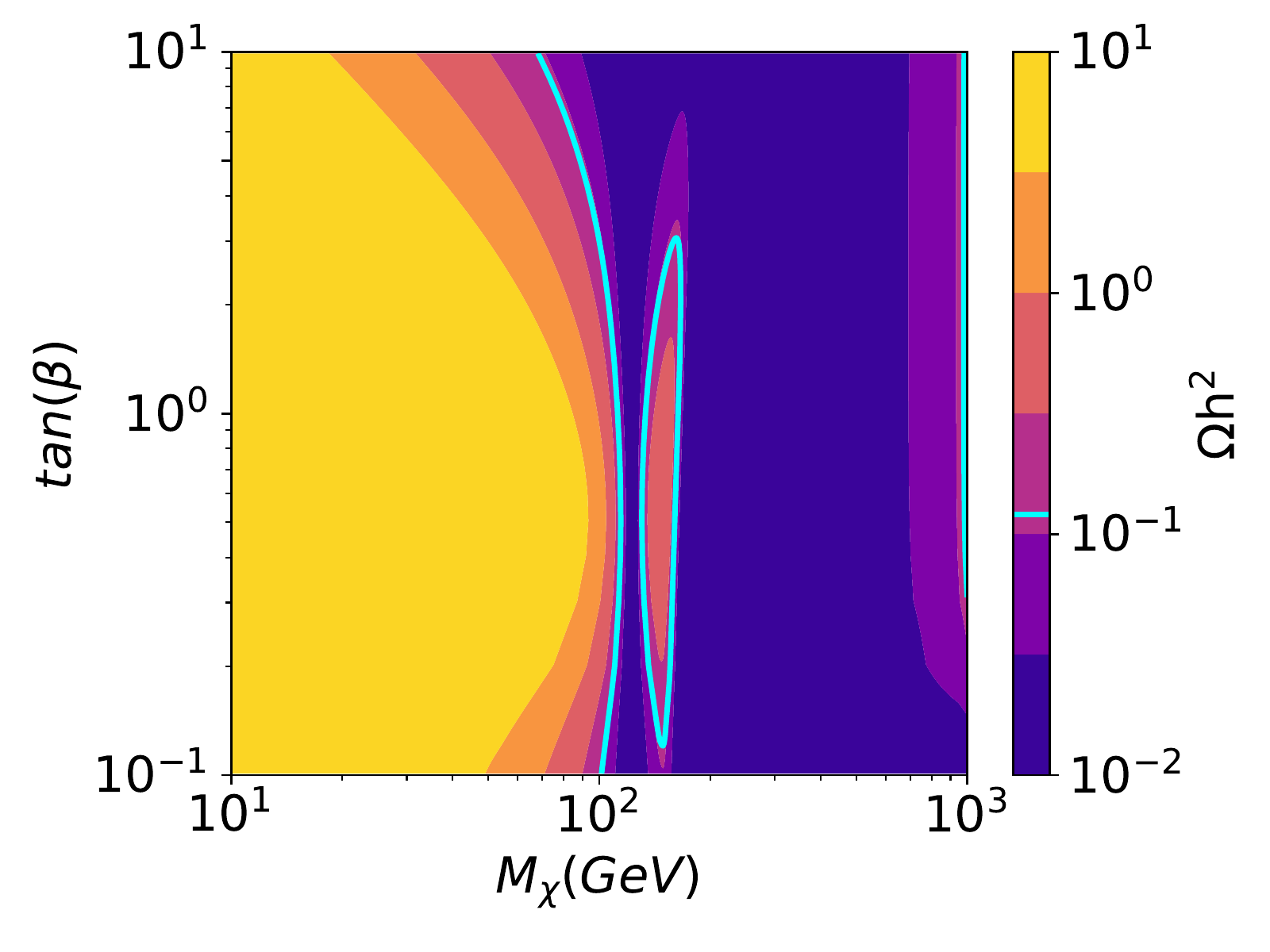}}
\vspace{2mm}
\caption{Predicted DM relic density in the \hdma model in the $\ma \hspace{0.5mm}$--$\hspace{0.5mm} \tan \beta$ (left panel) and $m_\chi \hspace{0.5mm}$--$\hspace{0.5mm} \tan \beta$ (right panel) plane, respectively. In the left (right) panel, $m_\chi = 10 \, {\rm GeV}$ ($m_a = 250 \, {\rm GeV})$ is employed as well as $\mH=\mA=\mHc= 600 \, {\rm GeV}$.  The color coding is identical to Figure~\ref{fig:relic_scan_mxd_ma}.}
\label{fig:relic_scan_mass_tanbeta}
\end{figure}

 In Figure~\ref{fig:relic_scan_mass_tanbeta} we display $\tan \beta$ scans as a function of $\ma$~(left panel) and $m_\chi$~(right panel). Both panels show that the values of $\ma$ ($m_\chi$) for which $\Omega h^{2} = 0.12$ do not depend strongly on the precise choice of $\tan \beta$. For choices of $\tan \beta \simeq 0.6$ the relic density becomes maximal and steadily decreases for larger and smaller values of $\tan \beta$. In the case of the $m_\chi \hspace{0.5mm}$--$\hspace{0.5mm} \tan \beta$ scan, the reduction of the DM relic density at $\tan \beta \simeq 0.1$ and $\tan \beta \simeq 3$  leads to the disappearance of the overabundant island around $m_\chi\simeq\ma/2$.
 
%%%%%%%%%%%%%%%%%%%%%%%%%%%%%%%%%%%%%%%%%%%%%%%%%%%%%%%%%%%%%%%%%%%%
%%%%%%%%%%%%%%%%%%%%%%%%%%%%%%%%%%%%%%%%%%%%%%%%%%%%%%%%%%%%%%%%%%%%
%%%%%%%%%%%%%%%%%%%%%%%%%%%%%%%%%%%%%%%%%%%%%%%%%%%%%%%%%%%%%%%%%%%%

\section{Proposed parameter scans}
\label{sec:scans}

The discussion of the theoretical motivations presented in Section~\ref{sec:constraints} together with our explicit studies  in Sections~\ref{sec:experimentbasics}, \ref{sec:nonMET}, \ref{sec:sensitivitystudies},  \ref{sec:DMdetection} and~\ref{sec:relic} suggest certain benchmarks for the parameters given in~\eqref{eq:inputparameters}. In this section, we describe how the parameter space of the \hdma model can be effectively explored through two-dimensional (2D) and one-dimensional (1D) scans of five input parameters: a common 2HDM heavy  spin-0 boson  mass \mH = \mA = \mHc, the  pseudoscalar mass \ma, the sine of the mixing angle $\sin \theta$, the ratio $\tan \beta$  of VEVs of the two 2HDM Higgs doublets  and the DM  mass $m_\chi$. The benchmark scenarios proposed in this white paper are a product of the work of the LHC Dark Matter Working Group members and have been agreed upon~\cite{LPCC-DMWG}.  They are not meant to provide an exhaustive scan of the entire \hdma parameter space, but are supposed to highlight many of the features that are special in the model, to showcase the complementarity of the various signatures and to ensure that the results of different analyses can be compared consistently.

\subsection[Scan in the $\ma$,  $\mH = \mA = \mHc$ plane]{Scan in the $\bm{\ma}$,  $\bm{\mH = \mA = \mHc}$ plane}
%\subsection{Scan in the $\bm{\ma}$,  $\bm{\mH = \mA = \mHc}$ plane}

The main 2D parameter grid proposed to explore the \hdma model with LHC data spans the combination of the pseudoscalar mass $\ma$ and a common heavy 2HDM  spin-0 boson  mass $\mH = \mA = \mHc$. The proposed values of the remaining \hdma parameters are given in \eqref{eq:benchmark}. Two example scans in the suggested mass-mass plane are given in the upper panels of Figures~\ref{fig:monoHbb_sensi} and~\ref{fig:monoZll_sensi}. These plots show the results of our sensitivity studies in the $h \, (b \bar b) + \MET$ and $Z \, (\ell^+ \ell^-) + \MET$ channel, respectively, and are based on $36 \, {\rm fb}^{-1}$ of $13 \, {\rm TeV}$ LHC data.  From the figures it is evident that in the benchmark scenario~\eqref{eq:benchmark}, one can already probe $M_a$ values  up to almost $350 \, {\rm GeV}$ and common heavy 2HDM  spin-0 boson  masses in the range of around $[300, 1000] \, {\rm GeV}$ with existing data.  The interpretation of other mono-$X$ channels such as $t W + \MET$, $t \bar t + \MET$ and $j + \MET$ (cf.~Section~\ref{sec:experimentbasics}) as well as non-$\MET$ searches for final states like $\tau^+ \tau^-$, $tb$ and  $t \bar t t \bar t$~(cf.~Section~\ref{sec:nonMET}) in this plane will allow to  illustrate the complementary of the different search strategies for the spin-0 \hdma states at the LHC. Furthermore, combinations of the results of different searches can be done consistently for~\eqref{eq:benchmark} and are expected to cover more parameter space than considering one signature at a time. 

\subsection[Scan in the $\ma\hspace{0.5mm}$--$\hspace{0.5mm} \tan \beta$ plane]{Scan in the $\bm{\ma}\hspace{0.5mm}$--$\hspace{0.5mm} \bm{\tan \beta}$ plane}
%\subsection{Scan in the $\bm{\ma}\hspace{0.5mm}$--$\hspace{0.5mm} \bm{\tan \beta}$ plane}

A 2D scan in the  $\ma\hspace{0.5mm}$--$\hspace{0.5mm} \tan \beta$ plane with  the common heavy 2HDM  spin-0 boson  masses fixed to $\mH = \mA = \mHc = 600 \, {\rm GeV}$ is proposed.  The remaining parameters should be chosen as in \eqref{eq:benchmark}. Two examples of such a scan can be found in the lower panels of Figures~\ref{fig:monoHbb_sensi} and~\ref{fig:monoZll_sensi}. With $36 \, {\rm fb}^{-1}$ of $13 \, {\rm TeV}$ LHC data, mono-Higgs and mono-$Z$ searches are already sensitive to $\tan \beta = {\cal O} (1)$ values for $M_a$ values up to around $300 \, {\rm GeV}$. Other mono-$X$ searches like $t \bar t +\MET$ and $j + \MET$ are at present only sensitive to $\tan \beta \lesssim 0.5$, which emphasises the special role that  resonant $\MET$ signatures such as $h + \MET$, $Z + \MET$ and $tW + \MET$ play in the \hdma model (see Section~\ref{sec:resonant}). Like  the mass-mass plane discussed before, also the $\ma\hspace{0.5mm}$--$\hspace{0.5mm} \tan \beta$ plane offers a nice way to compare and to contrast the LHC reach of $\MET$ and non-$\MET$ searches in the \hdma context. 

\subsection[Scans in $\sin \theta$]{Scans in $\bm{\sin \theta}$}
%\subsection{Scans in $\bm{\sin \theta}$}

Two 1D  scans in $\sin \theta$ are also proposed, one with $\mH = \mA = \mHc = 600 \, {\rm GeV}$ and $\ma = 200 \, {\rm GeV}$ and a second one with $\mH = \mA = \mHc = 1000 \, {\rm GeV}$ and $\ma = 350 \, {\rm GeV}$. In both scans, the remaining parameters should be set equal to \eqref{eq:benchmark}.  We recommend the scans in $\sin \theta$ because they will  allow for a further comparison of the sensitivities of the mono-Higgs and mono-$Z$ searches given that these two channels have a different $\sin \theta$ dependence (cf.~Figure~\ref{fig:svar}).  We add that for the two proposed scans only values of $\sin \theta < 0.75$ and $\sin \theta < 0.45$ will lead to a scalar potential that satisfies the BFB conditions. This follows from the inequality~\eqref{eq:BFB2}.

\subsection[Scan in $m_\chi$]{Scan in $\bm{m_\chi}$}
%\subsection{Scan in $\bm{m_\chi}$}

To make contact to DD, ID and DM relic density calculations, which are strongly dependent on the DM mass, we also recommend to perform 1D scans in $m_\chi$ spanning from $1\, {\rm GeV}$ to $500 \, {\rm GeV}$. The  spin-0 boson masses should be taken as $\mH = \mA = \mHc = 600 \, {\rm GeV}$ and $\ma = 250 \, {\rm GeV}$  in these scans and the other \hdma parameters set to~\eqref{eq:benchmark}. We recall that for masses $m_\chi \simeq \ma/2$ the observed DM relic density can be obtained~(cf.~Section~\ref{sec:relic}). Such fine-tuned \hdma scenarios are hence in agreement with cosmological observations (assuming a standard freeze-out picture) and  it should be possible to probe/exclude them with the help of the LHC, since the mono-$X$ signatures are largely insensitive to the precise choice of the DM mass as long as $m_\chi < \ma/2$  (cf.~Figure~\ref{fig:mdmvar}). Other interesting parameter choices are $M_a \gtrsim 2 m_t$ and $m_\chi \simeq m_t$  since in this parameter space $\Omega h^2 = 0.12$ can be obtained and such configurations can be tested with both $\MET$ and non-$\MET$ searches. \\[2mm]

\noindent While in all our scan recommendations we have employed a common 2HDM heavy  spin-0 boson  mass $\mH = \mA = \mHc$, in future \hdma interpretations of LHC data one may also want to consider cases with $M_H \neq M_A$, since having a mass splitting between the $H$, $A$ and $H^\pm$ can lead to interesting effects in the mono-Higgs and mono-$Z$ searches (see Figure~\ref{fig:mvar}) as well as the $t \bar t Z$ and $tbW$ final states (cf.~the discussion in Section~\ref{sec:others}).  

%%%%%%%%%%%%%%%%%%%%%%%%%%%%%%%%%%%%%%%%%%%%%%%%%%%%%%%%%%%%%%%%%%%%
%%%%%%%%%%%%%%%%%%%%%%%%%%%%%%%%%%%%%%%%%%%%%%%%%%%%%%%%%%%%%%%%%%%%
%%%%%%%%%%%%%%%%%%%%%%%%%%%%%%%%%%%%%%%%%%%%%%%%%%%%%%%%%%%%%%%%%%%%

\acknowledgments 

T.~Abe's work is supported by JSPS KAKENHI Grant Number 16K17715. The research of Y.~Afik and Y.~Rozen was supported by a grant from the United States-Israel Binational Science Foundation, Jerusalem, Israel, and by a grant from the Israel Science Foundation.  A.~Albert receives support from the German Federal Ministry of Education and Research under grant 05H15PACC1. The research of A.~Boveia and L.~M.~Carpenter is supported by the U.S. DOE grant  DE-SC0011726. K.~J.~Behr acknowledges the support of the Helmholtz Foundation. C.~Doglioni has received funding from the European Research Council under the European Union's Horizon 2020 research and innovation program (grant agreement 679305) and from the Swedish Research Council.  S.~Gori is supported by the NSF CAREER grant PHY-1654502 and grateful to  the Kavli Institute for Theoretical Physics in Santa Barbara, supported in part by the NSF under Grant No.~NSF~PHY11-25915, as well as the Aspen Center for Physics, supported by the National Science Foundation Grant No.~PHY-1066293, for  hospitality. U.~Haisch acknowledges the hospitality and support of the CERN Theoretical Physics Department. J.~Hisano's work is supported by Grant-in-Aid for Scientific research from the Ministry of Education, Science, Sports, and Culture~(MEXT), Japan, No.~16H06492, and also by the World Premier International Research Center Initiative (WPI Initiative), MEXT, Japan.  The work of J.~M.~No was partially supported by the European Research Council under the European Union's Horizon 2020 program (grant agreement 648680) and by the Programa Atraccion de Talento de la Comunidad de Madrid under grant n.~2017-T1/TIC-5202. DDP is supported by STFC under grant ST/M005437/1.  The work of T.~M.~P.~Tait is supported in part by NSF grant PHY-1316792. T.~Robens is supported in part by the National Science Centre, Poland, the HARMONIA project under contract UMO-2015/18/M/ST2/00518 (2016-2019), and by grant K~25105 of the National Research, Development and Innovation Fund in Hungary. We gratefully acknowledge the support by the U.S. DOE. 

%%%%%%%%%%%%%%%%%%%%%%%%%%%%%%%%%%%%%%%%%%%%%%%%%%%%%%%%%%%%%%%%%%%
%%%%%%%%%%%%%%%%%%%%%%%%%%%%%%%%%%%%%%%%%%%%%%%%%%%%%%%%%%%%%%%%%%%%
%%%%%%%%%%%%%%%%%%%%%%%%%%%%%%%%%%%%%%%%%%%%%%%%%%%%%%%%%%%%%%%%%%%%

\appendix

\section{Recasting procedure}
\label{app:recast}

In this appendix we discuss  a general strategy that can be used to reinterpret existing $t \bar t + \MET$, $b \bar b + \MET$ and $j + \MET$ results obtained in the DMF pseudoscalar model in terms of the \hdma model. Example diagrams that lead to these mono-$X$ signatures in the \hdma model are displayed in Figure~\ref{fig:nonresonant}. Only graphs involving the exchange of an~$a$ are depicted in this figure but similar diagrams  with an  $A$ are not explicitly shown. 

\begin{figure}[t!]
\centering
\includegraphics[width=0.5\textwidth]{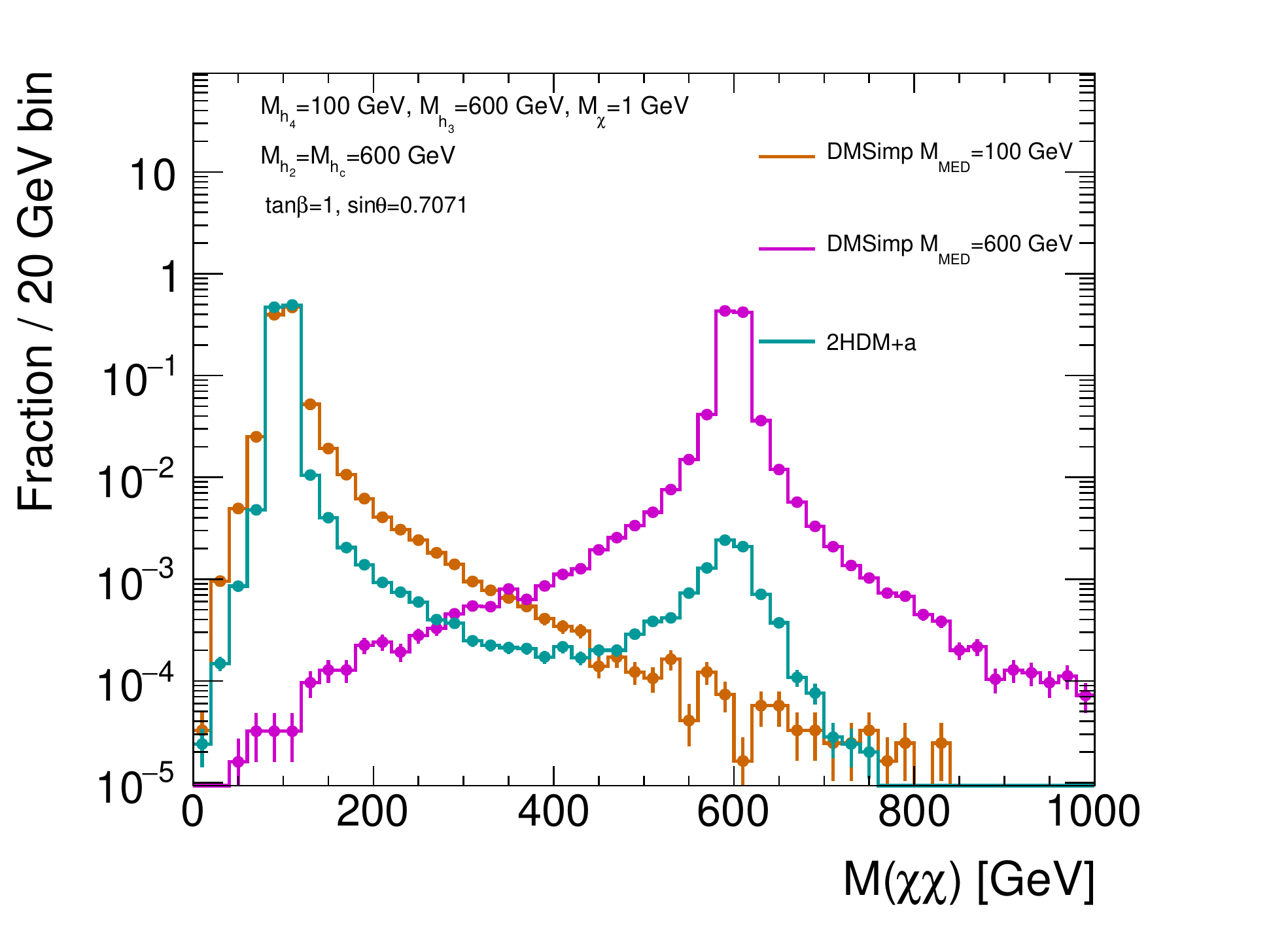}
\includegraphics[width=0.45\textwidth]{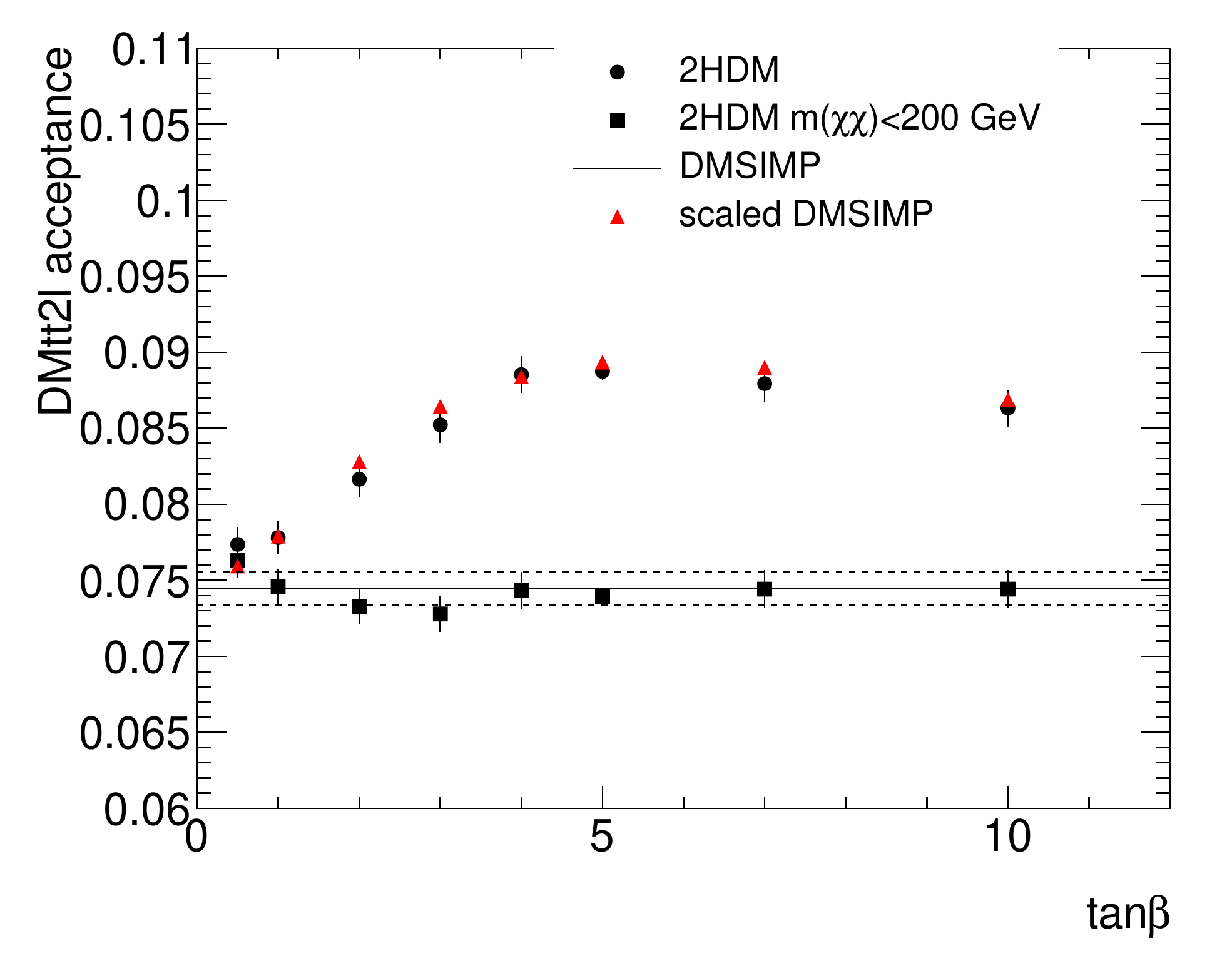}
\vspace{2mm}
\caption{ \label{fig:mchichi_DMsimpV2HDMa}
Left: Invariant mass of the $\chi\bar{\chi}$ system in $t \bar t + \MET$ production for the DMF pseudoscalar model with $\ma =100 \, {\rm  GeV}$~(brown) and $\ma =600 \, {\rm  GeV}$~(magenta) compared to the \hdma model with $\ma = 100 \, {\rm  GeV}$, $\mH = \mA = \mHc = 600 \, {\rm  GeV}$, $\sin\theta=1/\sqrt{2}=0.7071$ and $\tan\beta=1$~(cyan). Right: Acceptances of the   two-lepton $t \bar t + \MET$ analysis~\cite{Aaboud:2017rzf}  as a function of $\tan\beta$.  Shown are the predictions in the \hdma model without (round black markers) and with the cut $m_{\chi \bar \chi} <200 \, {\rm GeV}$ (square black markers), assuming $\ma = 150 \, {\rm GeV}$, $\mH= \mA = \mHc = 600 \, {\rm GeV}$ and $\sin\theta=0.35$. The DMF pseudoscalar model result (full black line) with its statistical uncertainty (dashed black lines) as well as the acceptance  ${\cal A}_{\hdma} \left (\ma, \mA \right )$ (red triangles) as defined in~\eqref{eq:recast} is also depicted.}
\end{figure}

 The relevance of the contributions from both the $a$ and $A$ in the \hdma model can be  demonstrated by considering the invariant mass $m_{\chi \bar \chi}$ of the $\chi \bar \chi$ system. Examples of~$m_{\chi \bar \chi}$ distributions in $t \bar t + \MET$ production are shown in the left panel of Figure~\ref{fig:mchichi_DMsimpV2HDMa}. The brown (magenta) histogram corresponds to the prediction in the DMF pseudoscalar model assuming a mediator mass of $\ma =100 \, {\rm  GeV}$ ($\ma =600 \, {\rm  GeV}$), while the cyan histogram illustrates the result in the \hdma model for the choices $\ma = 100 \, {\rm  GeV}$, $\mH = \mA = \mHc = 600 \, {\rm  GeV}$, $\sin\theta=1/\sqrt{2}=0.7071$ and $\tan\beta=1$. The predictions obtained in the  DMF pseudoscalar model both show a single  Breit-Wigner  peak  at $m_{\chi \bar \chi} = \ma$, which corresponds to the on-shell production of the mediator $a$ that subsequently decays to a pair of DM particles. The \hdma result instead features two mass peaks, one at $m_{\chi \bar \chi} = \ma$ and another one at $m_{\chi \bar \chi} = \mA$, because both pseudoscalars can be produced on-shell and then decay invisibly via either $a \to \chi \bar \chi$ or $A \to \chi \bar \chi$.  

The above discussion suggests that once the contributions from $a$ and $A$ production are separated, $t \bar t + \MET$, $b \bar b + \MET$ and $j + \MET$  results obtained in the DMF pseudoscalar model can be mapped into the $\hdma$ parameter space. In practice, the remapping is achieved by calculating the selection acceptances~${\cal A}_{\rm DMF} \left ( \ma \right )$ and~${\cal A}_{\rm DMF} \left ( \mA \right )$ in the DMF pseudoscalar model and the respective cross sections~$\sigma_{a, {\rm DMF}}$ and~$\sigma_{A, {\rm DMF}}$. The acceptance~${\cal A}_{\hdma} \left (\ma, \mA \right )$ in the \hdma model  is then obtained by computing the following weighted average
\begin{equation} \label{eq:recast}
{\cal A}_{\hdma} \left (\ma, \mA \right )=\frac{ {\cal A}_{\rm DMF} \left ( \ma \right )\, \sigma_{a, {\rm DMF}} + {\cal A}_{\rm DMF} \left ( \mA \right ) \, \sigma_{A, {\rm DMF}} }{\sigma_{a, {\rm DMF}}+\sigma_{A, {\rm DMF}}} \,.
\end{equation}

In the right panel of Figure~\ref{fig:mchichi_DMsimpV2HDMa} we show the results that are obtained by applying the latter equation to a parton-level implementation of the two-lepton $t \bar t + \MET$ analysis described in~\cite{Aaboud:2017rzf}. The round (square) black markers indicate the results of a direct calculation in the \hdma model without a $m_{\chi \bar \chi}$ cut (imposing the cut $m_{\chi \bar \chi} < 200 \, {\rm GeV}$), using $\ma = 150 \, {\rm GeV}$, $\mH= \mA = \mHc =600 \, {\rm GeV}$ and $\sin\theta=0.35$. The DMF pseudoscalar model result with its statistical uncertainty is represented by the solid and dashed black lines. The acceptance calculated from \eqref{eq:recast} is finally indicated by the red triangles. Two features are evident from the figure. First, the \hdma acceptance with cut agrees with uncertainties with the acceptance of the DMF pseudoscalar model. This is expected because the cut $m_{\chi \bar \chi} < 200 \, {\rm GeV}$ strongly suppresses the $A$ contribution in the \hdma model. Second, the acceptance estimated using  \eqref{eq:recast} agrees within uncertainties with the acceptance evaluated directly in the \hdma sample. 

\begin{figure}[t!]
\includegraphics[width=.5\textwidth]{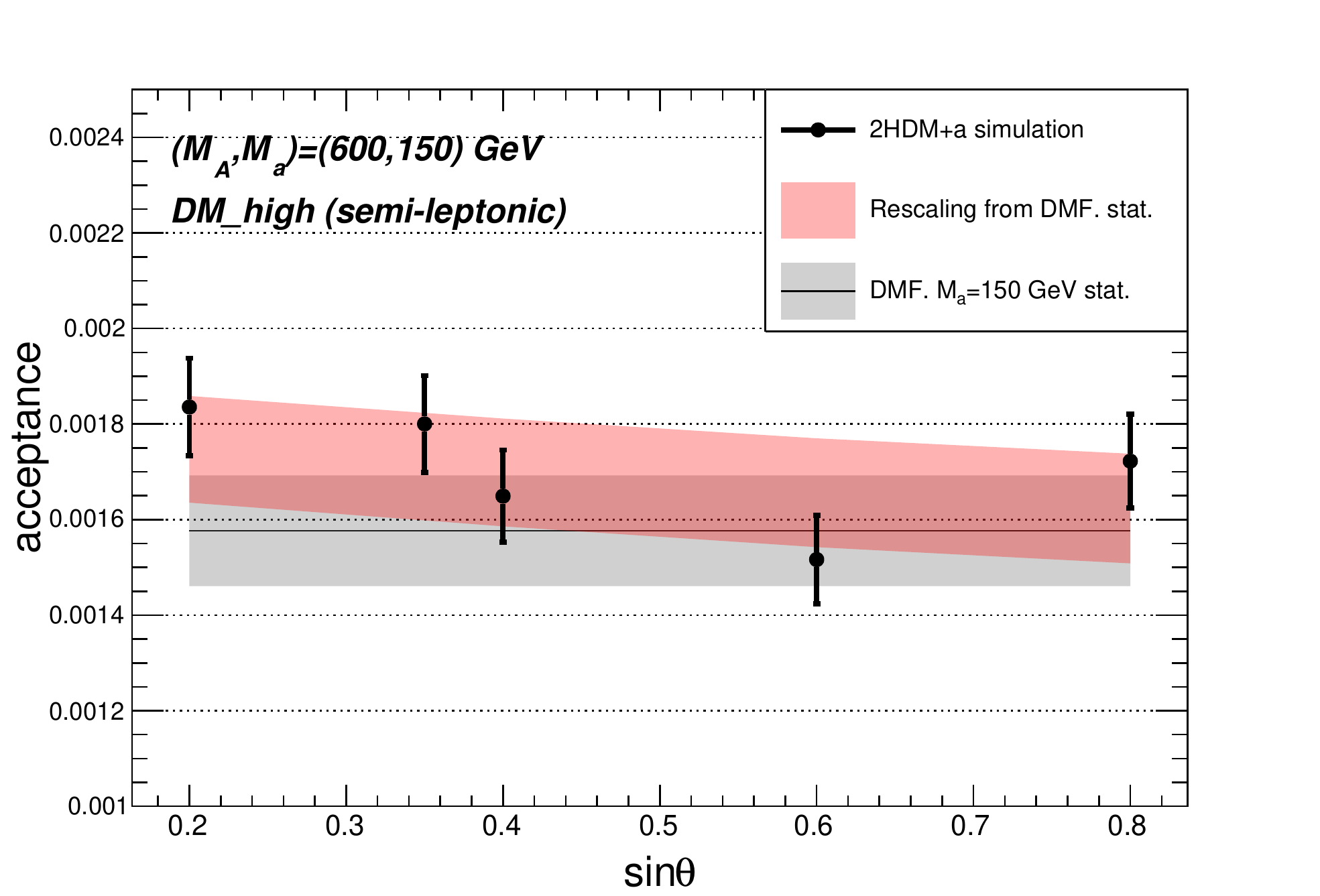}
\includegraphics[width=.5\textwidth]{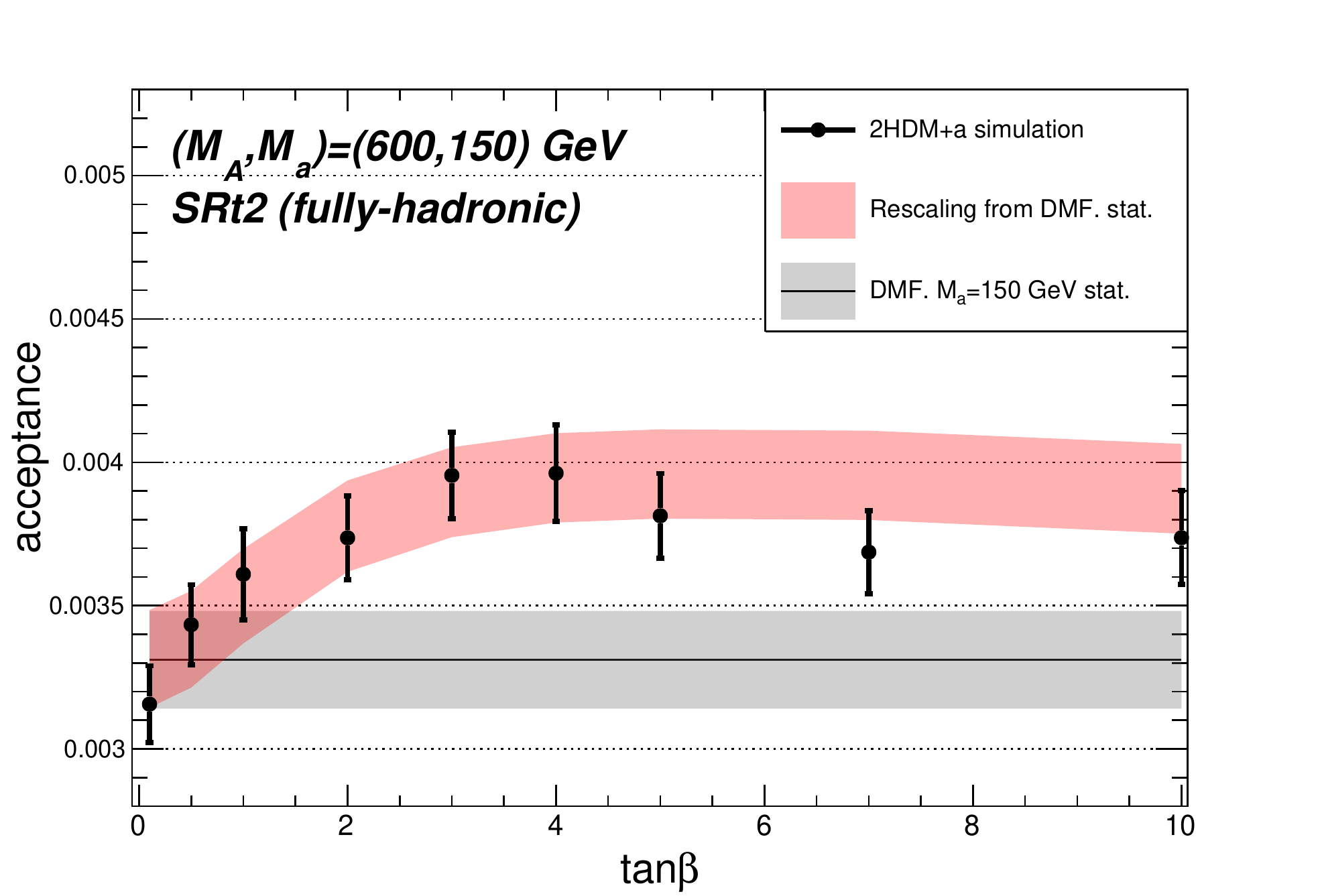}
\vspace{2mm}
\caption{Validation of~\eqref{eq:recast} in the case of the one-lepton (left panel) and the hadronic~(right panel)  final state arising from the $t \bar t+\MET$ signature. The direct \hdma calculations are indicated by the black dots and error bars, while the grey and red bands indicate the result in the DMF pseudoscalar model and the prediction obtained using the rescaling formula.  In the left (right) panel, the  chosen parameters are $\ma = 150 \, {\rm GeV}$, $\mH= \mA = \mHc = 600 \, {\rm GeV}$ and $\tan \beta = 1$ ($\sin\theta=0.35$).}
\label{DMHF:pof}
\end{figure}

Further validations of~\eqref{eq:recast} are presented in Figure~\ref{DMHF:pof}. In this figure we apply the rescaling formula to the case of the  one-lepton~\cite{Aaboud:2017aeu}~(left panel)   and the hadronic~\cite{Aaboud:2017rzf}~(right panel)  final state in $t \bar t+\MET$ production. The direct \hdma calculations are indicated by the black dots and error bars, while the grey and red  bands illustrate the result in the DMF pseudoscalar model and the prediction obtained using~\eqref{eq:recast} including statistical uncertainties.  In the left (right) panel, we have employed $\ma = 150 \, {\rm GeV}$, $\mH= \mA = \mHc = 600 \, {\rm GeV}$ and $\tan \beta = 1$ ($\sin\theta=0.35$). The rescaled results describe the $\sin \theta$ and $\tan \beta$ dependence of the \hdma result well with uncertainties.  We finally add that the formula~\eqref{eq:recast} has also been successfully tested in the case that $|\mA - \ma| \simeq 50 \, {\rm GeV}$, in which case the interference between the $a$ and $A$ contributions is relevant.  

%%%%%%%%%%%%%%%%%%%%%%%%%%%%%%%%%%%%%%%%%%%%%%%%%%%%%%%%%%%%%%%%%%%%
%%%%%%%%%%%%%%%%%%%%%%%%%%%%%%%%%%%%%%%%%%%%%%%%%%%%%%%%%%%%%%%%%%%%
%%%%%%%%%%%%%%%%%%%%%%%%%%%%%%%%%%%%%%%%%%%%%%%%%%%%%%%%%%%%%%%%%%%%

\section{Distributions of the $\bm{t \bar t + \MET}$ signal in the 2HDM+s model}
\label{app:ttMETscalar}

\begin{figure}
\centering
\includegraphics[width=.475\textwidth]{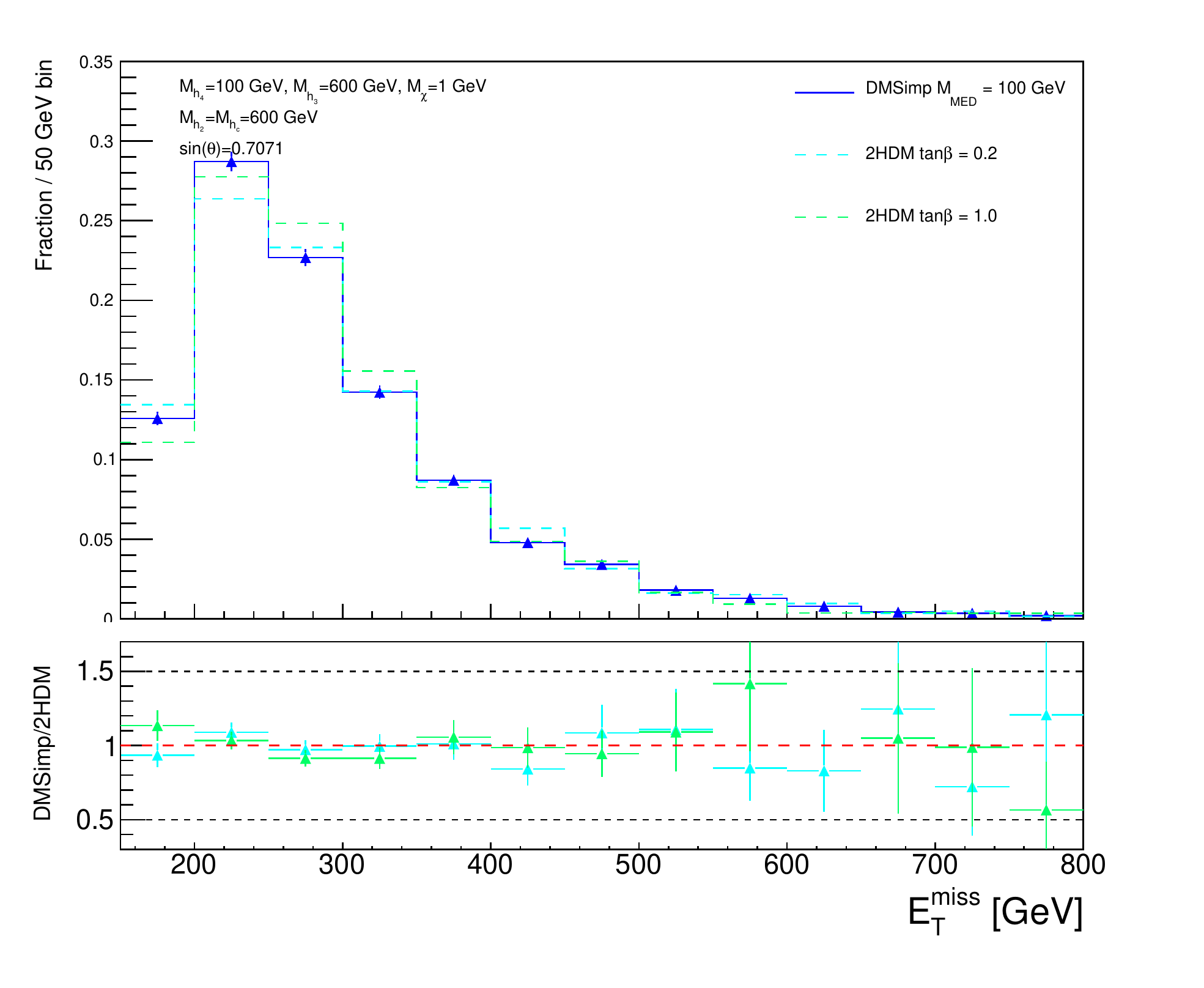} \quad 
\includegraphics[width=.475\textwidth]{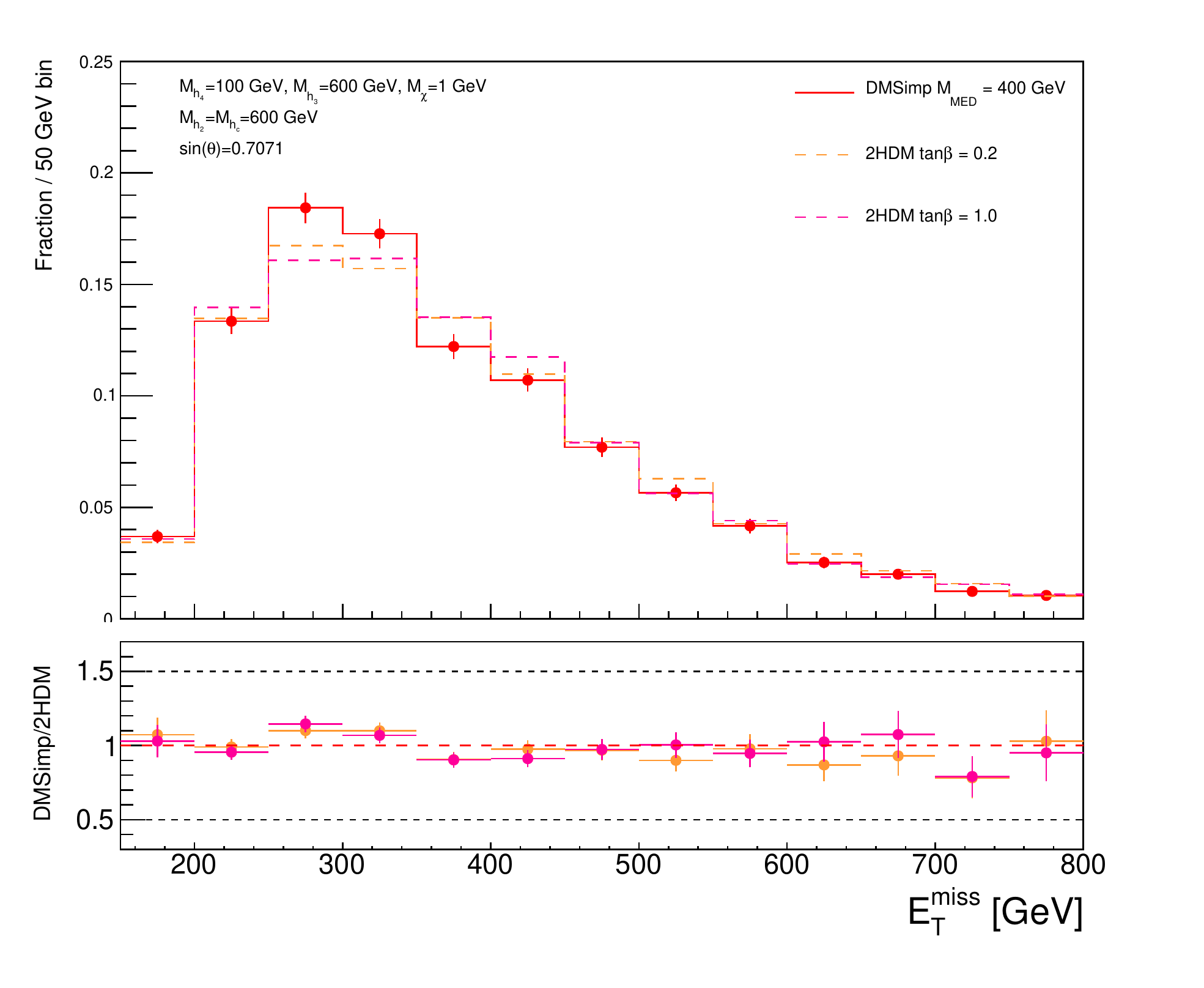}
\vspace{2mm}
\caption{ Normalised $\MET$ distributions for $t \bar t + \MET$ production in the 2HDM+s model. The black curves correspond to the prediction of the DMF scalar   model, while the coloured predictions illustrate the results in the 2HDM+s  model. In both panels the choices $M_H = M_A = M_{H^\pm} = 600 \, {\rm GeV}$, $\sin\theta=1/\sqrt{2}=0.7071$, $m_\chi = 1 \, {\rm GeV}$ and either $\tan \beta = 0.2$ or $\tan \beta = 1$ have been made. The mass of the scalar mediator $s$ is set to $M_s = 100 \, {\rm GeV}$ ($M_s = 400 \, {\rm GeV}$) in the left (right) panel.}
\label{fig:ttMETscalar}
\end{figure}

In this appendix, we present a concise study of the kinematic features of the  $t \bar t + \MET$ signature in the 2HDM+s model~\cite{Bell:2016ekl,Bell:2017rgi}, focusing like in the case of the  \hdma model on the $\MET$ spectrum (for the related studies in the \hdma model see~Section~\ref{sec:nonresonant}).  In Figure~\ref{fig:ttMETscalar}, we display normalised $\MET$ spectra corresponding to either  the 2HDM+s model~(coloured curves) or the DMF scalar   model~(black curves). In both panels the chosen parameters are $M_H = M_A = M_{H^\pm} = 600 \, {\rm GeV}$, $\sin\theta=1/\sqrt{2}=0.7071$, $m_\chi = 1 \, {\rm GeV}$ and $\tan \beta = 0.2$ as well as $\tan \beta = 1$, while in the left (right) plot we have employed $M_s = 100 \, {\rm GeV}$ ($M_s = 400 \, {\rm GeV}$). We observe that  the shape of the 2HDM+s distributions always  resembles  the corresponding one of the DMF model within uncertainties. This feature is expected because in the considered parameter benchmarks the 2HDM non-SM spin-0 states are significantly heavier than the additional scalar mediator, and thus decouple. We add that by studying simple observables like $\MET$ it is in principle not possible to disentangle DM-scalar from DM-pseudoscalar interactions. Angular correlations between  two visible final state objects in $X+\MET$ events can, however, serve such a purpose~\cite{Haisch:2016gry,Cotta:2012nj,Haisch:2013fla,Crivellin:2015wva}. 

%%%%%%%%%%%%%%%%%%%%%%%%%%%%%%%%%%%%%%%%%%%%%%%%%%%%%%%%%%%%%%%%%%%%
%%%%%%%%%%%%%%%%%%%%%%%%%%%%%%%%%%%%%%%%%%%%%%%%%%%%%%%%%%%%%%%%%%%%
%%%%%%%%%%%%%%%%%%%%%%%%%%%%%%%%%%%%%%%%%%%%%%%%%%%%%%%%%%%%%%%%%%%%

\section{Details on the MC generation}
\label{app:mcgeneration}

The studies presented in this white paper are all based on MC simulations that use an {\tt UFO} implementation of the type-II \hdma model as described in Section~\ref{sec:modeldescription}. The~{\tt UFO} implementation  called {\tt Pseudoscalar\_2HDM}~\cite{hdmaUFO} has been provided by the authors of~\cite{Bauer:2017ota} and a brief introduction to its basic usage can be found in {\tt README.txt}~\cite{hdmaREADME}. Below we give some details on   the signal generation of the $t \bar t t \bar t$ channel discussed in Section~\ref{sec:fourtopmain} as well as  the $h \, (b \bar b) + \MET$ and $Z \, (\ell^+ \ell^-)+ \MET$ signatures  considered in Section~\ref{sec:sensitivitystudies}.

\subsection{Four-top signature}

In Section~\ref{sec:fourtopmain} we have presented a study of the $t \bar t t \bar t$ channel, splitting the  total four-top production cross section into three different contributions: one that only includes the SM graphs ($|{\rm SM}|^2$), another one that is due to new-physics only ($|{\rm NP}|^2$) and finally a contribution that accounts for both the SM and the \hdma diagrams~($|{\rm SM} + {\rm NP}|^2$). In Table~\ref{tab-dmhf-4tops} we provide the {\tt MadGraph5\_aMC@NLO} syntax that has been used to generate the three different samples using the Pseudoscalar\_2HDM {\tt UFO}. 

\subsection{Mono-Higgs signature}

\begin{table}[t!]
\begin{tabular}{ccm{50mm}}
\toprule
 {\tt MadGraph5\_aMC@NLO} syntax & Legend symbol & Details \\\midrule
\verb| p p > t t~ t t~ / a z h1 QED<=2|& $|{\rm SM}+{\rm NP}|^2$ & Four-top
production including both SM and NP contributions and their
interference. \\\midrule
\verb| p p > t t~ t t~ / a z h1 QCD<=2|& $|{\rm NP}|^2$ & Four-top
production from NP processes, including interference terms among
$H,A,a$. \\\midrule
\verb| p p > t t~ t t~ / a z h1 QED<=0|& $|{\rm SM}|^2$ & Four-top 
production within the SM.\\
\bottomrule
\end{tabular}
\vspace{4mm} 
\caption{{\tt MadGraph5\_aMC@NLO} syntax used to obtain the different curves shown in the two panels in Figure~\ref{fig:4top}.}
\label{tab-dmhf-4tops}
\end{table}

The $h \, (b \bar b) + \MET$ sensitivity study presented in Section~\ref{sec:sensi_monohbb} is based on the generation of the signal using the Pseudoscalar\_2HDM {\tt UFO} together with {\tt MadGraph5\_aMC@NLO} and {\tt NNPDF23\_lo\_as\_0130} parton distribution functions (PDFs)~\cite{Ball:2012cx}. The {\tt MadGraph5\_aMC@NLO}  syntax used to generate the $gg$-fusion contribution reads 
\begin{eqnarray}
&& \verb| import model Pseudoscalar_2HDM | \hspace{2.25cm} \nonumber \\
&& \verb| g g > h1 xd xd~ [QCD] | \nonumber 
\end{eqnarray}
where  $\verb|[QCD]|$ indicates that one deals with a loop-induced process. 

The $b \bar b$-fusion channel is instead generated with 
\begin{eqnarray}
&& \verb| import model Pseudoscalar_2HDM-bbMET_5FS  | \nonumber \\
&& \verb| p p > h1 xd xd~  | \nonumber 
\end{eqnarray}
The first command loads the version of Pseudoscalar\_2HDM {\tt UFO}  corresponding to the five-flavour scheme (5FS). In this only the top quark is massive while the bottom quark is massless and thus 
can appear as a parton in the colliding protons. Both the top and bottom Yukawa coupling  are, however, non-zero.

\subsection[Mono-$Z$ signature]{Mono-$\bm{Z}$ signature}
%\subsection{Mono-$\bm{Z}$ signature}

The event samples that have been employed in the $Z \, (\ell^+ \ell^-) + \MET$ sensitivity study (see Section~\ref{sec:sensi_monozll}) have been obtained using the Pseudoscalar\_2HDM {\tt UFO} in conjunction with {\tt MadGraph5\_aMC@NLO}, {\tt NNPDF30\_lo\_as\_0130} PDFs~\cite{Ball:2014uwa} and  {\tt PYTHIA~8.2}~\cite{Sjostrand:2014zea} for parton showering.  The {\tt MadGraph5\_aMC@NLO}  syntax that should be used to generate the $gg$-fusion process including the decay to charge leptons is 
\begin{eqnarray}
&& \verb| import model Pseudoscalar_2HDM | \hspace{2.25cm} \nonumber \\
&& \verb| g g > l+ l- xd xd~ / h1 [QCD] | \nonumber 
\end{eqnarray}
with $l = e$ or $\mu$.  To increase the efficiency of the event generation, Feynman diagrams with an intermediate $s$-channel SM Higgs boson have been explicitly rejected using the  {\tt MadGraph5\_aMC@NLO} syntax~~$\verb|/ h1|$. 

In the case of the $b \bar b$-fusion channel the commands 
\begin{eqnarray}
&& \verb| import model Pseudoscalar_2HDM-bbMET_5FS  | \nonumber \\
&& \verb| p p > l+ l- xd xd~ / h1 a  | \nonumber 
\end{eqnarray}
should instead been used.  By loading the Pseudoscalar\_2HDM-bbMET\_5FS  {\tt UFO}  the calculation is again performed in the 5FS and the  {\tt MadGraph5\_aMC@NLO} syntax~~$\verb|/ h1 a|$ removes contributions with an intermediate Higgs or photon.  

\begin{figure}[t!]
\centering
\includegraphics[width=0.475\textwidth]{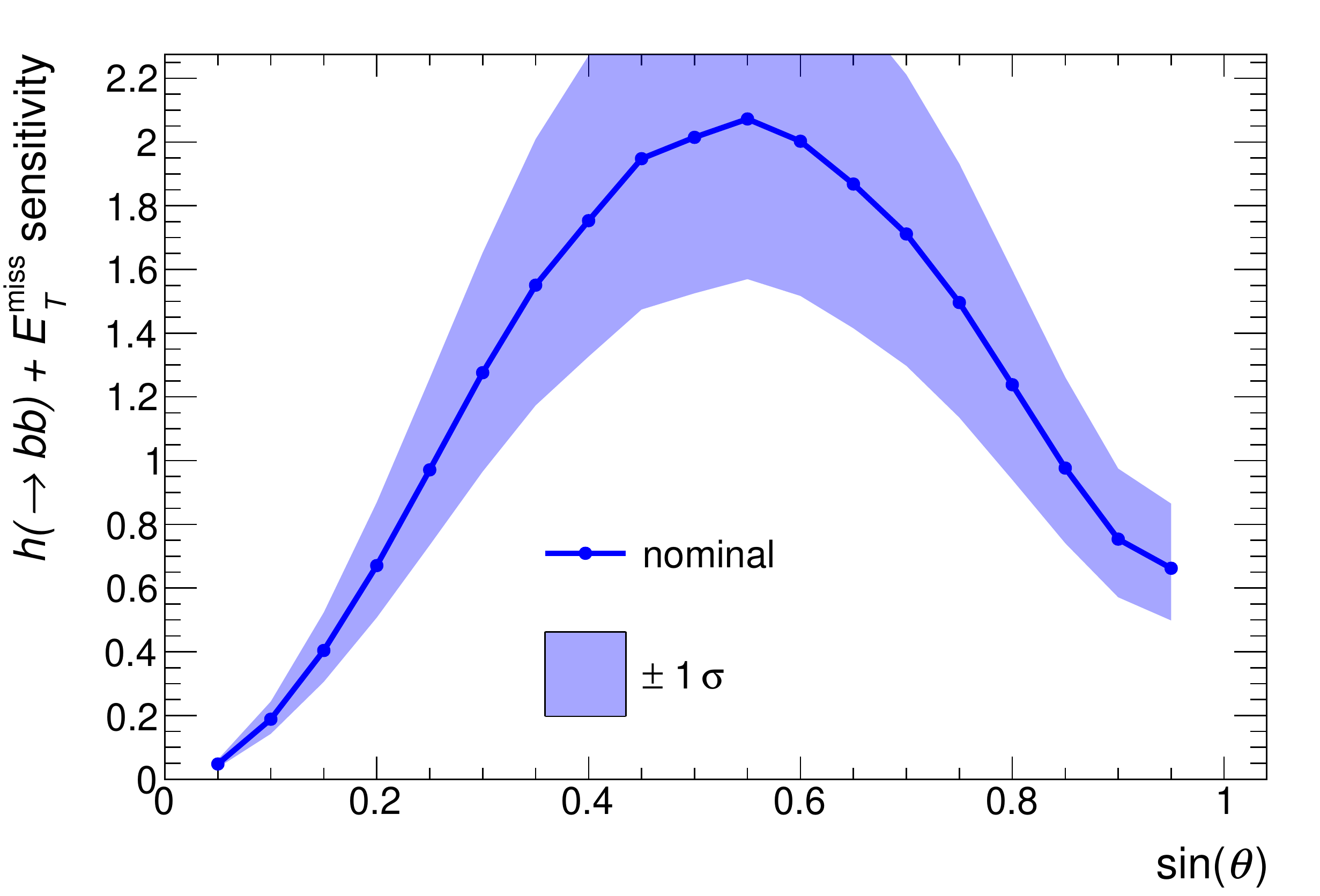} \quad 
\includegraphics[width=0.475\textwidth]{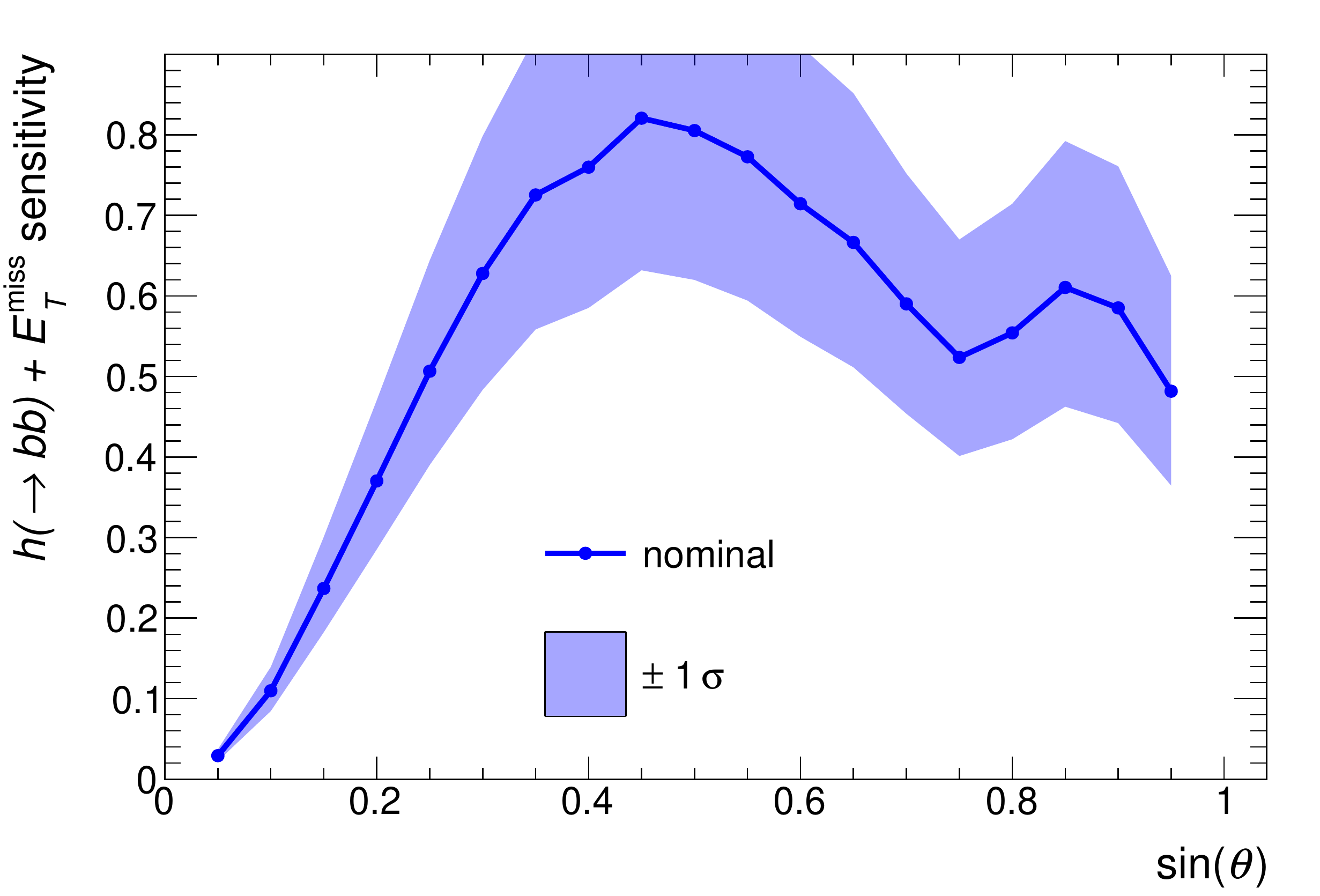}
\vspace{4mm}
\caption{Estimated sensitivities of  the $h \, (b \bar b)+\MET$ channel as a function of $\sin \theta$. The left (right) panel shows our results for $\mH = \mA = \mHc = 600 \, {\rm GeV}$ and $\ma = 200 \, {\rm GeV}$ ($\mH = \mA = \mHc = 1000 \, {\rm GeV}$ and $\ma = 350 \, {\rm GeV}$).  The remaining parameters are set equal to \eqref{eq:benchmark}. The sensitivities (points and curves), defined as the sum \eqref{eq:monoHbb_sensi} over \met bins, as well as the uncertainty on the sensitivities (shaded bands)  are based on the limits and uncertainties given in~\cite{Aaboud:2017yqz}. Bins with no content have a negligible sensitivity. }
\label{fig:monoHbb_appendix1}
\end{figure}

\begin{figure}[t!]
\centering
\includegraphics[width=0.7\textwidth]{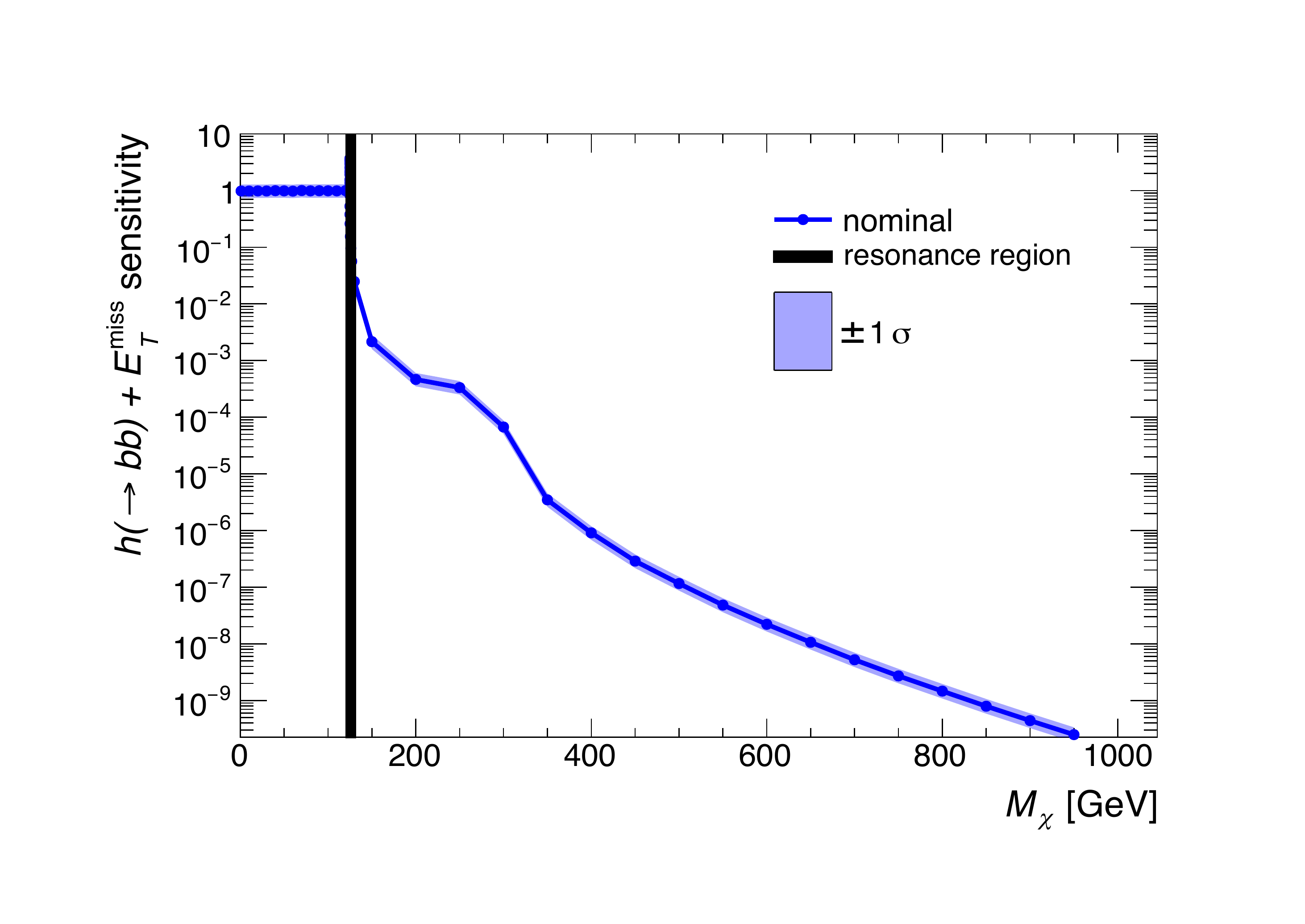}
\vspace{-2mm}
\caption{Estimated sensitivities of  the $h \, (b \bar b)+\MET$ channel as a function of $m_\chi$. The  results shown correspond to $\mH = \mA = \mHc = 600 \, {\rm GeV}$, $\ma = 250 \, {\rm GeV}$ and the parameter choices made in \eqref{eq:benchmark}. The colour coding resembles that used in Figure~\ref{fig:monoHbb_appendix1}. It is recommended to stay at least $1 \, {\rm GeV}$ away from the region where $\ma = 2 m_\chi$ to avoid numerical effects from the resonance in the generation. }
\label{fig:monoHbb_appendix2}
\end{figure}

\subsection{Heavy flavour signatures}

In case of the generation of heavy flavour signatures, one must consider which flavour scheme to employ between the 5FS and four-flavour scheme (4FS). The 5FS is preferred to keep the model predictions simpler to generate and to use, while a 4FS scheme may be more suitable if the mediator $a$ is not much heavier that the bottom quark (i.e.~$m_a \lesssim 20 \, {\rm GeV}$) and the $\MET$ requirement imposed in the search is not large. 

%%%%%%%%%%%%%%%%%%%%%%%%%%%%%%%%%%%%%%%%%%%%%%%%%%%%%%%%%%%%%%%%%%%%
%%%%%%%%%%%%%%%%%%%%%%%%%%%%%%%%%%%%%%%%%%%%%%%%%%%%%%%%%%%%%%%%%%%%
%%%%%%%%%%%%%%%%%%%%%%%%%%%%%%%%%%%%%%%%%%%%%%%%%%%%%%%%%%%%%%%%%%%%

\section{Details on the mono-Higgs sensitivity study}
\label{app:extramonoh}

In this appendix, we  show additional results  of our sensitivity study of the $h \, (b \bar b)+\MET$ signature presented in Section~\ref{sec:sensi_monohbb}. Figure~\ref{fig:monoHbb_appendix1} displays the estimated sensitivities for the two $\sin \theta$ benchmarks recommended in Section~\ref{sec:scans},~i.e.~$\mH = \mA = \mHc = 600 \, {\rm GeV}$ and $\ma = 200 \, {\rm GeV}$ (left panel) and $\mH = \mA = \mHc = 1000 \, {\rm GeV}$ and $\ma = 350 \, {\rm GeV}$~(right panel). From the panels, one observes that for the benchmark value $\sin \theta = 0.35$ introduced in \eqref{eq:benchmark} the sensitivity of the mono-Higgs is enhanced compared to the choices  $\sin \theta = 0.15$ and  $\sin \theta = 0.7$ employed in Figure~\ref{fig:svar}. In Figure~\ref{fig:monoHbb_appendix2} we furthermore  plot the expected sensitivity of the $h \, (b \bar b)+\MET$ search as a function of the DM mass $m_\chi$. The  results shown correspond to $\mH = \mA = \mHc = 600 \, {\rm GeV}$, $\ma = 250 \, {\rm GeV}$ and the choices made in \eqref{eq:benchmark}. With the present data set mono-Higgs searches have already sensitivity to DM masses up to around $m_\chi \simeq \ma/2 = 125 \, {\rm GeV}$. Recalling that the  observed DM relic density can be obtained for $m_\chi \simeq \ma/2$ ~(see~Section~\ref{sec:relic}), the latter finding implies that the LHC can already test  \hdma scenarios that predict the correct value of $\Omega h^2$. 

\newpage 

%\bibliography{2HDMa-whitepaper}
%\bibliographystyle{JHEP}

\providecommand{\href}[2]{#2}\begingroup\raggedright\endgroup

\end{document}